%% file: ANA-HION-2021-09-PAPER.tex
\documentclass[cernpreprint, atlasdraft=false, texlive=2023, USenglish, texmf, orcidlogo]{atlasdoc}
 
\usepackage{atlaspackage}
\usepackage{atlasbiblatex}
 
\usepackage{atlasphysics, atlasheavyion, atlasprocess, atlasparticle, atlasjetetmiss, atlasmisc}

\graphicspath{{logos/}{figures/}}
 
\usepackage{ANA-HION-2021-09-PAPER-defs}

\AtlasTitle{Measurement of substructure-dependent jet suppression in Pb+Pb collisions at 5.02 TeV with the ATLAS detector}

\AtlasAbstract{
The ATLAS detector at the Large Hadron Collider has been used to measure jet substructure modification and suppression in Pb+Pb collisions at a nucleon--nucleon center-of-mass energy $\sqrt{s_{_\mathrm{NN}}}=5.02~\mathrm{TeV}$ in comparison with $pp$ collisions at $\sqrt{s}=5.02~\mathrm{TeV}$. The Pb+Pb data, collected in 2018, have an integrated luminosity of $1.72~\mathrm{nb^{-1}}$, while the $pp$ data, collected in 2017, have an integrated luminosity of $260~\mathrm{pb}^{-1}$. Jets used in this analysis are clustered using the anti-$k_{t}$ algorithm with a radius parameter $R=0.4$.
The jet constituents, defined by both tracking and calorimeter information, are used to determine the angular scale $r_\mathrm{g}$ of the first hard splitting inside the jet by reclustering them using the Cambridge--Aachen algorithm and employing the soft-drop grooming technique.
The nuclear modification factor, $R_\mathrm{AA}$, used to characterize jet suppression in Pb+Pb collisions, is presented differentially in $r_\mathrm{g}$, jet transverse momentum, and in intervals of collision centrality. The $R_\mathrm{AA}$ value is observed to depend significantly on jet $r_\mathrm{g}$. Jets produced with the largest measured $r_\mathrm{g}$ are found to be twice as suppressed as those with the smallest $r_\mathrm{g}$ in central Pb+Pb collisions. The $R_\mathrm{AA}$ values do not exhibit a strong variation with jet $p_\mathrm{T}$ in any of the $r_\mathrm{g}$ intervals.  The $r_\mathrm{g}$ and $p_\mathrm{T}$ dependence of jet $R_\mathrm{AA}$ is qualitatively consistent with a picture of jet quenching arising from coherence and provides the most direct evidence in support of this approach.
}

\AtlasRefCode{HION-2021-09}
 
\PreprintIdNumber{CERN-EP-2022-182}

\AtlasJournalRef{Phys. Rev. C 107 (2023) 054909}
\AtlasDOI{10.1103/PhysRevC.107.054909}

\hypersetup{pdftitle={ATLAS document},pdfauthor={The ATLAS Collaboration}}
 
\begin{document}
 
\maketitle

\section{Introduction}
 
In ultrarelativistic heavy-ion collisions a hot and dense state of matter known as a quark-gluon plasma (QGP) is produced. The temperatures attained in the fireball are such that quarks and gluons are no longer confined within their parent hadrons~\cite{STAR:2005gfr,PHENIX:2004vcz}. Producing a QGP in the laboratory provides unique opportunities to study quantum chromodynamics (QCD). Experimental results from the Relativistic Heavy Ion Collider (RHIC) and the Large Hadron Collider (LHC) comparing proton--proton (\pp) and heavy-ion collisions have led to the conclusion that the QGP expands hydrodynamically.
The remarkably small viscosity needed to describe the data suggests that this matter exhibits strong-coupling behavior at large distance scales~\cite{Gyulassy:2004zy}. In contrast, at short distance scales the interactions among quarks and gluons are expected to become weaker because QCD exhibits asymptotic freedom.  It is not fully understood how such hydrodynamic behavior arises from the interactions among elementary quarks and gluons~\cite{Busza:2018rrf}. To address this challenge, the dynamics of the QGP must be characterized over a range of length scales.
 
Jets are highly collimated sprays of particles resulting from the point-like, hard scattering of quarks and gluons, collectively referred to as partons. Jets are produced at the earliest stages of a high-energy heavy-ion collision and become attenuated while passing through the plasma, a phenomenon known as jet quenching~\cite{Gyulassy:1990ye,Bjorken:1982tu}. The energy loss suffered by a jet and the modification of its radiation pattern provide information about the plasma's microscopic structure. Scattered partons in the vacuum, studied using \pp collisions, experience a cascade of radiative processes and thus the modification of the jet evolution process in the QGP is inherently a multiscale problem.
 
Techniques that enable the identification of the hardest splitting in the parton shower of a jet, and more generally the substructure of the jet and its overall radiation pattern, have been developed~\cite{Krohn:2009th,Dasgupta:2013ihk}.
In theoretical calculations the \SoftDrop grooming procedure~\cite{Larkoski:2014wba}, a generalization of the modified mass-drop tagger~\cite{Dasgupta:2013ihk}, shows less sensitivity to the effects of initial-state radiation, multiparton interactions, and nonperturbative contributions to substructure observables. Since these effects typically contribute large uncertainties, their mitigation allows more rigorous comparisons with experimental data~\cite{STDM-2017-04, STDM-2017-33}. The potential of such techniques to characterize jet substructure in heavy-ion collisions has also been recognized~\cite{Mehtar-Tani:2016aco,Chien:2016led,Andrews:2018jcm}, and a first series of measurements have been performed indicating a mild modification of jet-substructure variables in \PbPb collisions relative to \pp results ~\cite{CMS-HIN-16-006,CMS-HIN-16-024,ALICE:2019ykw,ALICE:2021obz,STAR:2021kjt}. However these initial studies have not yet elucidated the relationship between the observed modifications and the energy loss of the jet.
 
Measurements of jet quenching in heavy-ion collisions performed at RHIC and the LHC~\cite{Foka:2016zdb,Cunqueiro:2021wls} typically fall into one of two categories: quantifying the total energy loss or studying the modification of the jet's radiation pattern.
The rate of jets produced in central heavy-ion collisions at a given jet transverse momentum, \ptjet, is observed to be approximately a factor of two lower than in \pp collisions~\cite{HION-2011-02,HION-2013-06,HION-2017-10,CMS-HIN-13-005, CMS-HIN-18-014}. Additionally, back-to-back dijet and photon/$\Zboson$--jet pairs are observed to have poorer \pt balance in \PbPb collisions than in \pp collisions~\cite{HION-2010-02,HION-2018-06,CMS-HIN-19-013, CMS-HIN-15-013}. These measurements provide information about the total jet energy loss and its parametric dependence on the in-medium path length and flavor of the initiating parton. They also constrain how energy loss may fluctuate on a jet-by-jet basis.
Another class of measurements which studies the fragmentation properties of quenched jets, typically through the particle-momentum distributions within jets~\cite{HION-2017-04,HION-2017-08,CMS-HIN-12-013,CMS-HIN-16-014,CMS-HIN-16-020}, is less sensitive to the energy loss but can directly assess how the radiation pattern is modified. This class includes the current heavy-ion jet-substructure measurements~\cite{CMS-HIN-16-006,CMS-HIN-16-024,ALICE:2019ykw,STAR:2020ejj,ALICE:2021obz} because the distributions of substructure quantities are normalized per jet. Currently missing from the experimental program are analyses that correlate a jet's energy loss with its radiation pattern.
 
A common theoretical framework for describing jet quenching is the coherence picture in which interference effects, crucial for determining the structure of the vacuum parton shower~\cite{Dokshitzer:1987nm}, become disrupted by the medium, resulting in energy loss in the form of additional radiation~\cite{Baier:1996kr}.  Recent studies have also shown the emergence of a critical angle in the first hard splitting of a jet, above which the jet loses energy incoherently, i.e. as multiple color entities~\cite{Casalderrey-Solana:2012evi, Milhano:2017nzm,Caucal:2019uvr, Caucal:2020uic, JETSCAPE:2022jer}. This approach leads to the expectation that a wide jet with two prongs, each of which may act as a separate emitter of radiation, will lose more energy than a narrow jet which acts as a single source of radiation. More generally, jets with different substructure are expected to experience different amounts of energy loss according to the degree to which the medium resolves the jet and induces decoherence of its radiation pattern~\cite{Mehtar-Tani:2016aco}.
 
This paper describes a measurement of jet suppression, a measure of the energy loss, as a function of the observed substructure of the jet. The measurements are based on $1.72~\inb$ of \PbPb collision data collected in 2018 and $260~\ipb$ of \pp collision data collected in 2017, both at $\sqn = 5.02~\TeV$. The jets are reconstructed from energy deposits in the ATLAS calorimeters with the \antikt algorithm with a radius parameter of $R=0.4$~\cite{Cacciari:2008gp} and by applying a subtraction procedure that removes the underlying event contribution to the jet kinematics on an event-by-event basis~\cite{HION-2011-02}. The \SoftDrop grooming procedure is applied with parameters $\beta=0$ and $z_{\mathrm{cut}}=0.2$ to jet constituents formed from charged-particle tracks and calorimeter energy deposits.
The substructure is quantified through \rg,
the angle subtended by the subjets chosen using the \SoftDrop procedure to tag the first hard splitting of a jet.
Jets which fail the \SoftDrop requirement are considered as single-prong jets and assigned $\rg=0$.
The per-event jet yields as a function of \ptjet and \rg are measured in four centrality intervals, which characterize the overlap of the colliding nuclei in \PbPb collisions.
The analogous differential cross-sections are measured in \pp collisions, and the suppression is quantified through the nuclear modification factor, \RAA. The cross-sections, yields, and \RAA are reported for jets with $\ptjet > 158~\GeV$ and $0 \leq \rg < 0.4$.
The measured cross-sections and yields are unfolded to a particle-level phase space.
 
\section{ATLAS detector}
 
\newcommand{\AtlasCoordFootnote}{
ATLAS uses a right-handed coordinate system with its origin at the nominal interaction point (IP)
in the centre of the detector and the \(z\)-axis along the beam pipe.
The \(x\)-axis points from the IP to the centre of the LHC ring,
and the \(y\)-axis points upwards.
Cylindrical coordinates \((r,\phi)\) are used in the transverse plane,
\(\phi\) being the azimuthal angle around the \(z\)-axis.
The pseudorapidity is defined in terms of the polar angle \(\theta\) as \(\eta = -\ln \tan(\theta/2)\).
The rapidity is defined as \(y\) = \(0.5 \ln[(E+p_{z})/(E-p_{z})]\) where \(E\) and \(p_{z}\) are the energy and \(z\)-component of the momentum along the beam direction, respectively.
Angular distance is measured in units of \(\Delta R \equiv \sqrt{(\Delta\eta)^{2} + (\Delta\phi)^{2}}\).}
 
The ATLAS detector~\cite{PERF-2007-01} at the LHC is a multipurpose particle detector with a forward--backward symmetric cylindrical geometry and a near-\(4\pi\) coverage in solid angle.\footnote{\AtlasCoordFootnote}
It consists of an inner tracking detector surrounded by a thin superconducting solenoid, electromagnetic and hadron calorimeters, and a muon spectrometer.
The inner-detector system is immersed in a \SI{2}{\tesla} axial magnetic field and provides charged-particle tracking within \(|\eta| < 2.5\).
The high-granularity silicon pixel detector covers the vertex region and typically provides four measurements per track, with the first hit typically being in the insertable B-layer installed before Run~2~\cite{ATLAS-TDR-2010-19,PIX-2018-001}.
It is followed by the silicon microstrip tracker (SCT) which usually provides eight measurements per track.
These silicon detectors are complemented by the transition radiation tracker, a drift-tube-based detector, which surrounds the SCT and has coverage up to \(|\eta| = 2.0\).
 
The calorimeter system covers the pseudorapidity range \(|\eta| < 4.9\).
Within the region \(|\eta|< 3.2\), electromagnetic calorimetry is provided by barrel and
endcap high-granularity lead/liquid-argon (LAr) calorimeters,
with an additional thin LAr presampler covering \(|\eta| < 1.8\)
to correct for energy loss in material upstream of the calorimeters.
Hadronic calorimetry is provided by the steel/scintillator-tile calorimeter,
segmented into three barrel structures within \(|\eta| < 1.7\), and two copper/LAr hadronic endcap calorimeters.
The solid angle coverage is completed with copper/LAr and tungsten/LAr calorimeter modules (FCal),
covering the forward regions of $3.1 < |\eta| < 4.9$.
The zero-degree calorimeters (ZDC) consist of layers of alternating quartz rods and tungsten plates and are located symmetrically at $z=\pm140$~m and cover $|\eta| \geq 8.3$. In \PbPb\ collisions, the ZDCs primarily measure `spectator' neutrons: neutrons that do not interact hadronically when the incident nuclei collide.
 
Events of interest are selected for recording and offline analysis
by the first-level (L1) trigger system implemented in custom hardware,
followed by selections made by algorithms implemented in software in the high-level trigger~\cite{TRIG-2016-01}.
 
An extensive software suite~\cite{ATL-SOFT-PUB-2021-001} is used in data simulation, in reconstruction and analysis of real and simulated events, in detector operations, and in the trigger and data acquisition systems of the experiment. The events used in this analysis were selected by a jet trigger~\cite{TRIG-2016-01}. The L1 trigger identified jet candidates by applying a sliding-window algorithm and selecting events passing a \pt threshold of 30~\GeV. These events were then passed to the high-level jet trigger, which uses a jet reconstruction and background subtraction procedure similar to that used in the offline analysis and requires a minimum \ptjet of $100~\GeV$ for \antikt $R=0.4$ jets. The jet trigger was fully efficient for the \ptjet range considered in this measurement.

\section{Data samples and event selection}
The data used in this analysis are taken from the 2018 \PbPb\ and 2017 \pp\ runs, both at  $\sqn = 5.02~\TeV$.
The average number of collisions per bunch-crossing (pileup) was 0.003 in the \PbPb\ data and ranges from 1.4 to 4.4 in the \pp\ data depending on the data-taking run. Events from both the \PbPb and \pp collisions were collected using the jet triggers with the same \pt threshold as described above.
The events are required to have been collected during stable beam conditions, and to satisfy detector and data-quality requirements~\cite{DAPR-2018-01}.
Both the \pp\ and \PbPb\ events are required to have at least one reconstructed primary vertex, and the \PbPb events are also required to satisfy nominal offline \MB \PbPb collision criteria, identical to those used in Ref.~\cite{HION-2019-11}. This additional requirement, based on a combination of the total transverse energy \ET measured in the FCal, denoted by \sumETFcal, and the total energy deposited in the ZDC identifies and rejects $0.2\%$ of the selected events as pile-up events.
 
The degree of geometric overlap of the colliding Pb nuclei is characterized by `event centrality', and each collision event is assigned to a centrality interval.
The procedure used to experimentally define centrality intervals in this paper follows that used in other measurements of \PbPb collisions performed by the ATLAS Collaboration~\cite{HION-2018-06}. The distribution of the total transverse energy deposited in the FCal is divided into successive quantiles of the total inelastic cross-section for \PbPb collisions, with the highest \sumETFcal events corresponding, on average, to the events with the largest geometric overlap between the colliding nuclei.
The centrality intervals used are: 0--10\% (highest \sumETFcal), 10--30\%, 30--50\%, and 50--80\%.  A Glauber model analysis of the \sumETFcal distribution is used to evaluate the mean nuclear thickness function, \avgTAA, for different centrality intervals~\cite{Glauber,Loizides:2014vua}.
The \avgTAA values and uncertainties, which are discussed in Ref.~\cite{HION-2018-25}, are listed in Table~\ref{tab:lumi} for each centrality selection considered in this measurement.
 
\begin{table}[ht]
\centering
\caption{\label{tab:lumi} The  \avgTAA\ values and uncertainties for the centrality selections used in this measurement.}
\begin{tabular}{|c|c|c|}
\hline
Centrality selection & $\avgTAA \pm \delta \avgTAA[1/mb] $ & $\delta \avgTAA/\avgTAA[\%]$ \\
\hline
0--10\%  & 23.21 $\pm$ 0.06 & 0.26 \\
10--30\% & 11.57 $\pm$ 0.12 & 1.04 \\
30--50\% & 3.92  $\pm$ 0.11 & 2.73 \\
50--80\% & 0.73  $\pm$ 0.04 & 5.67 \\
\hline
\end{tabular}
\end{table}
 
The Monte Carlo (MC) simulations used in this analysis are multijet events modeled with the \PYTHIA~{8}~\cite{Sjostrand:2014zea} MC generator with leading-order matrix elements for dijet production which were matched to the parton shower. The A14 set of tuned parameters~\cite{ATL-PHYS-PUB-2014-021} and the \NNPDF[23lo] parton distribution functions~\cite{Ball:2012cx} were used. The events generated by \PYTHIA[8] were passed through a \GEANT-based simulation~\cite{AGOSTINELLI2003250,SOFT-2010-01} of the ATLAS detector and its response. By reproducing the detector conditions in the data-taking run, this allows the simulations to account for underlying-event (UE) effects in the jet reconstruction.
For \PbPb collisions, the simulated dijet events were overlaid with \MB \PbPb data, and this `overlay' sample was reweighted on an event-by-event basis to obtain the same centrality distribution as in the jet-triggered \PbPb data sample.
The simulated events are digitized and reconstructed in the same way as data events.

\section{Analysis procedure}
 
\subsection{Jet reconstruction}
\label{sec:jetreco}
 
The jet reconstruction procedures follow those used by the ATLAS Collaboration for previous jet measurements in \PbPb\ collisions~\cite{HION-2017-10,ATLAS:2021ktw}.
Jets are reconstructed using the \antikt\ algorithm~\cite{Cacciari:2008gp} implemented in the FastJet software package~\cite{Fastjet}. In both \pp\ and \PbPb\ collisions, jets with $R = 0.4$ are formed by clustering calorimeter towers with $\Delta\eta \times\Delta\phi=0.1\times\pi/32$ in the pseudorapidity range $|\eta|< 4.9$ over the full azimuthal range.
The energies in the towers are obtained by summing the energies of calorimeter cells at the electromagnetic energy scale~\cite{PERF-2011-03} within the tower boundaries. The UE contribution to each tower is subtracted on an event-by-event basis by estimating the average local energy density, $\rho(\eta,\phi)$. This estimate is obtained by measuring the average energy in calorimeter towers as a function of $\eta$ and including the azimuthal modulation due to harmonic flow characterized by the second-, third-, and fourth-order flow coefficients~\cite{HION-2016-06}. The $\phi$-dependent background subtraction is performed only for jet reconstruction in \PbPb events, as the event activity is observed to be aziumthally symmetric in \MB \pp\ data\cite{HION-2015-09}. An iterative procedure is applied to prevent the jets from biasing the determination of $\rho$~\cite{ATLAS:2021ktw}.
Following the UE subtraction, new estimates of $\rho$ are obtained by excluding the contributions coming from towers within jets from the average energy density and flow parameterization.
Additionally, the UE is also corrected for $\eta$- and $\phi$-dependent nonuniformities in the detector response to soft particles by applying correction factors derived in \MB \PbPb\ data. In \pp\ collisions, the same background subtraction procedure is used, but without the nonuniformity correction and no harmonic modulation of the UE.

Following the UE subtraction, $\eta$- and $\pT$-dependent multiplicative calibration factors derived in MC simulations are applied to the jet four-momentum vector to account for the noncompensating hadronic response of the calorimeter~\cite{ATLAS-CONF-2015-016}.
An additional correction based on in situ studies of jets recoiling against photons, \Zboson\ bosons, and jets in other regions of the calorimeter is applied to account for known differences between data and the MC sample used to derive the calibration~\cite{HION-2018-06}. This calibration is followed by a cross-calibration which relates the jet energy scale (JES) of jets reconstructed using the procedure outlined in this section to the JES in 13~\TeV\ \pp\ collisions~\cite{ATLAS-CONF-2015-016}. The performance of the jet reconstruction in \pp and \PbPb collisions has been studied in detail as described in Ref.~\cite{HION-2017-10}.
The jet transverse momentum, denoted by \ptjet, is defined as the calibrated \pt of $R = 0.4$ jets formed from calorimeter towers. Calibrations are applied to the \pt\ of the jet and not individually to the constituent calorimeter towers in this procedure. Fully calibrated jets having $\ptjet > 158~\GeV$ and $|y|<2.1$ are used in this measurement.
The \ptjet binning was designed to match previous jet measurements by ATLAS~\cite{HION-2011-02,HION-2013-06,HION-2017-10} and have uniformly spaced bins on the logarithmic scale.
MC-truth jets are defined in simulated events at particle level by applying the \antikt\ algorithm with $R = 0.4$ to stable particles with a proper lifetime greater than 30~ps in the generator's event record, but excluding muons and neutrinos, which do not leave significant energy deposits in the calorimeter.
 
\subsection{Jet constituents}
 
The granularity of the ATLAS calorimeter towers, which are used as constituents in the initial jet reconstruction, is too coarse to resolve collimated prongs within a jet that have small opening angles comparable to the expected resolving power of the QGP
~\cite{Caucal:2019uvr,Caucal:2020uic}. In order to improve the angular and energy resolution of the jet constituents, new constituents are defined by combining the superior angular resolution of the tracker with calorimeter information. These objects, referred to as track-calo-clusters (TCCs), are built using the energies of topological cell clusters (topo-clusters)~\cite{PERF-2014-07} and the spatial coordinates of charged-particle tracks~\cite{ATL-PHYS-PUB-2017-015}. TCCs were used in previous ATLAS studies to improve the resolution of jet-substructure variables used in $\Wboson$/$\Zboson$--boson tagging~\cite{ATL-PHYS-PUB-2020-008}. They were also used in combination with particle-flow~\cite{PERF-2015-09} objects to optimize jet reconstruction in \pp collisions~\cite{JETM-2018-06}. In this measurement, TCCs are reconstructed in a manner similar to that in Ref.~\cite{ATL-PHYS-PUB-2017-015} but with some adjustments to account for the UE in \PbPb collisions as detailed below.
 
Topo-clusters, used to define the energy scale of the TCCs, are built from calorimeter cells using a noise-suppression algorithm~\cite{PERF-2014-07} and have been used extensively in jet-substructure measurements in \pp collisions~\cite{STDM-2017-04, STDM-2017-33}. In \PbPb collisions, a $\phi$-modulated background subtraction, similar to the UE subtraction procedure employed for jet reconstruction (Section~\ref{sec:jetreco}), is applied to the calorimeter cells before using them to build topo-clusters. Charged-particle tracks are reconstructed from hits in the inner detector using standard optimization algorithms for \PbPb collisions as detailed in Refs.~\cite{PERF-2015-08,HION-2011-03}.
Reconstructed tracks used as seeds in the TCC reconstruction are required to have $\pt > 3~\GeV$ and to also meet several criteria designed to select primary charged particles~\cite{HION-2017-04}.
Tracks in an event are extrapolated to the calorimeter and are matched to one or more topo-clusters, depending on the topo-cluster size and the track extrapolation uncertainty.
 
Following the track and topo-cluster matching, the procedure to determine TCC four-momentum vectors is described briefly below and is detailed further in Ref.~\cite{ATL-PHYS-PUB-2017-015}. For an isolated match between a track from the selected primary vertex and a topo-cluster,
the topo-cluster energy and the track direction are used to form a single TCC. In the case of topo-clusters which do not match any tracks, the topo-cluster's four-momentum vector is directly used to create a TCC. Conversely, instances where tracks are not matched to any topo-clusters, observed to occur in less than $1\%$ of the cases, are treated analogously by creating a TCC using the track's four-momentum vector.
In cases where multiple tracks are matched to one or more topo-clusters, the TCC algorithm is designed to create exactly one TCC object per track originating from the primary vertex. Such multiple matches are handled by using the track's angular coordinates, while splitting the topo-cluster energy between the corresponding TCC objects to account for energy sharing between the different matches.
 
The topo-cluster energy sharing procedure for tracks matched to clusters makes use of three general concepts and is detailed in Ref.~\cite{ATL-PHYS-PUB-2017-015}. First, each cluster matched to the seed track should contribute to the resulting TCC object in proportion to its share of the total \pt of all matched clusters. Second, if a cluster is matched to multiple TCC objects, its contribution to a given TCC should be proportional to the share of the seed track of the total \pt of all tracks matched to the cluster. Third, the proportion of topo-cluster energy assigned to each of those tracks should itself be weighted by that cluster’s share of the total energy of all clusters matching the track. The $3~\GeV$ \pt threshold for charged tracks in the TCC reconstruction is optimized to prevent low-\pt UE tracks from sharing the energy of high-\pt topo-clusters.
 
All TCCs having $\pt > 4~\GeV$ are used in this measurement. Jets considered in this measurement with $\ptjet >158~\GeV$ and $|y|$ < 2.1 have an average of 6--7 associated TCCs above the $4~\GeV$ threshold in \pp and \PbPb collisions. TCCs built by matching tracks to topo-clusters account for ${\approx}77\%$ of the total number of TCCs within a jet in \pp collisions. Due to the lower tracking efficiency in \PbPb compared to \pp collisions, the fraction of TCCs within a jet built by matching tracks to topo-clusters drops to ${\approx}59\%$ in central \PbPb collisions. The remaining TCCs are mostly cases where topo-clusters are not matched to any tracks.
 
\subsection{Jet Grooming and \rg}
 
Jet grooming algorithms are used to isolate the prongs of a jet that correspond to a hard splitting in the evolution of the parton shower by systematically removing the contributions of soft wide-angle radiation. The \SoftDrop grooming procedure~\cite{Larkoski:2014wba}, a generalization of the modified mass-drop tagger~\cite{Dasgupta:2013ihk} is employed in this measurement to distinguish perturbative radiation from soft, mostly nonperturbative, components of the jet~\cite{CMS-HIN-16-006,CMS-HIN-16-024,ALICE:2019ykw,STAR:2020ejj}. The angular distance \rg between the subjets that tag the first hard splitting of a jet is measured in this analysis.
 
The procedure starts with reclustering the constituents of a jet using the \ca (\CamKt) algorithm~\cite{Wobisch:1998wt, Dokshitzer:1997in} to form a clustering tree with a purely angle-ordered structure.
In this measurement, jets are initially reconstructed from calorimeter towers with the \antikt algorithm as described in Section~\ref{sec:jetreco}, and the calorimeter's response is well-defined~\cite{ATLAS:2021ktw}. Calorimeter-tower jets satisfying the kinematic thresholds listed in Section~\ref{sec:jetreco} are selected to be reclustered using the TCCs associated with the jet.
TCCs having \pt above $4~\GeV$ and within $\Delta R < 0.4$ of a reconstructed jet's axis are utilized in the jet reclustering process of the \SoftDrop procedure. The $4~\GeV$ \pt threshold for the TCCs, applied only at the detector level, is used to further reduce the influence of the UE on the measurement of \rg.
When reclustering MC-truth particle-level jets, all stable charged and neutral particles within the jets are used in the \SoftDrop reclustering procedure.
The \CamKt algorithm clusters the nearest constituents first, working outwards towards the wider-angle constituents, independent of their \pt. The \CamKt reclustered jet is then recursively declustered into two `subjets' and it is determined whether the subjets $sj_1$ and $sj_2$ with transverse momenta $\pt^{sj_{1}}$ and $\pt^{sj_{2}}$ satisfy the \SoftDrop condition:
 
\begin{equation}
\label{eq:sd}
\frac{\mathrm{min}(\pt^{sj_{1}},\pt^{sj_{2}})}{\pt^{sj_{1}}+\pt^{sj_{2}}} > z_\mathrm{cut}\left(\frac{\Delta R_{12}}{R}\right)^\beta,
\end{equation}
where $z_\mathrm{cut}$ and $\beta$ are algorithm parameters, \textit{R} is the radius parameter value used to cluster the jets ($R = 0.4$), and $\Delta R_{12} = \sqrt{(\Delta y_{12})^2+(\Delta \phi_{12})^2}$ is the distance in $y$--$\phi$ between the subjets, where $y$ is the rapidity. If the \SoftDrop condition in Eq.~(\ref{eq:sd}) is not satisfied, then the subjet with the smaller \pt is dropped and the procedure is iterated on the remaining subjets. If the \SoftDrop condition is satisfied at any point, the algorithm terminates and \rg is set to the $\Delta R_{12}$ value between the subjets $sj_1$ and $sj_2$.
 
If the condition described in Eq.~(\ref{eq:sd}) cannot be satisfied, $\rg = 0$ is assigned to the jet to indicate that the \SoftDrop procedure cannot find two subjets that satisfy Eq.~(\ref{eq:sd}).
Jets with $\rg = 0$ are primarily those with a single TCC object or MC-truth particle carrying a significant fraction of the jet energy.
For all $z_\mathrm{cut} > 0$ and $\beta > 0$, the procedure is infrared- and collinear-safe, while it is infrared-safe when $\beta = 0$~\cite{Larkoski:2014wba}.
The \SoftDrop parameter values $z_\mathrm{cut} = 0.2$ and $\beta = 0$, corresponding to an angle-independent grooming setting, are used for the results shown. These values are chosen to maximize the efficiency of selecting the two hardest prongs of the jet against the competing effect of jets failing the \SoftDrop grooming with a high $z_\mathrm{cut}$ value~\cite{Mulligan:2020tim}. At the particle level, less than $4\%$ of the jets fail the \SoftDrop condition when using $z_\mathrm{cut} = 0.2$ and $\beta = 0$, thereby minimizing a possible bias from looking at different populations of jets in \pp and \PbPb collisions.
The \rg value of a jet measured using $z_\mathrm{cut} = 0.2$ and $\beta = 0$ is also observed to be insensitive to lowering the $4~\GeV$ threshold for the TCCs.
 
\begin{figure}[t]
\centering
\includegraphics[width=1\textwidth]{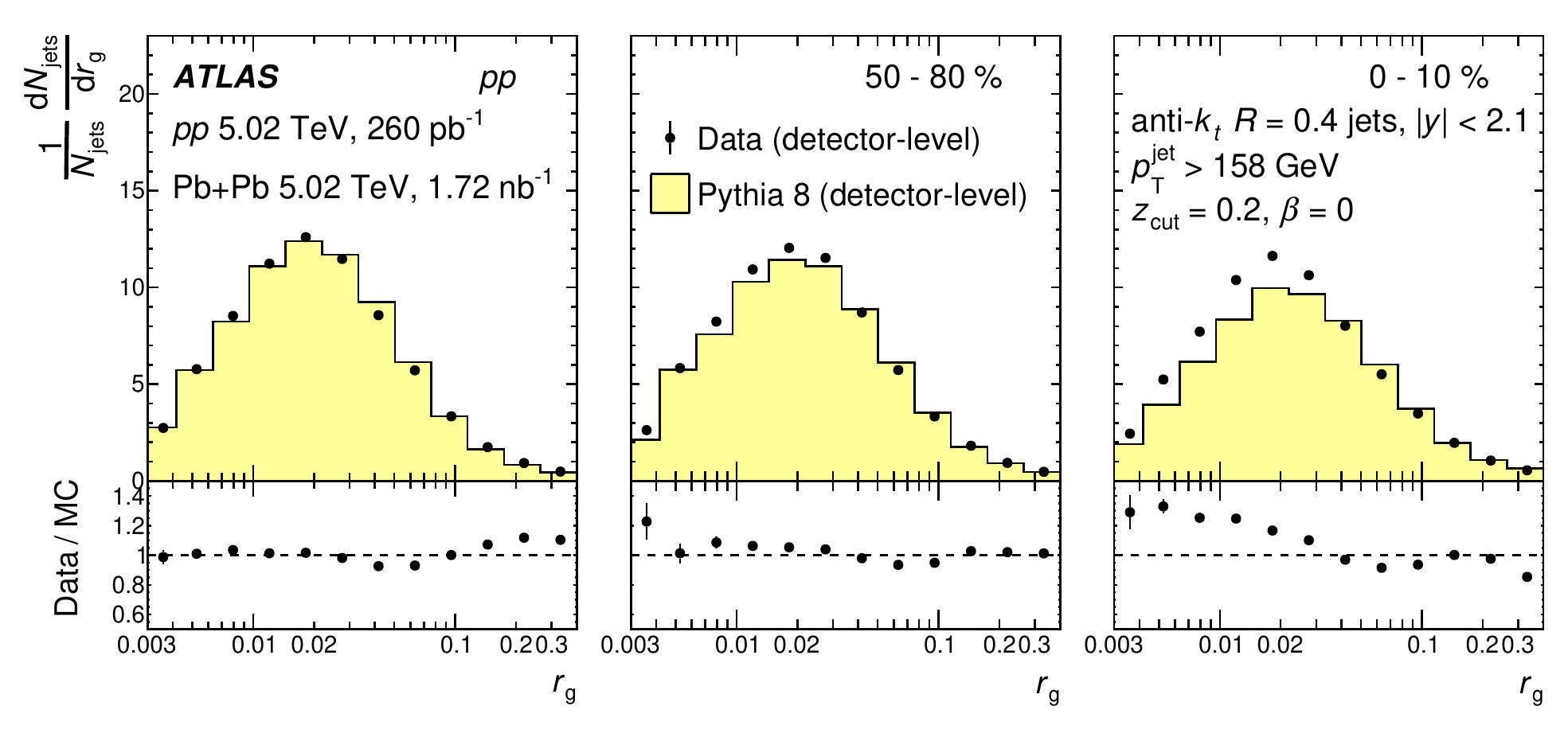}\\
\caption{Top: Self-normalized \rg distributions measured using detector-level TCCs from \pp data at $\sqrt{s} = 5.02~\TeV$ (left) and \PbPb data at $\sqn = 5.02~\TeV$ (second and third panels) for different centrality intervals, compared with the detector-level distributions predicted by the \PYTHIA[8] generator. The error bars represent statistical uncertainties. Bottom: Data-to-MC ratios of \rg distributions at detector level. The legend applies to all the panels.}
\label{fig:jetrg_reco}
\end{figure}

Distributions of \rg measured using detector-level TCCs from \pp and \PbPb data are shown in Figure~\ref{fig:jetrg_reco}, and are compared with the detector-level distributions from the \PYTHIA[8] MC events. The detector-level \rg distributions measured using TCCs show good agreement between data and MC events in \pp and peripheral \PbPb collisions. The \rg distribution differences between data and simulation for central \PbPb collisions can plausibly be attributed to effects from jet quenching, not modeled in simulation, and are discussed in detail in Section~\ref{sec:result}.
No systematic uncertainties are shown in Figure~\ref{fig:jetrg_reco} because it is only intended to demonstrate the performance of TCCs in measuring \rg.
 
\subsection{Unfolding}
 
The jet yields are unfolded in both jet \ptjet and \rg to the particle level in order to account for nonuniformities in detector performance and bin migration due to the finite resolution in jet energy, TCC energy, and TCC position, including migration in and out of the measurement phase space. The measurement phase space in this analysis corresponds to $\ptjet>158~\GeV$, $|y|<2.1$, and $0<\rg<0.4$ for the jets.
 
At the particle level, the \SoftDrop procedure is applied to MC-truth jets using all stable charged and neutral particles within the jet as constituents.
Utilizing the RooUnfold package~\cite{Adye:2011gm},
the yields of groomed jets are unfolded using a two-dimensional iterative Bayesian unfolding algorithm~\cite{DAgostini:1995} simultaneously in \ptjet and \rg,
while the inclusive jet yields are unfolded using a one-dimensional unfolding in \ptjet.
The migration matrices are reweighted simultaneously in MC-truth \ptjet and \rg by the ratios of the corresponding distributions in data to those in the reconstructed MC sample, separately for \pp collisions and each centrality interval in \PbPb collisions.
The reweighting factors are smoothed as a function of \rg and \ptjet, and their effect on bin values of the migration matrices is typically less than 25\%.
The uncertainty from the reweighting procedure is estimated by unfolding with and without the reweighting factors and is detailed in Section~\ref{sec:systematics}.
The effects of inefficiencies and misidentified jets in the measurement are accounted for in the analysis by applying an efficiency correction and subtraction procedure, respectively, through the unfolding procedure.
In the iterative Bayesian unfolding method implemented in RooUnfold, the inefficiencies are handled by applying a multiplicative factor to each MC-truth bin of the migration matrix.
The inefficiencies in this measurement arise mainly from jets migrating out of the measurement phase space at the detector level because of energy resolution effects and from jets that fail the \SoftDrop grooming procedure.
MC-truth jets having significantly lower \pt than $158~\GeV$ (the \ptjet threshold) that migrate into the measurement phase space at the detector level are treated as misidentified jets. The rate of such misidentified jets is observed to be negligible in this measurement, and are accounted for by treating them as an additive background.
The effect of jets reconstructed entirely from overlapping UE particles in this measurement is insignificant.
 
The migration matrices used in the unfolding procedure are binned uniformly on a logarithmic scale and have 13 bins along \rg from $0$ to $0.4$ and 16 bins along \ptjet from $158~\GeV$ to $1~\TeV$. Additionally, in the MC-truth phase space, \ptjet has one overflow bin and six underflow bins down to $80~\GeV$ for a smooth inefficiency correction, and \rg has one overflow and one underflow bin in the migration matrix. There is no significant migration between centrality intervals in this measurement.
A separate migration matrix is generated for each centrality interval in \PbPb collisions, using the \PYTHIA[8] sample overlaid with \MB \PbPb data and for \pp collisions, using the \PYTHIA[8] sample.
The number of iterations in the unfolding procedure is optimized independently for different centrality intervals to balance minimizing any bias from the unfolding against fluctuations in MC samples. The optimal number of iterations ranges from four to six depending on the centrality interval. The remnant unfolding bias evaluated in MC samples is better than $1\%$.
 
The differential jet cross-section in \pp collisions is defined as:
 
\begin{equation*}
\frac{\dif^2\sigma_\mathrm{jet}}{\dif\ptjet \dif\rg} = \frac{1}{L_\mathrm{int}} \frac{N_\mathrm{jets}}{\Delta\ptjet\Delta\rg},
\end{equation*}
 
where $N_\mathrm{jets}$ is the unfolded jet yield, $\Delta\ptjet$ and $\Delta\rg$ are the widths of the \ptjet and \rg bins, respectively, and $L_\mathrm{int}$ is the integrated luminosity.
 
The jet yield per \MB event jet yield in \PbPb collisions, referred to as the per-event yield for simplicity, is defined as:
\begin{equation*}
\frac{1}{N_\mathrm{evt}^\mathrm{cent}}\frac{\dif^2 N_\mathrm{jet}^\mathrm{cent}}{\dif\ptjet \dif\rg} = \frac{1}{N_\mathrm{evt}^\mathrm{cent}} \frac{N^\mathrm{cent}_\mathrm{jets}}{\Delta\ptjet\Delta\rg},
\end{equation*}
 
where $N_\mathrm{jet}^\mathrm{cent}$ is the unfolded jet yield in \PbPb collisions within intervals of \ptjet and \rg for a given centrality (denoted by the superscript `cent'), and $N_\mathrm{evt}^\mathrm{cent}$ is the the number of \MB events in the centrality interval.
 
Modification of the jet yield in \PbPb collisions in a given centrality interval relative to the jet yield in \pp collisions is quantified using the nuclear modification factor, $\RAA$, defined as:
 
\begin{equation}
\RAA(\ptjet, \rg) = \left. \frac{1}{N_\mathrm{evt}^\mathrm{cent}}\frac{1}{\avgTAA}\frac{\dif^2 N_\mathrm{jet}^\mathrm{cent}}{\dif\ptjet \dif\rg} \middle/ \frac{\dif^2\sigma_\mathrm{jet}}{\dif\ptjet \dif\rg}\right.,
\end{equation}
where \avgTAA is the mean nuclear thickness function, presented in Table~\ref{tab:lumi}. At a given \ptjet, \rg, and centrality interval, $\RAA < 1$ indicates a suppression of jet production in \PbPb collisions relative to that in \pp collisions.
 
\section{Systematic uncertainties}
\label{sec:systematics}
 
Several sources of systematic uncertainty are considered for this analysis. The systematic uncertainties in the jet cross-sections in \pp collisions and per-event yields in \PbPb collisions arise from the jet energy scale and resolution, the constituent TCC energy scale and resolution, the unfolding procedure, integrated \pp luminosity, and $\avgTAA$ (\PbPb only). The jet- and TCC-related uncertainties are evaluated in the analysis by modifying the migration matrix for each uncertainty contribution. The deviations from the nominal unfolded result for each variation are then combined in quadrature to calculate the total contribution.
 
\begin{figure}[ht]
\centering
\includegraphics[width=1\textwidth]{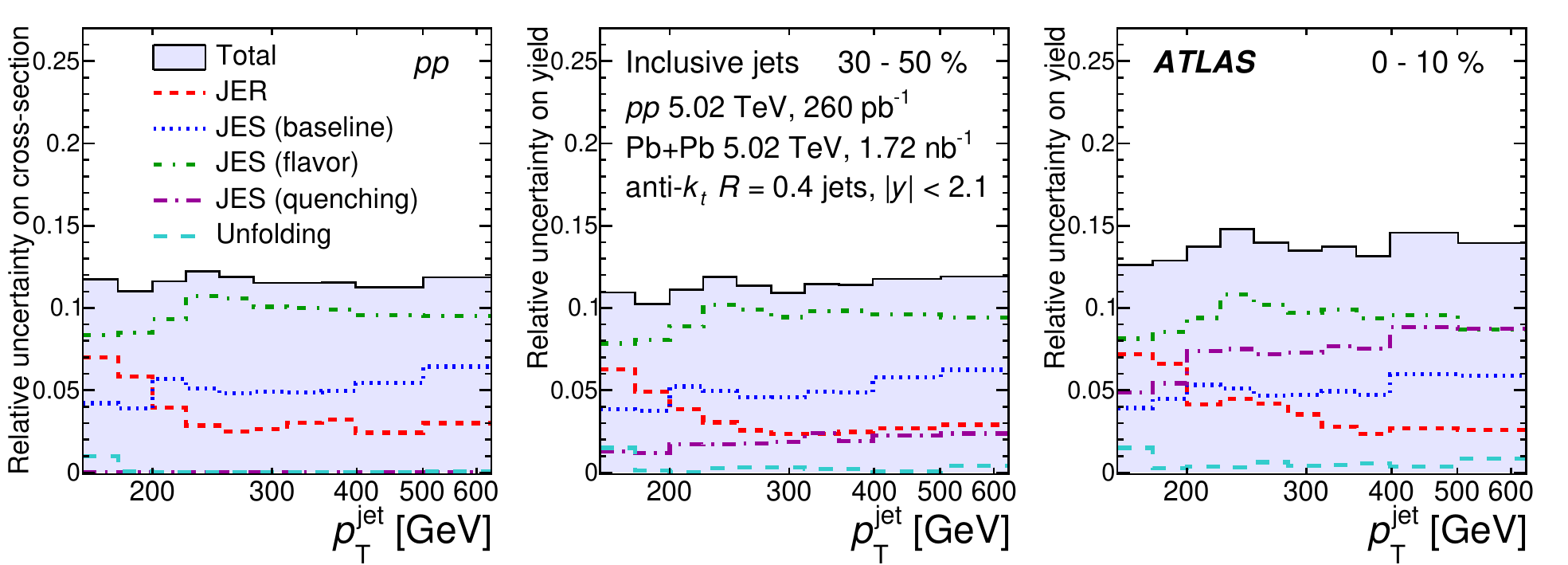}
\caption{The relative systematic uncertainties in inclusive \ptjet cross-sections in \pp collisions at $\sqrt{s} = 5.02~\TeV$ (left) and in per-event jet yields in \PbPb collisions at $\sqn = 5.02~\TeV$ for different event centralities (middle and right panels). The legend applies to all of the panels.}
\label{fig:jetpt_syst}
\end{figure}

\begin{figure}[ht]
\centering
\includegraphics[width=1\textwidth]{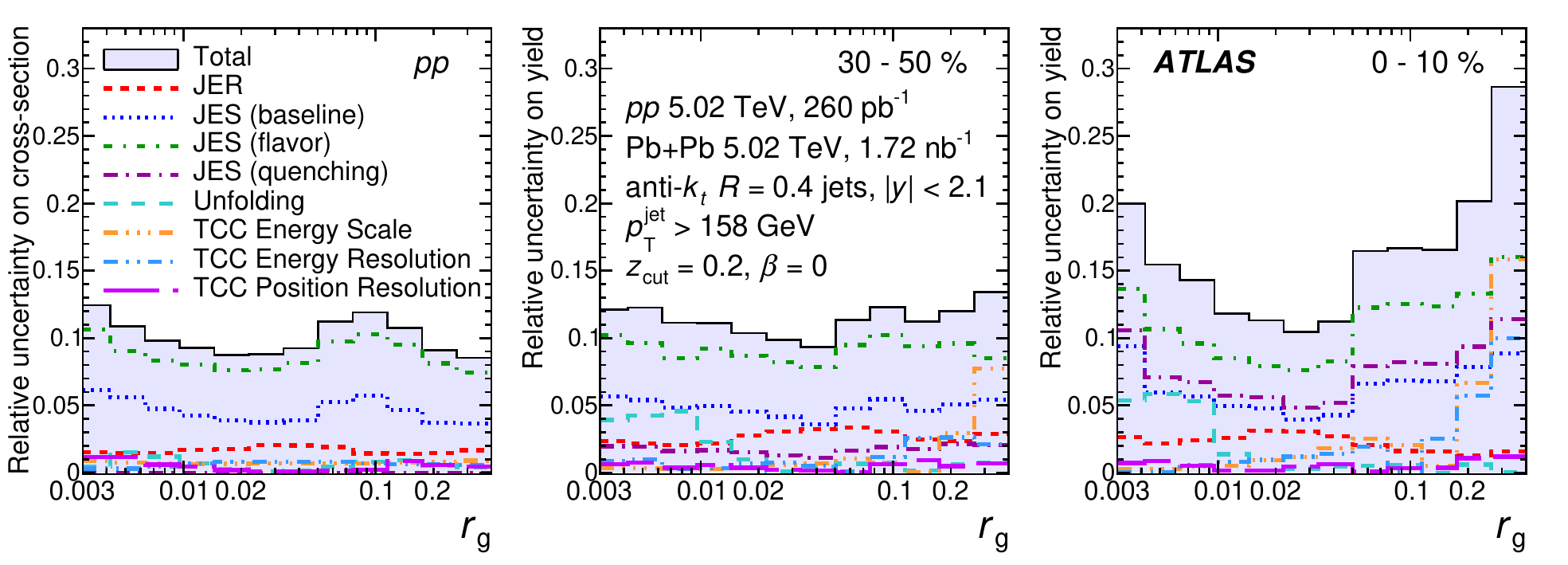}
\caption{The relative systematic uncertainties in inclusive \rg cross-sections in \pp collisions at $\sqrt{s} = 5.02~\TeV$ (left) and in per-event jet yields in \PbPb collisions at $\sqn = 5.02~\TeV$ for different event centralities (middle and right panels) shown for \SoftDrop parameters $z_\mathrm{cut} = 0.2$ and $\beta=0$. The legend applies to all of the panels.}
\label{fig:thetag_syst}
\end{figure}

\begin{figure}[t!]
\centering
\includegraphics[width=1\textwidth]{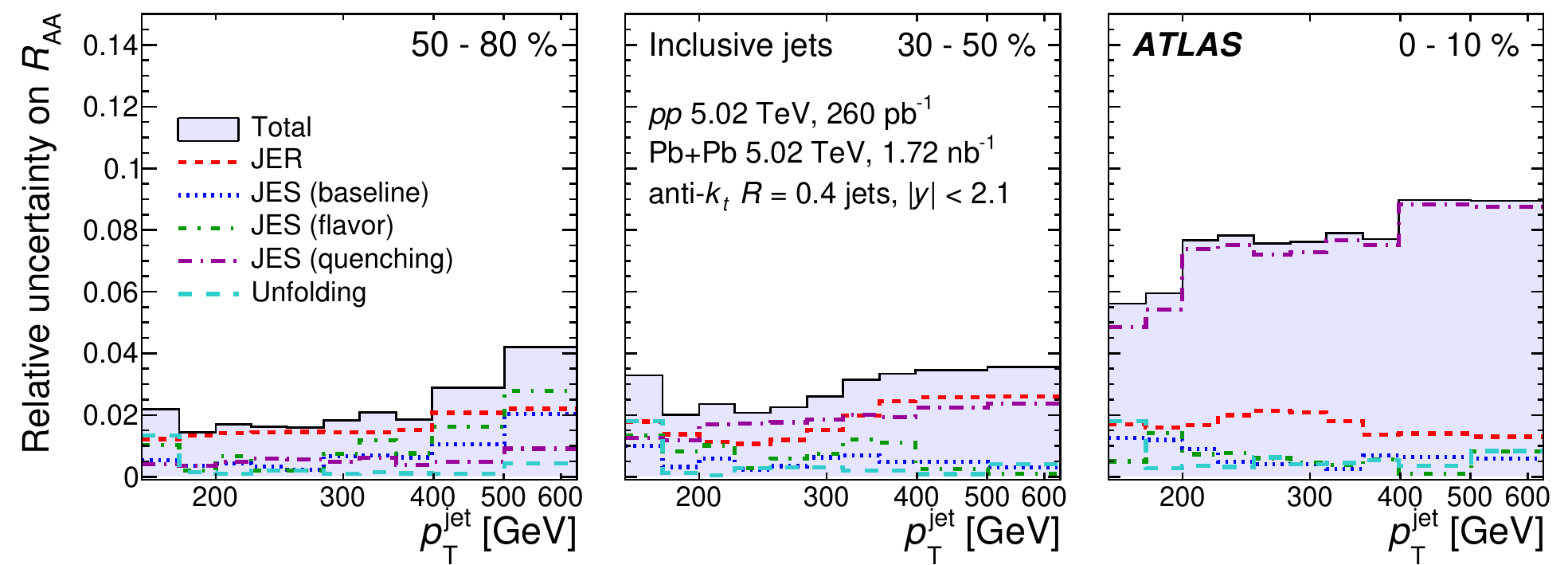}
\caption{The relative systematic uncertainties in the $R_\mathrm{AA}$ measurements as a function of \ptjet in different centrality intervals of \PbPb collisions at $\sqn = 5.02~\TeV$. The legend applies to all of the panels.}
\label{fig:jetraa_syst}
\end{figure}

\begin{figure}[t!]
\centering
\includegraphics[width=1\textwidth]{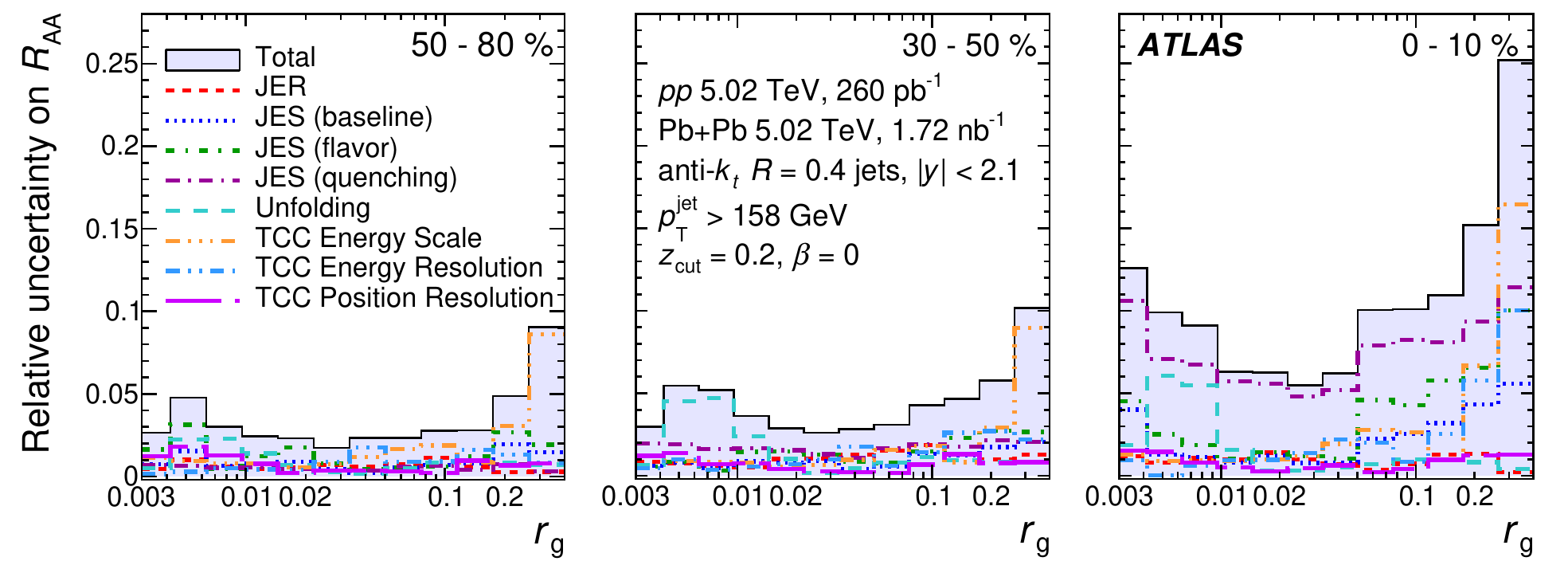}
\caption{The relative systematic uncertainties in the $R_\mathrm{AA}$ measurements as a function of \rg in different centrality intervals of \PbPb collisions at $\sqn = 5.02~\TeV$ shown for \SoftDrop parameters $z_\mathrm{cut} = 0.2$ and $\beta=0$. The legend applies to all of the panels.}
\label{fig:thetagraa_syst}
\end{figure}

The systematic uncertainty in the JES has three parts.
The first (labeled `JES baseline' in Figures~\ref{fig:jetpt_syst}--\ref{fig:thetagraa_syst}) is a centrality-independent component that is determined from in situ studies of the calorimeter response to jets reconstructed with the procedure used in 13~\TeV\ \pp\ collisions~\cite{PERF-2011-03,PERF-2016-04}, with an additional component which accounts for the relative energy-scale difference between the jet reconstruction procedures used in 13~\TeV \pp collisions and in this measurement~\cite{ATLAS-CONF-2015-016}.
The potential mismodeling of relative abundances of jets initiated by quarks and gluons in the MC events, and of the calorimetric response to quark and gluon jets, are accounted for by the second component (labeled `JES flavor' in Figures~\ref{fig:jetpt_syst}--\ref{fig:thetagraa_syst}).
The third, centrality-dependent component (applicable in \PbPb collisions only and labeled `JES quenching' in Figures~\ref{fig:jetpt_syst}--\ref{fig:thetagraa_syst}) accounts for modifications of the parton shower due to quenching~\cite{HION-2017-04, CMS-HIN-16-020} resulting in a different detector response to jets in \PbPb\ collisions that is not modeled by the MC simulation. It is evaluated by the method used for data collected by ATLAS in 2011 and 2015~\cite{ATLAS-CONF-2015-016}, which compares the jet \pt\ measured in the calorimeter with the sum of the \pt of charged particles within the jet, in both the data and MC samples. The selected charged-particle tracks have $\pt > 4$~\GeV\ to exclude particles from the UE. The ratio of the sum of the charged-particle \pt to the \ptjet provides a data-driven estimate of the centrality-dependence of the JES. The centrality-dependent JES component, pertinent only to jets in \PbPb\ collisions, is the dominant source of uncertainty in the jet \RAA measurement, contributing to a relative uncertainty of ${\approx}8\%$ in central collisions.
 
The uncertainty due to the jet energy resolution (JER) is evaluated by applying a Gaussian smearing factor to the reconstructed \ptjet in the MC sample.
The smearing factor is evaluated using an in situ technique in 13~\TeV\ \pp\ data that involves studies of dijet \pt balance~\cite{PERF-2011-04,PERF-2016-04}.
Further, an uncertainty is included to account for differences between the tower-based jet reconstruction and the jet reconstruction used in 13~\TeV\ \pp\ data analyses, as well as differences between calibration procedures~\cite{ATLAS-CONF-2015-016}.
The jet energy scale and resolution are the dominant sources of uncertainty in measuring the jet cross-sections and \RAA in this analysis and contribute a maximum relative uncertainty of about 14\% and 9\%, respectively.

Although the TCC energies are not utilized in measuring \ptjet, they are used in the measurement of \rg via the \SoftDrop procedure.
The TCC energy response is primarily determined by the topo-cluster energy response.
The uncertainties in the topo-cluster energy scale and resolution, measured using 13~\TeV\ \pp data~\cite{STDM-2017-33}, are applied directly to \pp data.
The cluster uncertainties from 13~\TeV\ \pp data~\cite{STDM-2017-33} are scaled conservatively by a factor of two for \PbPb data to cover the uncertainty in the topo-cluster energy response in the dense
\PbPb collision environment.
The inefficiency of matching between tracks and topo-clusters is driven by the angular resolution of the topo-clusters, which is used to match these objects in the TCC reconstruction.
A random Gaussian smearing of 5-mrad is applied independently in $\eta$ and $\phi$, following previous ATLAS measurements~\cite{STDM-2012-17,STDM-2017-33}, to account for potential mismodeling of topo-cluster positions and is observed to have a negligible effect on the resulting TCCs and the \rg distributions.
TCCs are reconstructed using topo-clusters with energies and positions varied according to their uncertainties in MC events and are used to unfold the data.
The deviations from the nominal unfolded result, due to individual topo-cluster uncertainty sources, are combined in quadrature.
 
The effects of topo-cluster splitting and merging were studied in detail, following the method used in previous measurements~\cite{STDM-2017-33}.
The MC generator predictions precisely match each of the splitting and merging scenarios tested in data and thus no systematic uncertainty is assigned to these effects.
 
The angular coordinates ($\eta$, $\phi$) of a TCC corresponding to a charged particle are determined by the track seeding the TCC object.
Systematic variations of these coordinates are evaluated to account for
tracking mismodeling in MC events.
These differences are decomposed into two components: one from the uncertainty in the inner-detector material derived from studies in \pp collisions~\cite{PERF-2015-07}, and the other from the modeling of pixel cluster merging inside dense environments~\cite{PERF-2015-08, ATL-PHYS-PUB-2016-007}, such as inside the core of high-energy jets and in central \PbPb collisions.
These track uncertainties are propagated through the TCC reconstruction and are observed to have a negligible effect on the final \rg distributions.
 
In this measurement the TCC-related uncertainties typically have a more modest effect than jet-related uncertainties as only the angular distance between the subjets (\rg) is measured. They become significant, contributing a relative uncertainty of up to $15\%$, at large \rg (${>}0.3$) in central \PbPb collisions as UE fluctuations in the outer region of the jet lead to poorer performance in identifying the hard subjets.
 
The differences between the unfolded results with and without the migration-matrix reweighting factors are assigned as systematic uncertainties arising from the unfolding procedure. The unfolding uncertainty is observed to have a smaller effect than other sources, contributing a relative uncertainty of ${<}5\%$ to the cross-section, yield, and \RAA results.
 
The statistical uncertainties in the unfolding due to the statistical uncertainties associated with the size of the accumulated dataset are evaluated using the pseudo-experiment technique with 500 separate stochastic variations of the input spectrum as described in Ref.~\cite{ATL-PHYS-PUB-2021-011}. The contributions from statistical fluctuations in the response matrix are similarly evaluated using the same number of stochastic variations. The two contributions to the statistical uncertainty are combined in quadrature and are observed to contribute ${<}1\%$ to the total systematic uncertainty in the results. 
 
For the 2017 \pp data, the LUCID-2 detector~\cite{LUCID2} was used for the primary luminosity measurement. The uncertainty in the integrated luminosity is estimated to be $1.6\%$ using the methods described in Ref.~\cite{ATLAS-CONF-2019-021}.
For \PbPb collisions, the systematic uncertainty in the mean nuclear thickness function, $\avgTAA$, is estimated by varying the MC Glauber model parameters as detailed in Ref.~\cite{HION-2016-06}, and is listed in Table~\ref{tab:lumi}.
 
A summary of the relative uncertainties in the unfolded inclusive jet cross-sections, yields, and nuclear modification factor are shown in Figures~\ref{fig:jetpt_syst}--\ref{fig:thetagraa_syst}.
Figures~\ref{fig:jetpt_syst} and~\ref{fig:thetag_syst} present a summary of the total and individual relative uncertainties in the jet cross-sections and per-event yields as a function of \ptjet and \rg, respectively. The individual uncertainties are summed in quadrature to obtain the total uncertainty. The jet energy scale is the dominant source of uncertainty in the cross-sections and per-event jet yields except at large \rg in central \PbPb collisions, where the TCC energy-scale uncertainties have a comparable effect. A summary of the total and individual relative uncertainties in the \RAA is shown in Figures~\ref{fig:jetraa_syst} and~\ref{fig:thetagraa_syst} as a function of \ptjet and \rg, respectively. The uncertainties which are common to \pp and \PbPb collisions, such as the centrality-independent JES and JER  uncertainties, are treated as correlated when determining the uncertainty in \RAA. However, as the centrality-dependent component of the JES uncertainty (only for \PbPb collisions) does not cancel out in the cross-section ratios, it becomes the dominant source of uncertainty in the \RAA results.
 
\vspace*{1in}
 
\clearpage
 
\section{Results}
\label{sec:result}
 
\begin{figure}[t]
\centering
\includegraphics[width=0.44\textwidth]{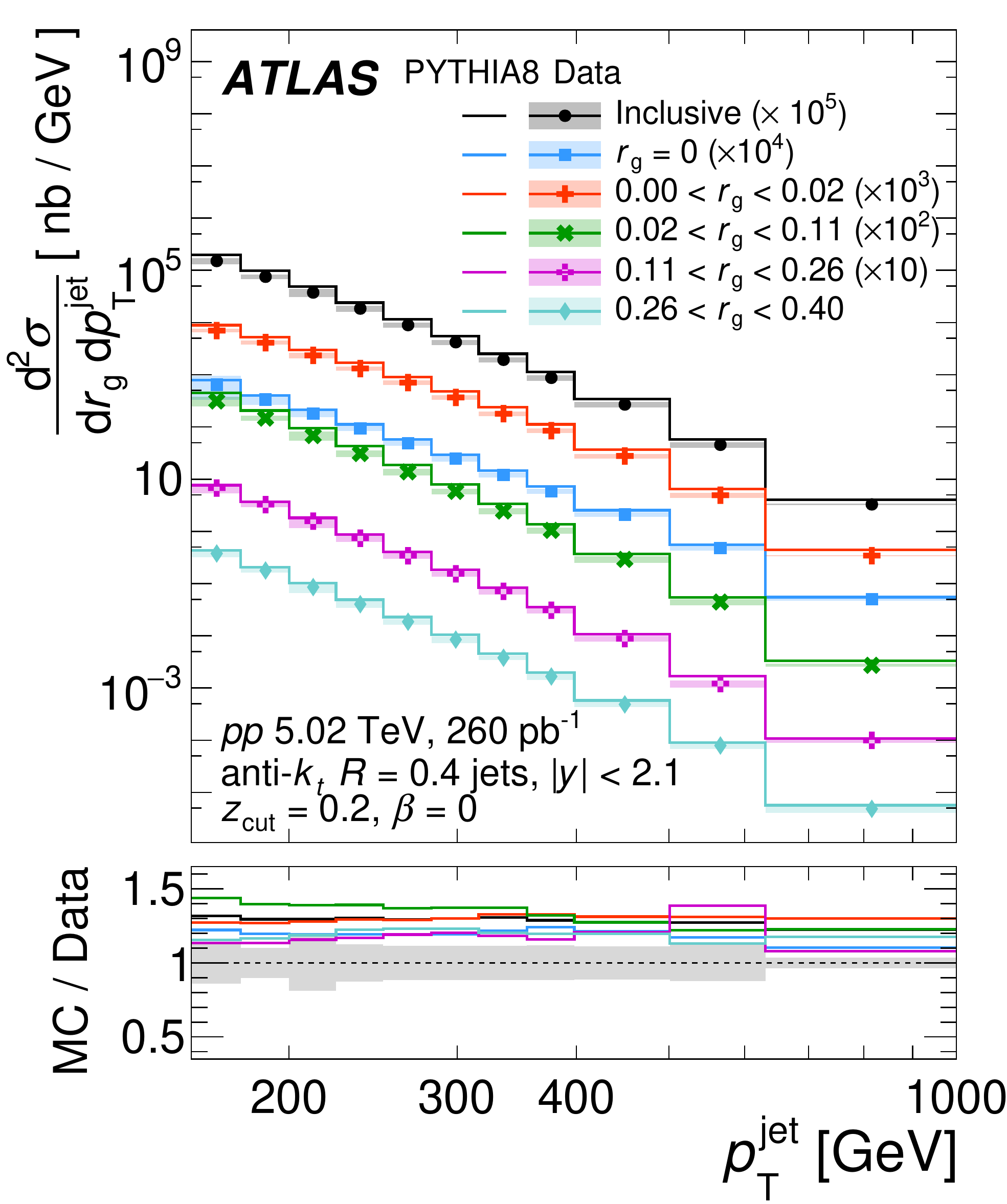}
\linebreak
\includegraphics[width=0.44\textwidth]{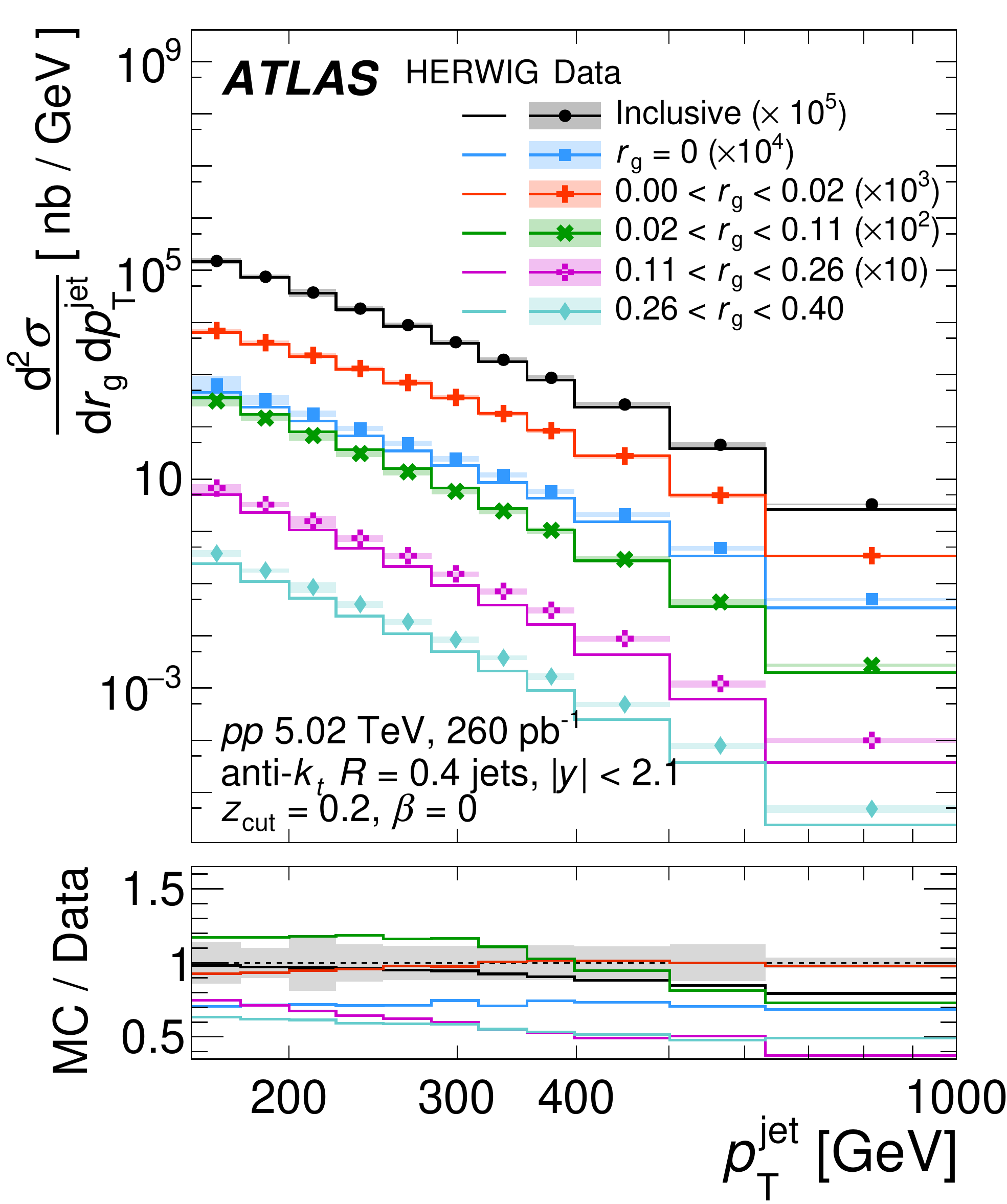}
\includegraphics[width=0.44\textwidth]{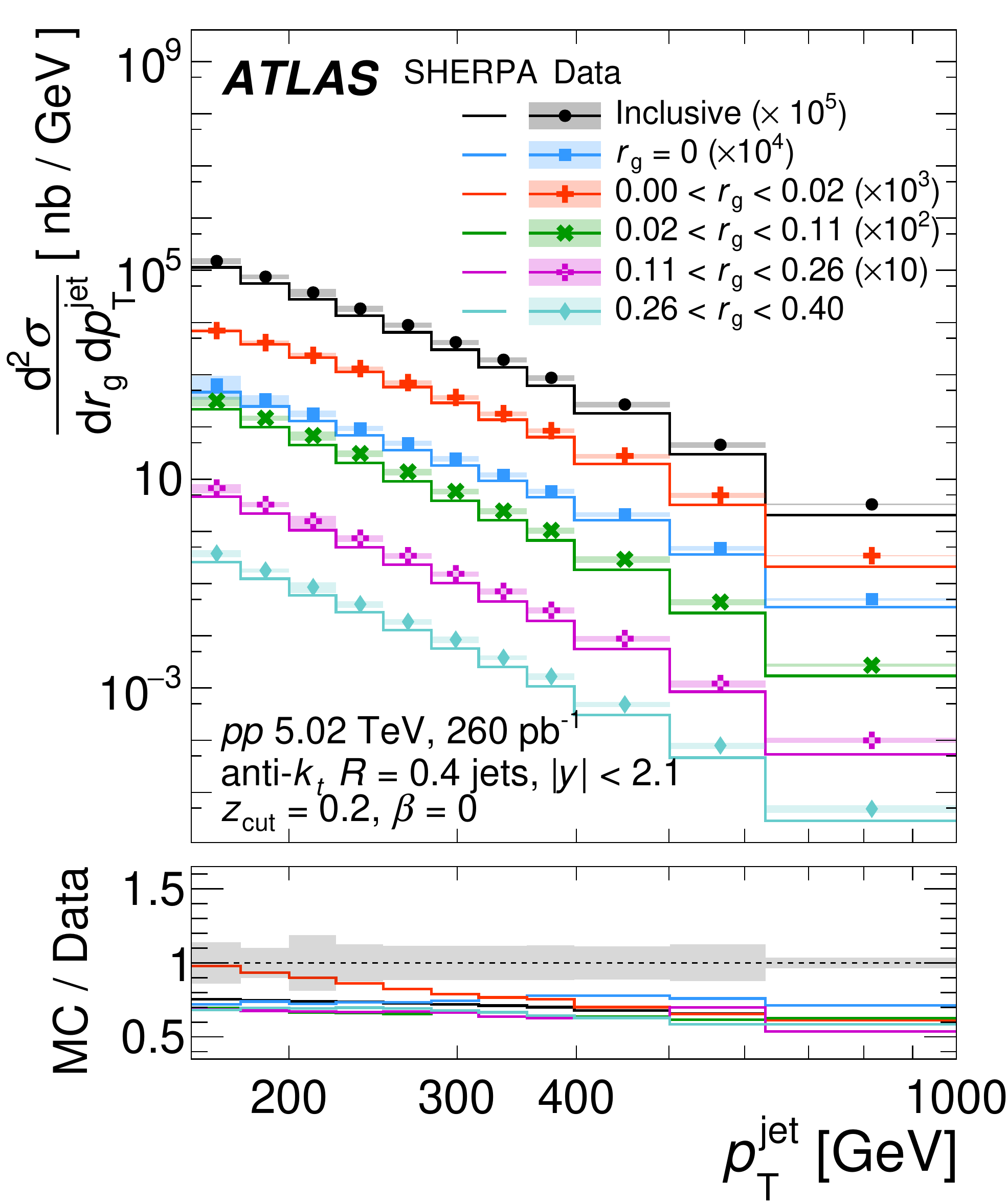}
\caption{Differential cross-section of jets passing the \SoftDrop grooming condition in different \rg intervals in \pp collisions  at $\sqrt{s} = 5.02~\TeV$ as a function of \ptjet.
Distributions for inclusive jets (no grooming) and for those jets that fail the \SoftDrop requirement ($\rg=0$) are also shown.
For visual clarity, the distributions are scaled by factors specified in the legend. The statistical uncertainties are smaller than the symbols, while shaded bars represent systematic uncertainties. Note that the highest \ptjet bin does not contain the overflow and that the 1.6\% luminosity uncertainty is not shown.
The \pp jet cross-sections are compared with predictions from three MC generators, \PYTHIA[8] (top), \HERWIG (bottom left) and \SHERPA (bottom right).
The ratios of jet cross-section predictions from different MC generators to the unfolded data are shown in the lower panels. The statistical uncertainties on the ratios are smaller than the line widths, while shaded bars in the lower panels represent systematic uncertainties in inclusive cross-section ratios.
}
\label{fig:jetspectra_pp}
\end{figure}

The unfolded differential jet cross-sections obtained from \pp collision data are shown in Figure~\ref{fig:jetspectra_pp} as a function of \ptjet and \rg.
The differential cross-sections are reported for four intervals in $r_\mathrm{g}$, each scaled up by successive powers of 10. The cross-sections are also reported for jets that fail the \SoftDrop grooming condition.
These are assigned a value of $\rg=0$ in this analysis and they make up about $3.5\%$ of the inclusive \pp jet cross-section, and are also shown in Figure~\ref{fig:jetspectra_pp}.
 
\begin{figure}[t]
\centering
\includegraphics[width=0.42\textwidth]{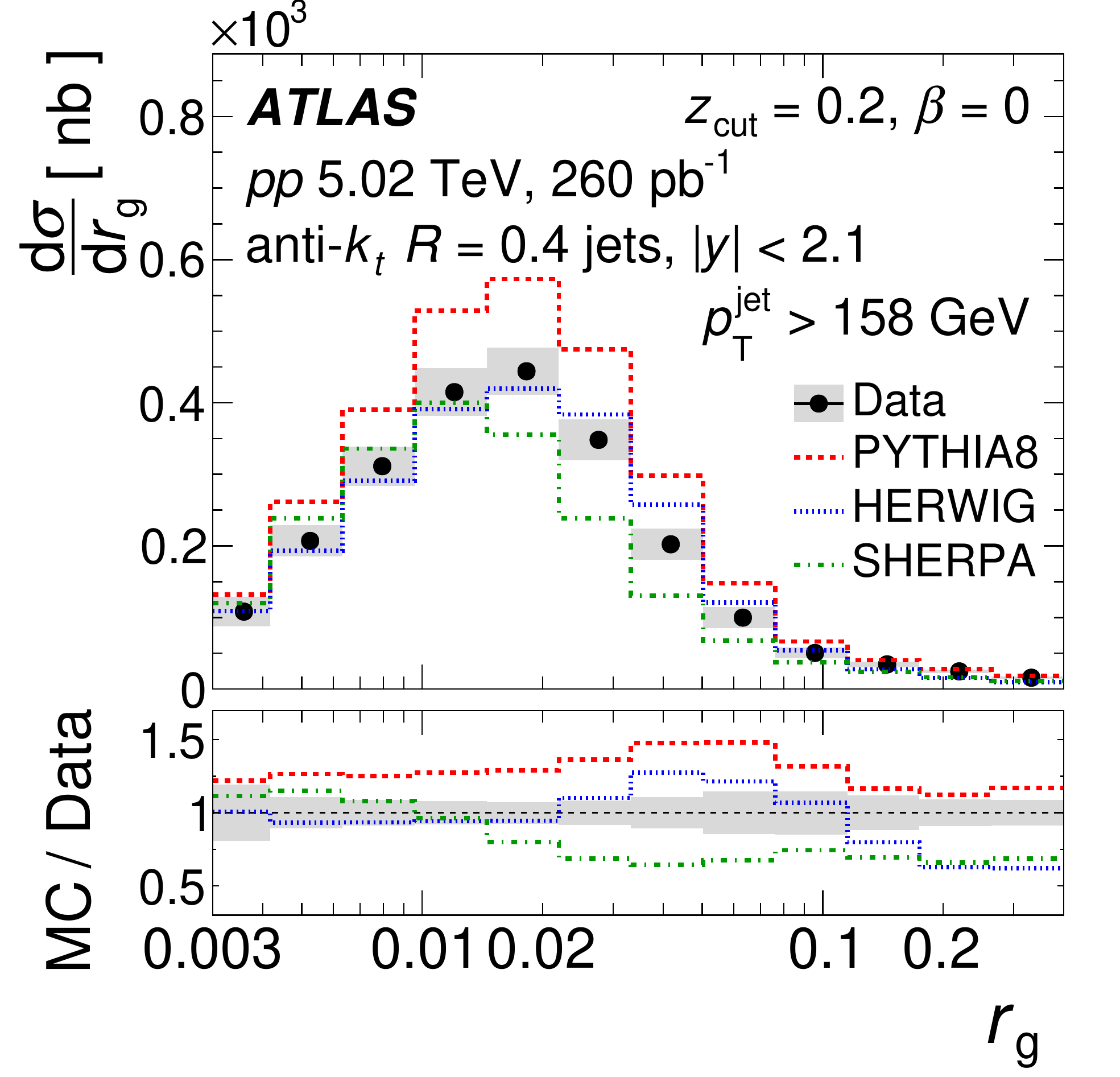}
\linebreak
\includegraphics[width=0.42\textwidth]{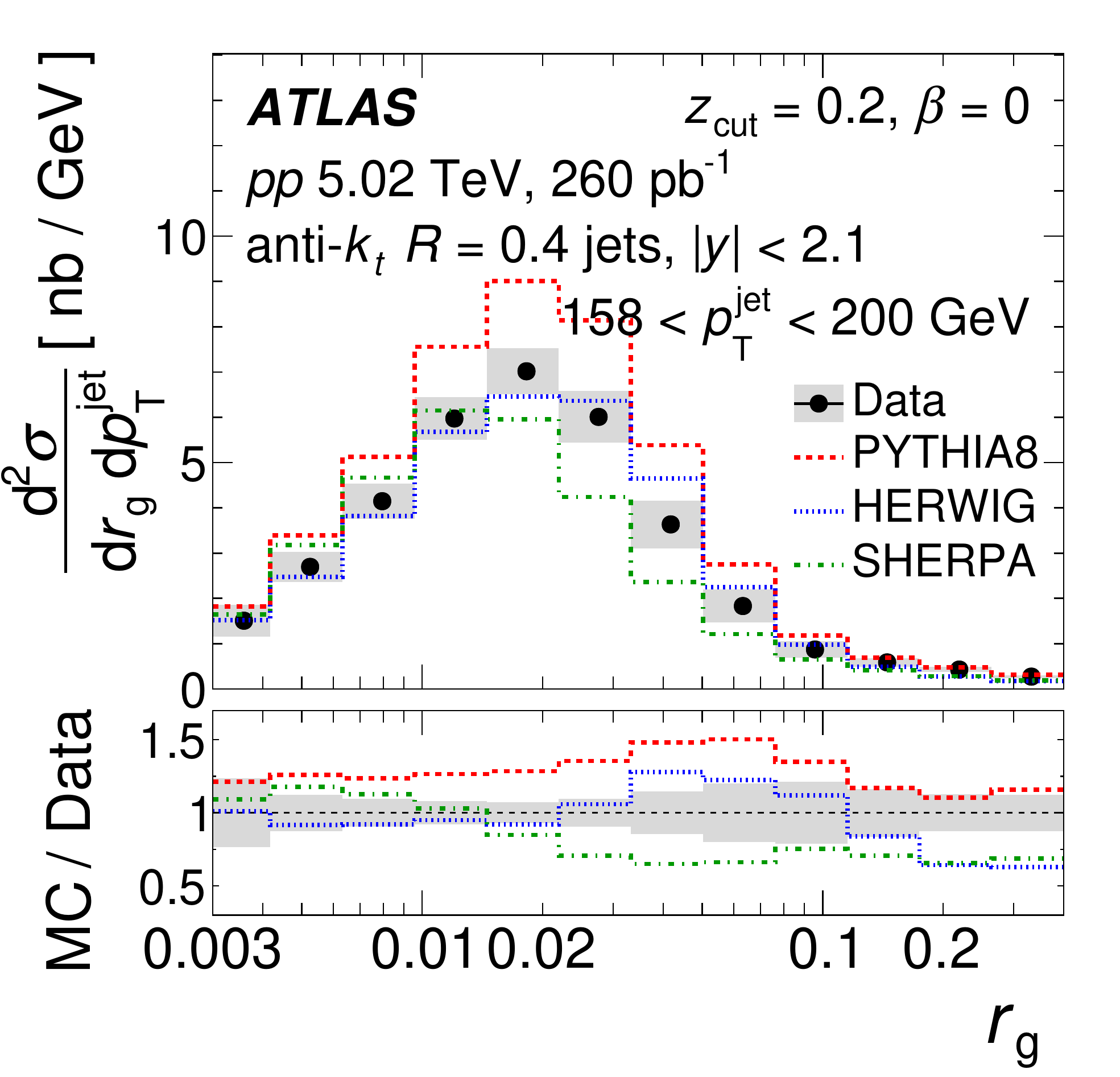}
\includegraphics[width=0.42\textwidth]{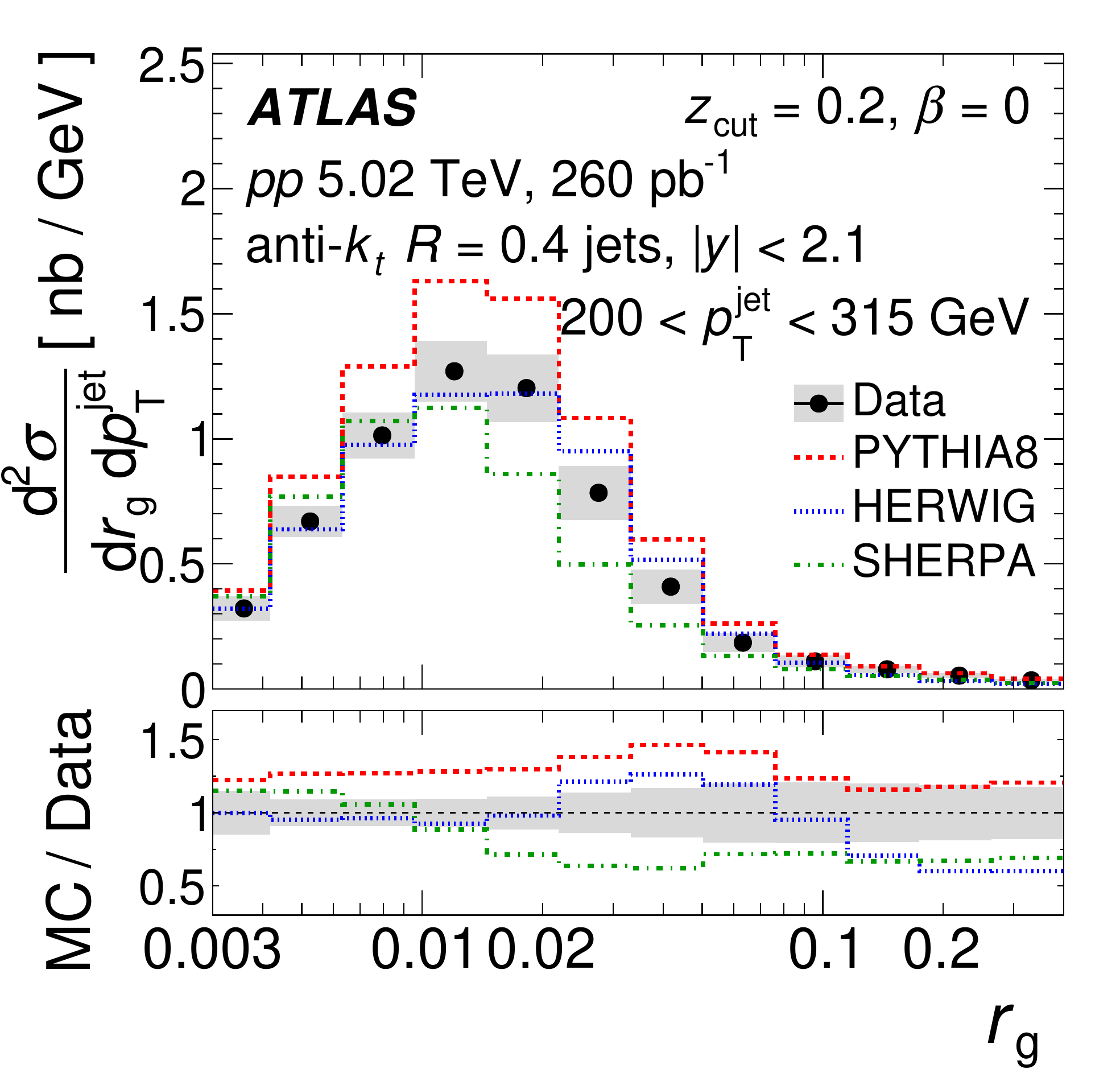}
\includegraphics[width=0.42\textwidth]{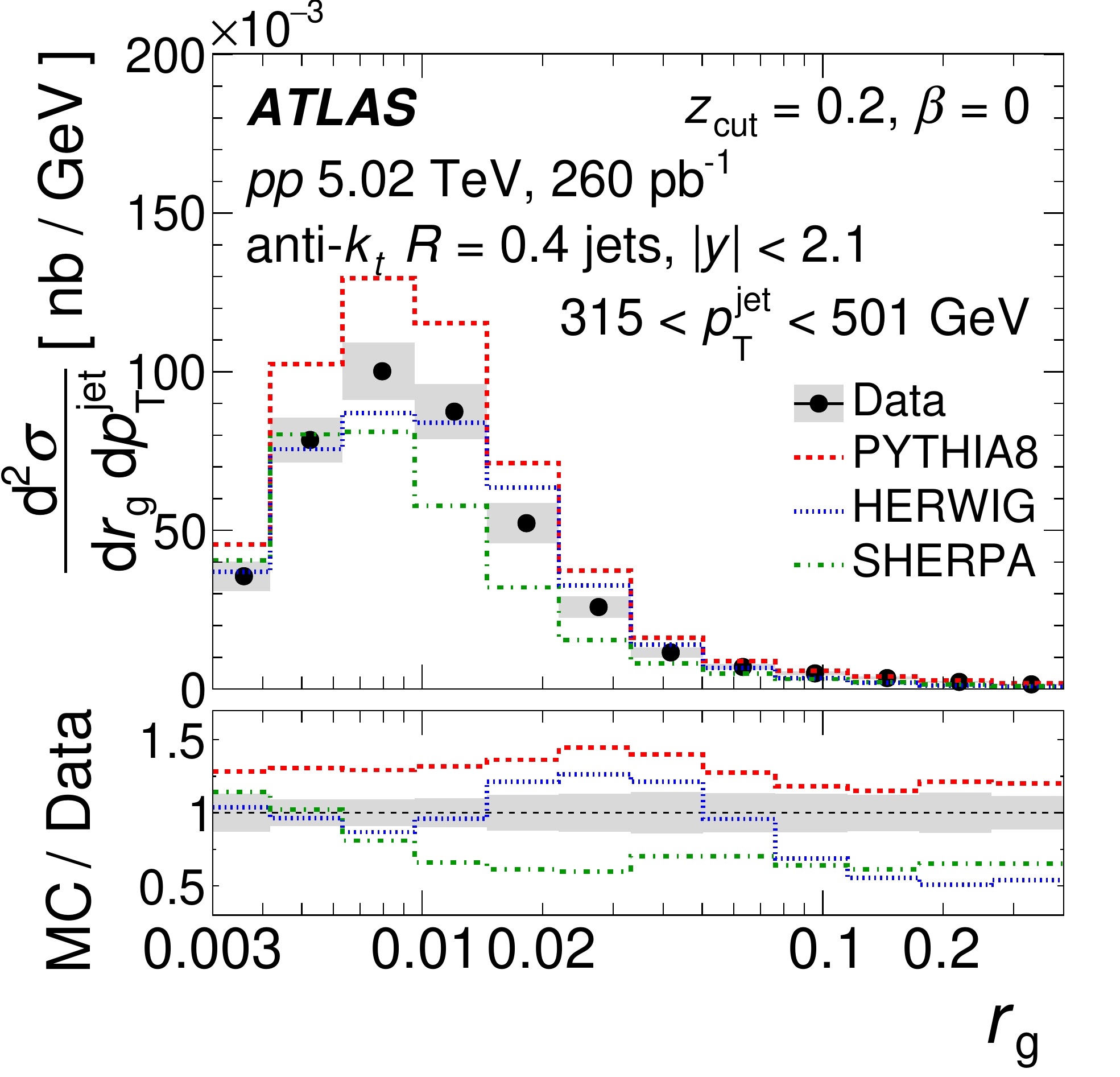}
\includegraphics[width=0.42\textwidth]{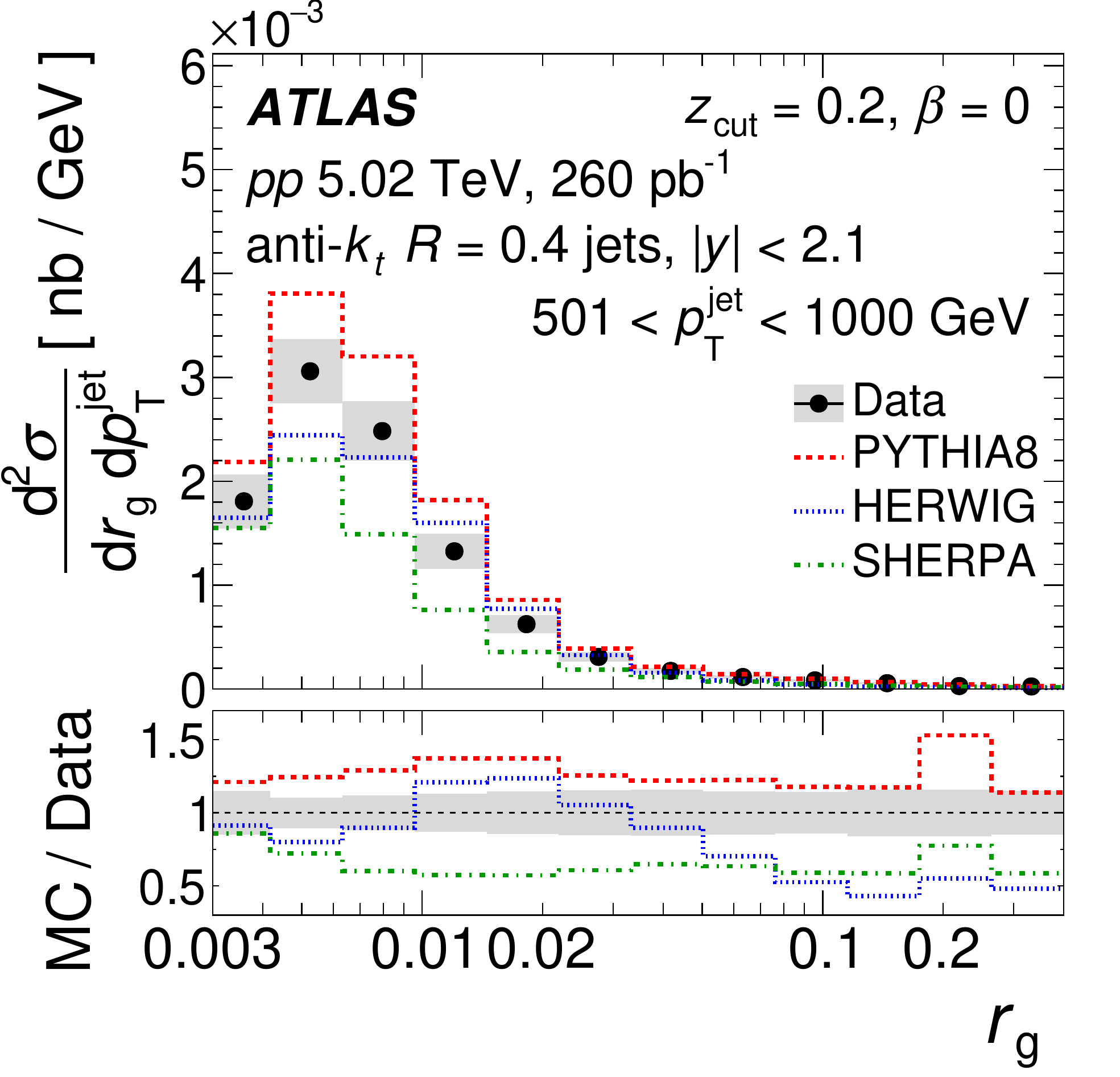}
\caption{Differential cross-section of jets passing the \SoftDrop grooming condition in \pp collisions  at $\sqrt{s} = 5.02~\TeV$ as a function of \rg is shown for different \ptjet intervals and also for $\ptjet > 158$~\GeV.
The error bars represent statistical uncertainties, and are smaller than the marker size here, while the shaded bars represent systematic uncertainties. The $1.6\%$ uncertainty in the \pp luminosity is not shown. The last bin does not contain the overflow.
The ratios of jet cross-section predictions from different MC generators to the unfolded data are shown in the lower panels. The shaded bars in the lower panels represent systematic uncertainties in cross-section ratios.}
\label{fig:jetrg_pp}
\end{figure}

Figure~\ref{fig:jetrg_pp} shows the unfolded inclusive differential cross-section in \pp collision data as a function of \rg.
The differential cross-sections in \rg are also reported for four bins in \ptjet. The distributions are observed to peak at lower values of \rg with increasing \ptjet. Tighter collimation of jets with increasing \ptjet is expected from the larger boost of the fragmenting parton and an increased quark-initiated jet fraction~\cite{ATL-PHYS-PUB-2017-015}.
 
The jet differential cross-sections measured in \pp data are compared with particle-level predictions from the \PYTHIA[8], \HERWIG, and \SHERPA generators in Figures~\ref{fig:jetspectra_pp} and~\ref{fig:jetrg_pp}.
The \HERWIG predictions are obtained from multijet events generated at next-to-leading order by \HERWIG[7.1.3]~\cite{Bellm:2017jjp} with the \NNPDF[3.0] PDF set~\cite{Ball:2014uwa} for the matrix element calculation.
The \SHERPA predictions are from multijet events generated using \SHERPA[2.2.5]~\cite{Bothmann:2019yzt}.
The matrix element calculation was included for the \(2\rightarrow2\) process at leading order, and the default \SHERPA parton shower~\cite{Schumann:2007mg} based on Catani--Seymour dipole factorization was used for the showering with \pT ordering, using the \CT[14nnlo] PDF set~\cite{Dulat:2015mca}.
The \PYTHIA[8] predictions overestimate the \pp data jet cross-sections but describe the shape of the \ptjet and \rg distributions well, both inclusively and differentially in \ptjet and \rg.
The \HERWIG and \SHERPA predictions underestimate the cross-sections 
but describe the shape of both the inclusive jet differential cross-section as a function of \pTjet and the differential cross-section in different \rg bins.
The \HERWIG generator predicts a slightly wider \rg distribution than seen in \pp data across all \pTjet regions studied here. The \Sherpa generator predicts a significantly narrower \rg distribution than seen in \pp data at low \ptjet and underestimates the jet cross-sections at high \ptjet. The trend of tighter collimation of jets with increasing \ptjet observed in \pp data is reproduced by all three MC generators with varying levels of accuracy.
 
\begin{figure}[ht]
\centering
\includegraphics[width=0.475\textwidth]{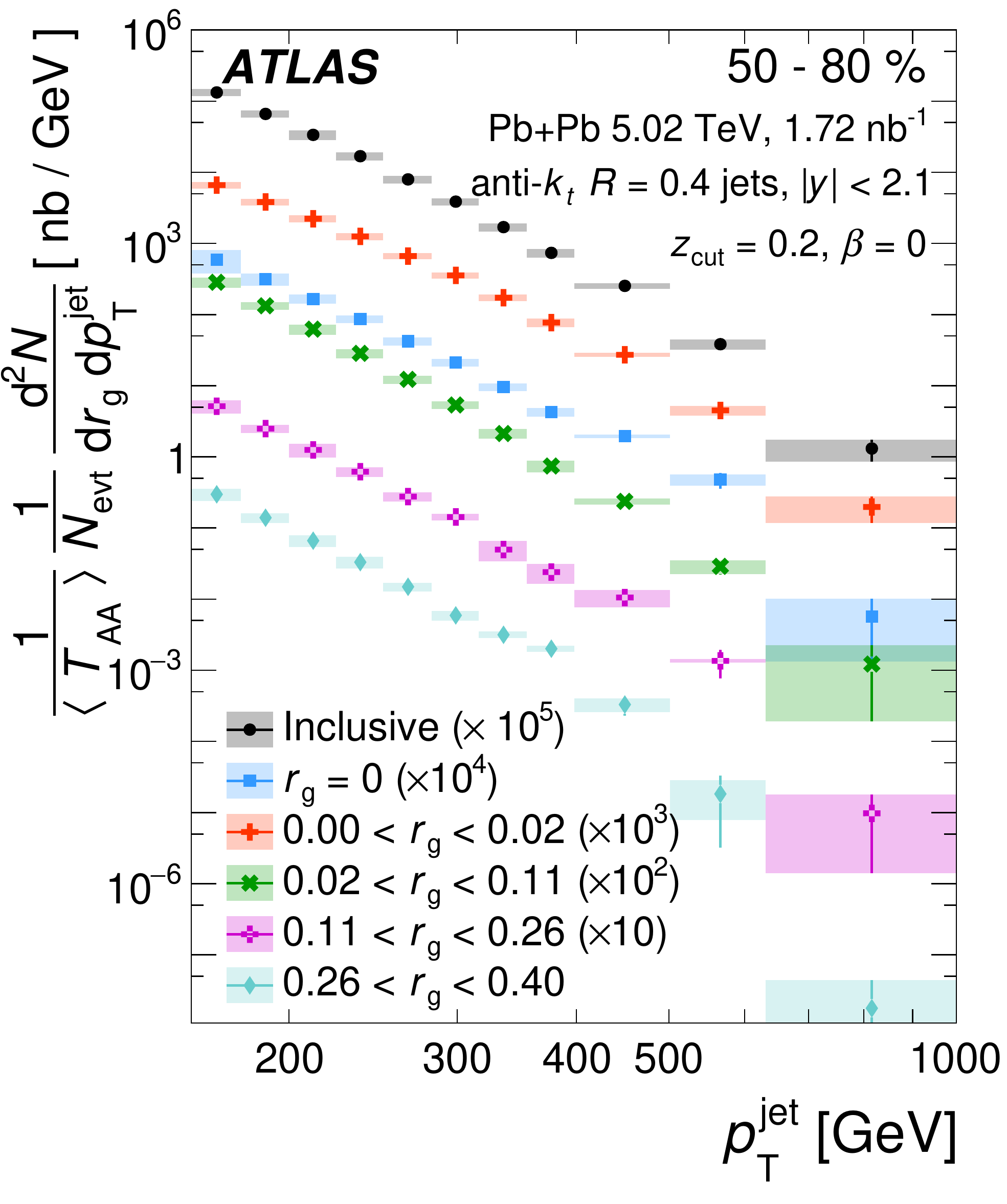}
\includegraphics[width=0.475\textwidth]{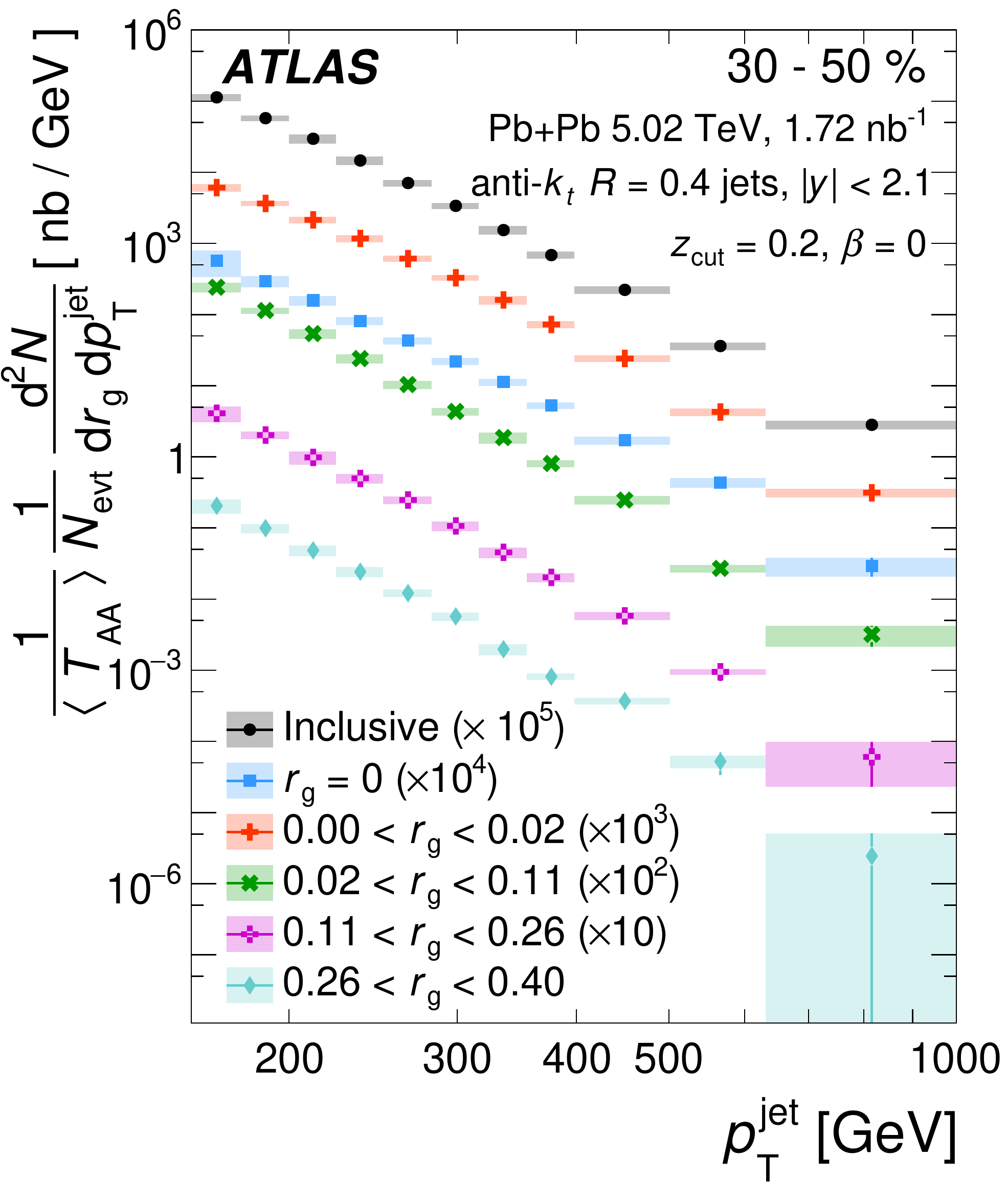}
\includegraphics[width=0.475\textwidth]{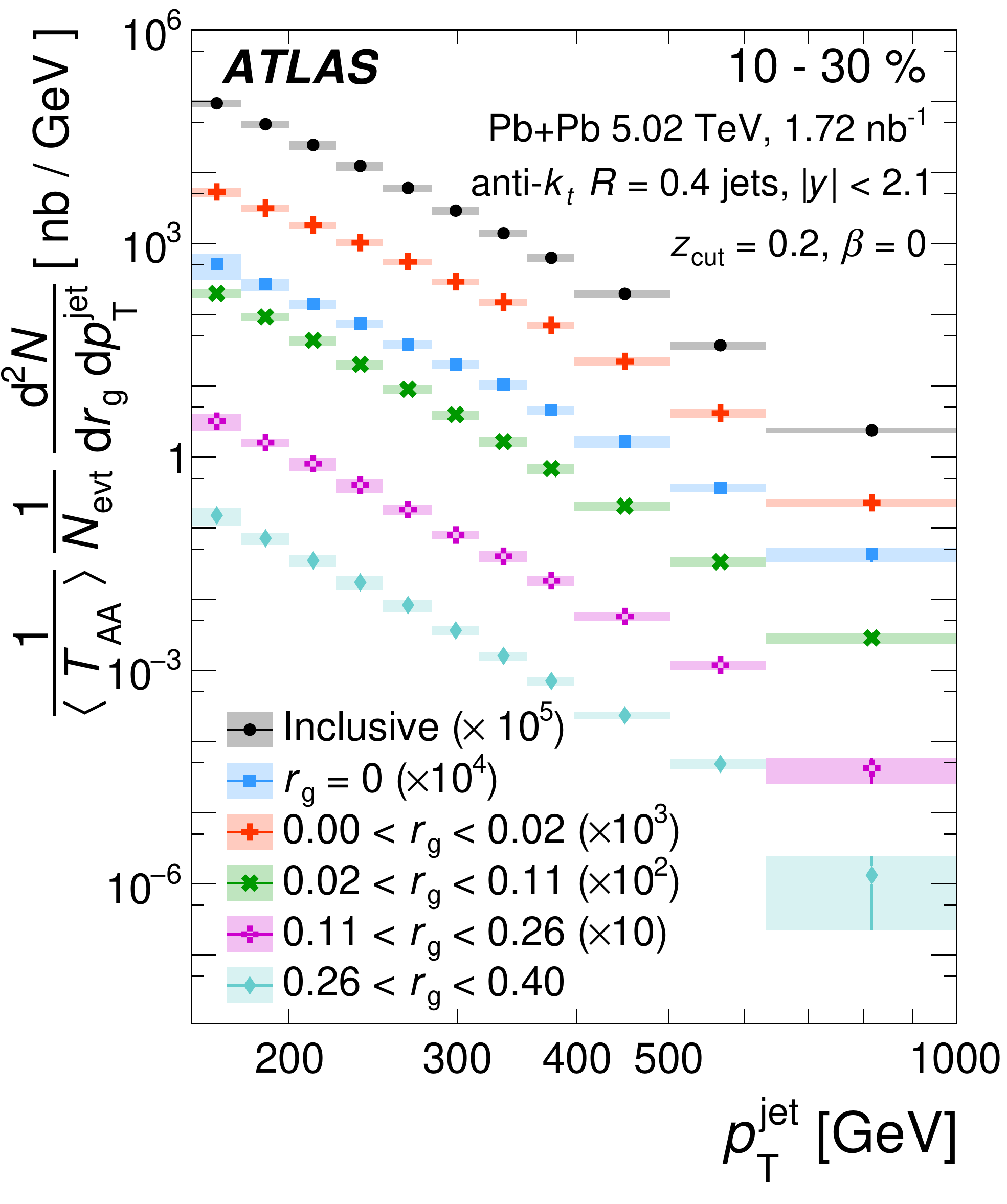}
\includegraphics[width=0.475\textwidth]{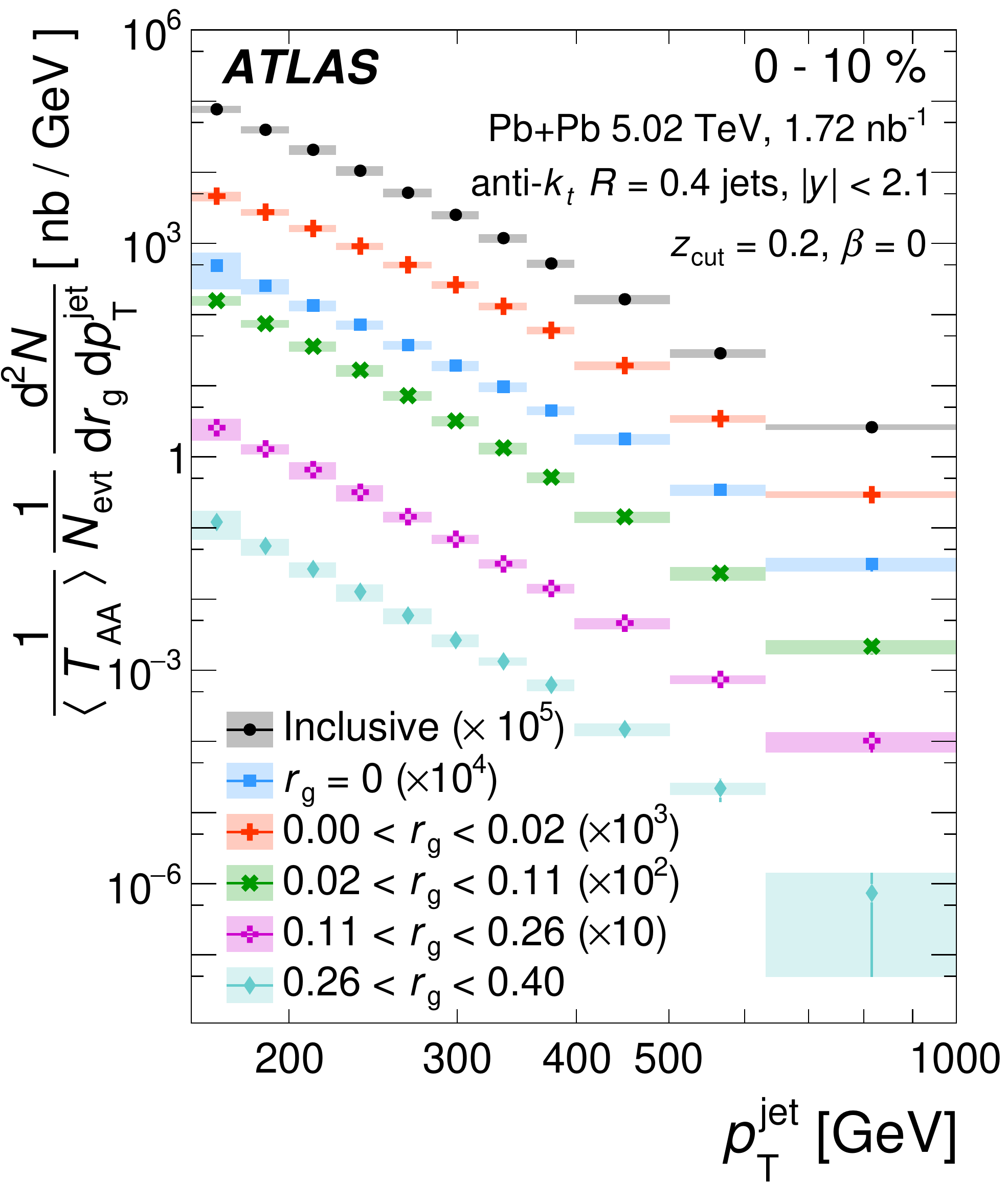}
\caption{The per-event inclusive jet yields in \PbPb collisions  at $\sqn = 5.02~\TeV$ normalized by \avgTAA as a function of \ptjet in four centrality intervals. Distributions for inclusive jets (no grooming) and for those jets that fail the \SoftDrop requirement (denoted by $\rg=0$) are also shown.
For visual clarity, the distributions are scaled by factors specified in the legend.
The last bin does not contain the overflow.
The statistical uncertainties are indicated by the error bars while the systematic uncertainties are indicated by the shaded bars. The normalization uncertainties of \avgTAA for each centrality interval are not shown, but are listed in Table~\ref{tab:lumi}.
}
\label{fig:jetspectra_PbPb}
\end{figure}

\begin{figure}[ht]
\centering
\includegraphics[width=0.475\textwidth]{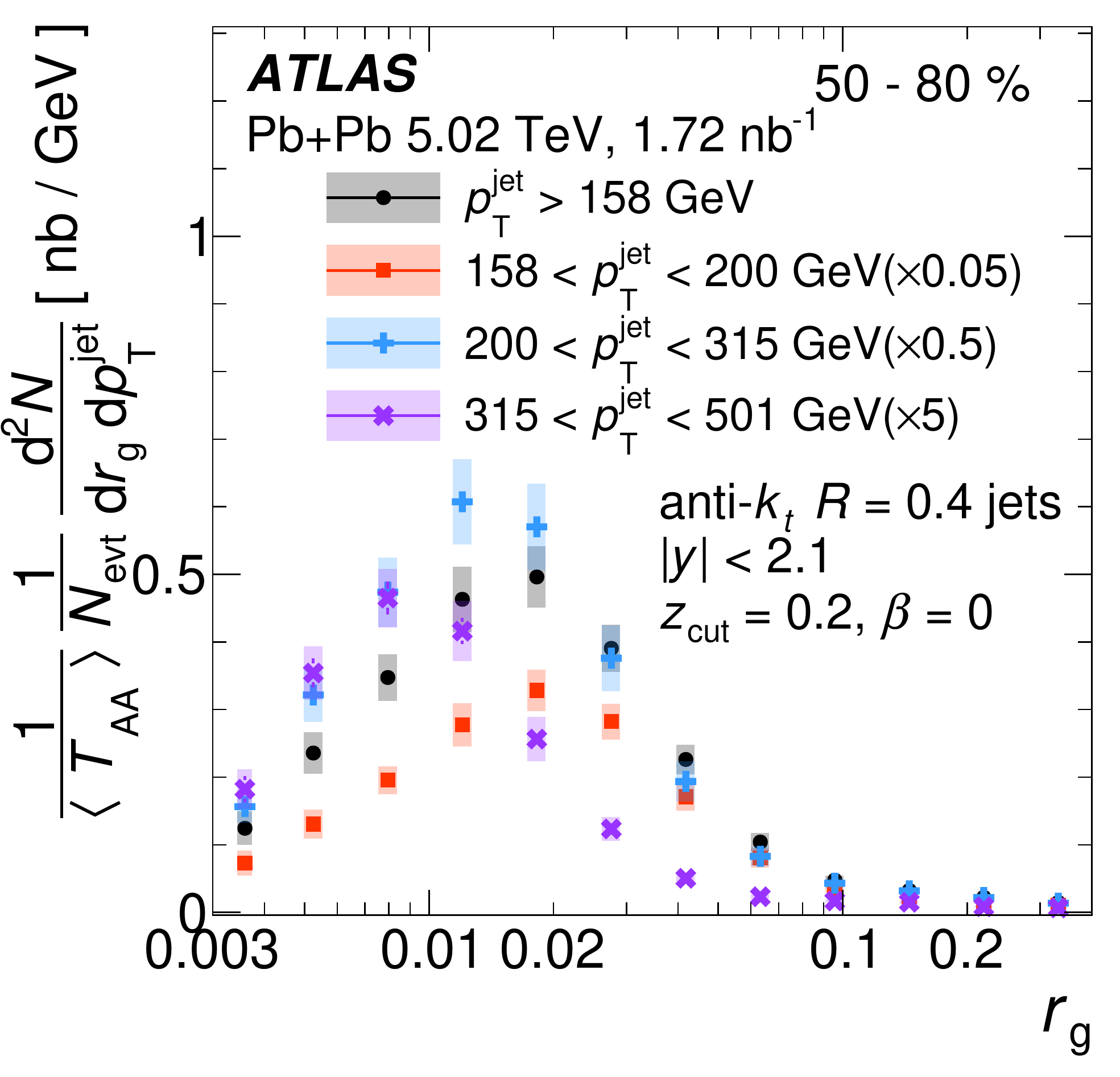}
\includegraphics[width=0.475\textwidth]{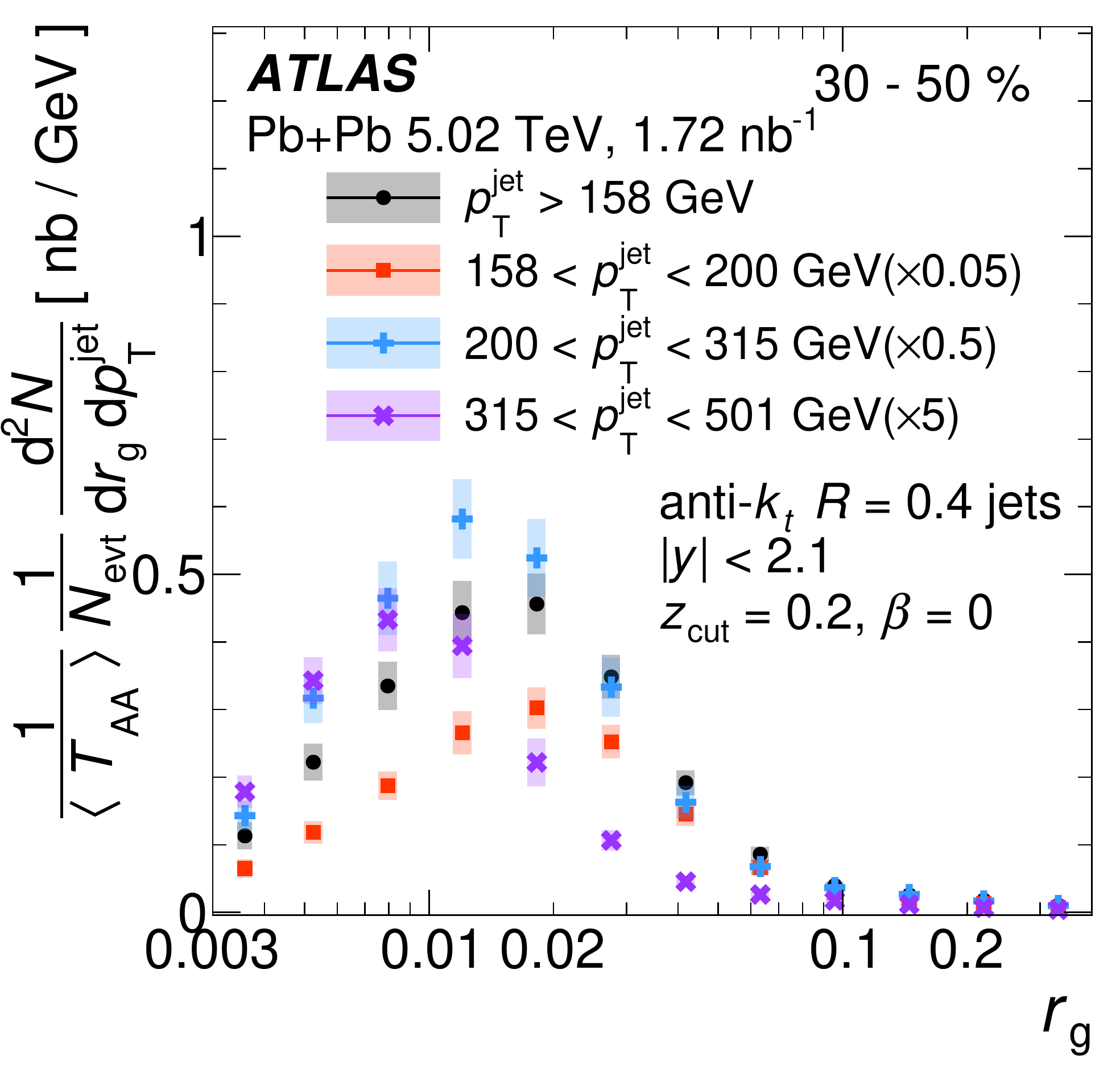}
\includegraphics[width=0.475\textwidth]{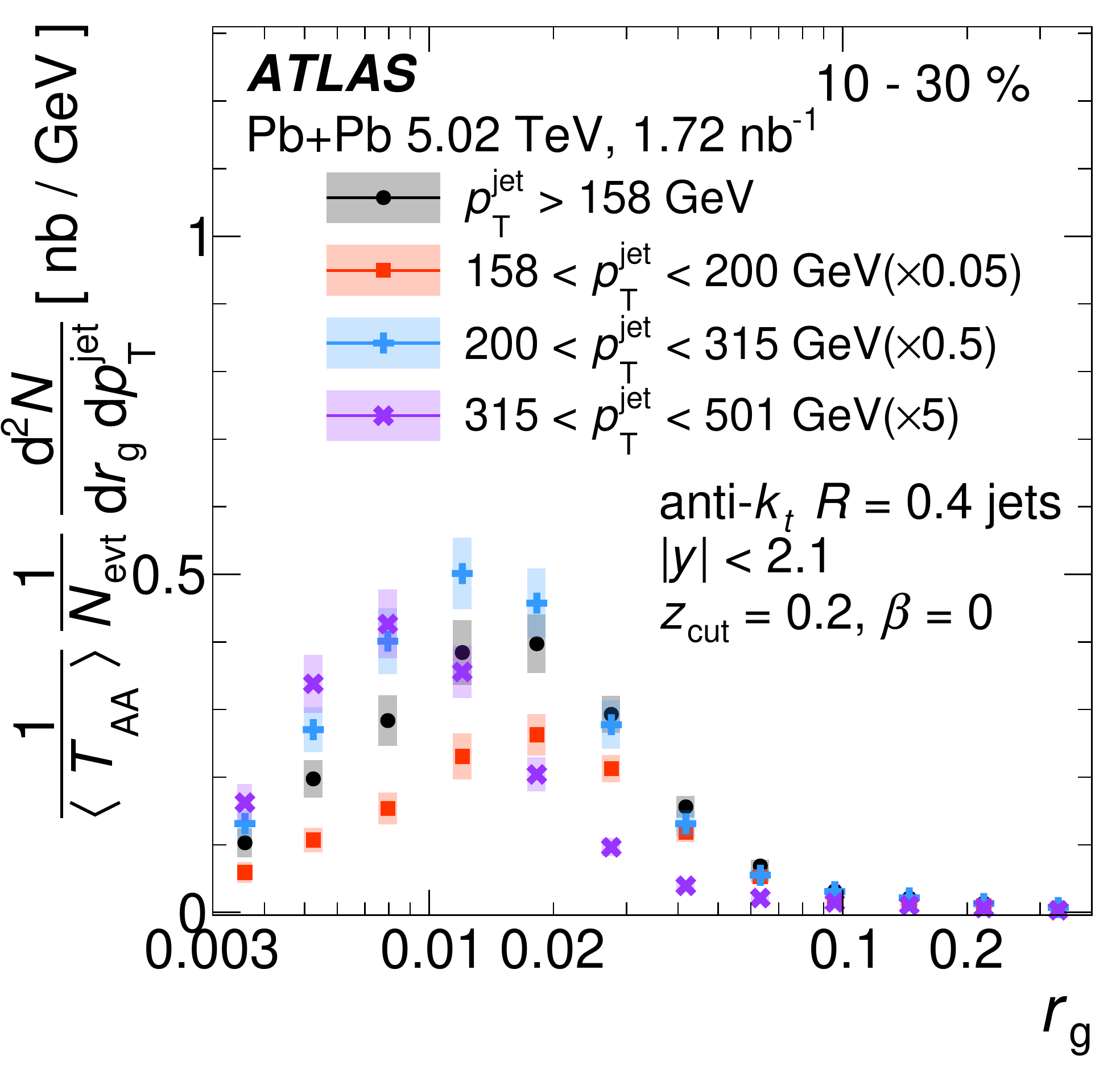}
\includegraphics[width=0.475\textwidth]{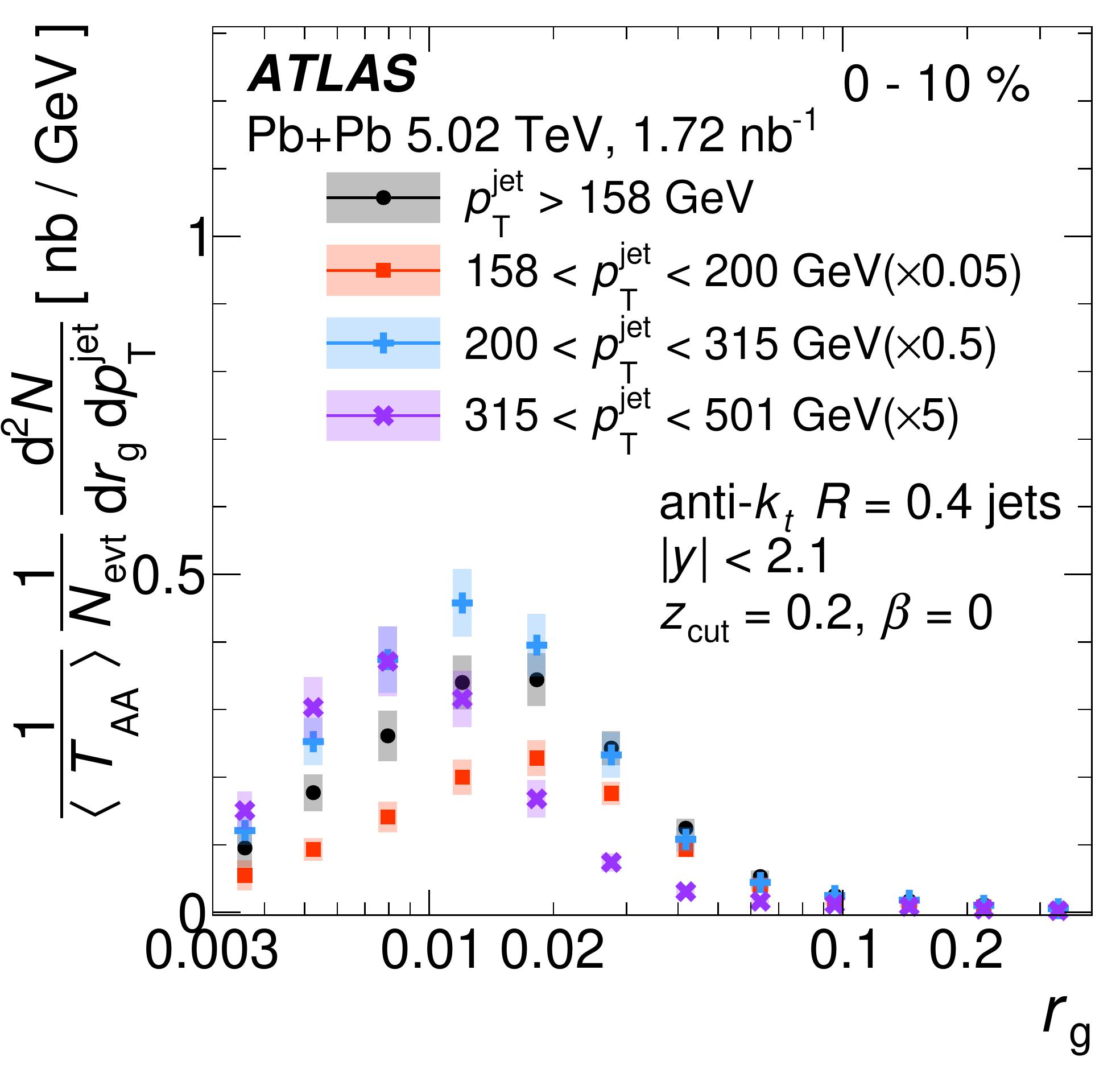}
\caption{The per-event jet yields in \PbPb collisions  at $\sqn = 5.02~\TeV$ normalized by \avgTAA as a function of \SoftDrop \rg in four centrality intervals and three \ptjet intervals as well for $\ptjet > 158$~\GeV. The last bin does not contain the overflow. The statistical uncertainties are indicated by the error bars while the systematic uncertainties are indicated by the shaded bars. The normalization uncertainties of \avgTAA for each centrality interval are not shown, but are listed in Table~\ref{tab:lumi}.
}
\label{fig:jetspectra_vs_rg_PbPb}
\end{figure}

Figures~\ref{fig:jetspectra_PbPb} and~\ref{fig:jetspectra_vs_rg_PbPb} show the \PbPb per-event yields normalized by \avgTAA for four centrality intervals as a function of \ptjet and \rg, respectively.
The yields are also shown for jets that fail the \SoftDrop grooming condition (denoted by $\rg=0$) as a function of \ptjet; they account for about 4--5.5\% of the inclusive \PbPb jet yield.
The per-event jet yields in \PbPb data normalized by \avgTAA fall below the corresponding \pp jet cross-sections for a given \rg or \ptjet interval, indicating jet suppression, and this difference increases for more central events and for increasing \rg.
The \rg distributions are also observed to peak at lower values of \rg with increasing \ptjet in both \pp collisions and all centrality intervals in \PbPb collisions. The \ptjet spectra are observed to have similar slopes in the different \rg regions for both \pp and \PbPb collisions.


\begin{figure}[ht]
\centering
\includegraphics[width=0.475\textwidth]{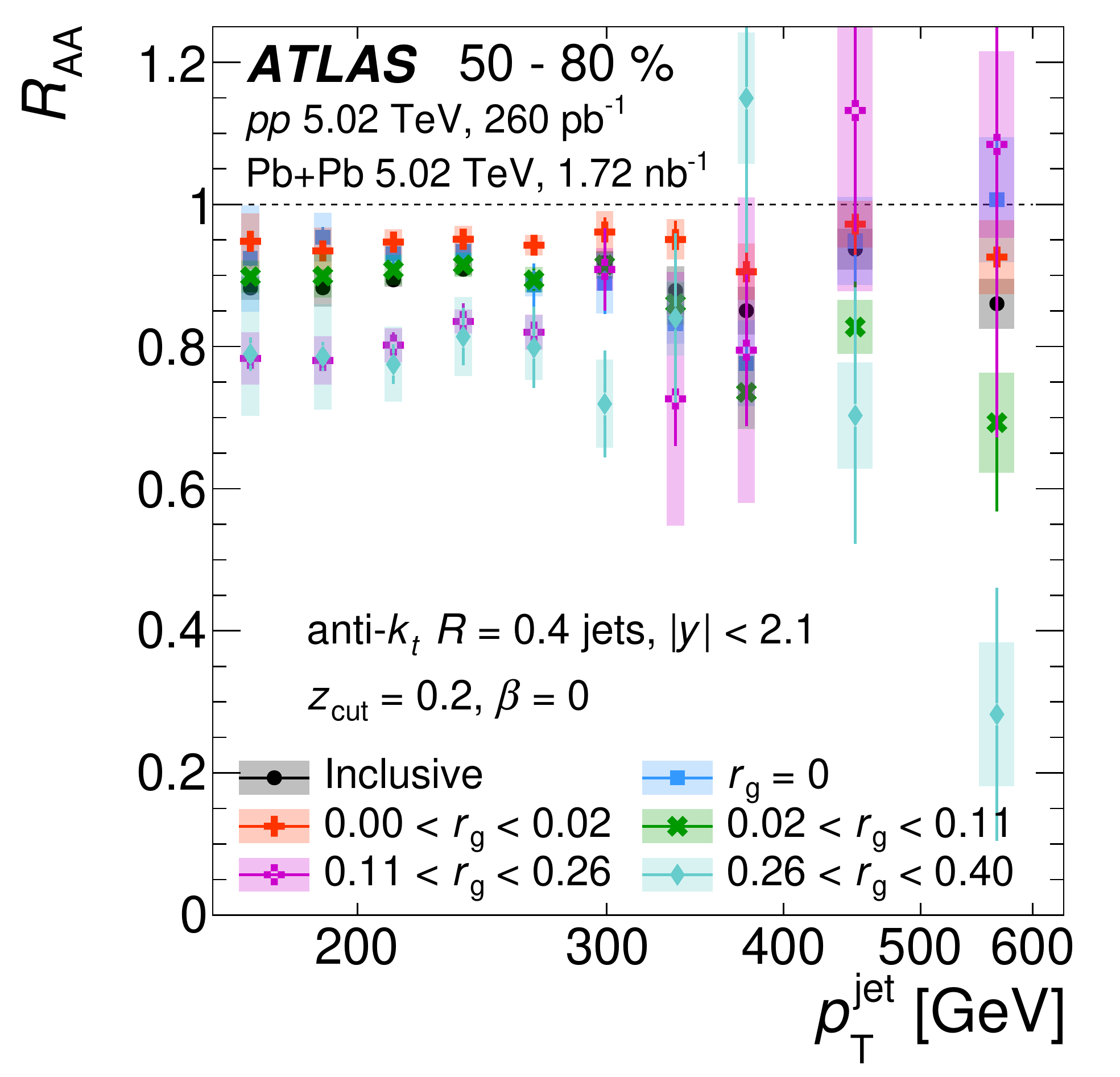}
\includegraphics[width=0.475\textwidth]{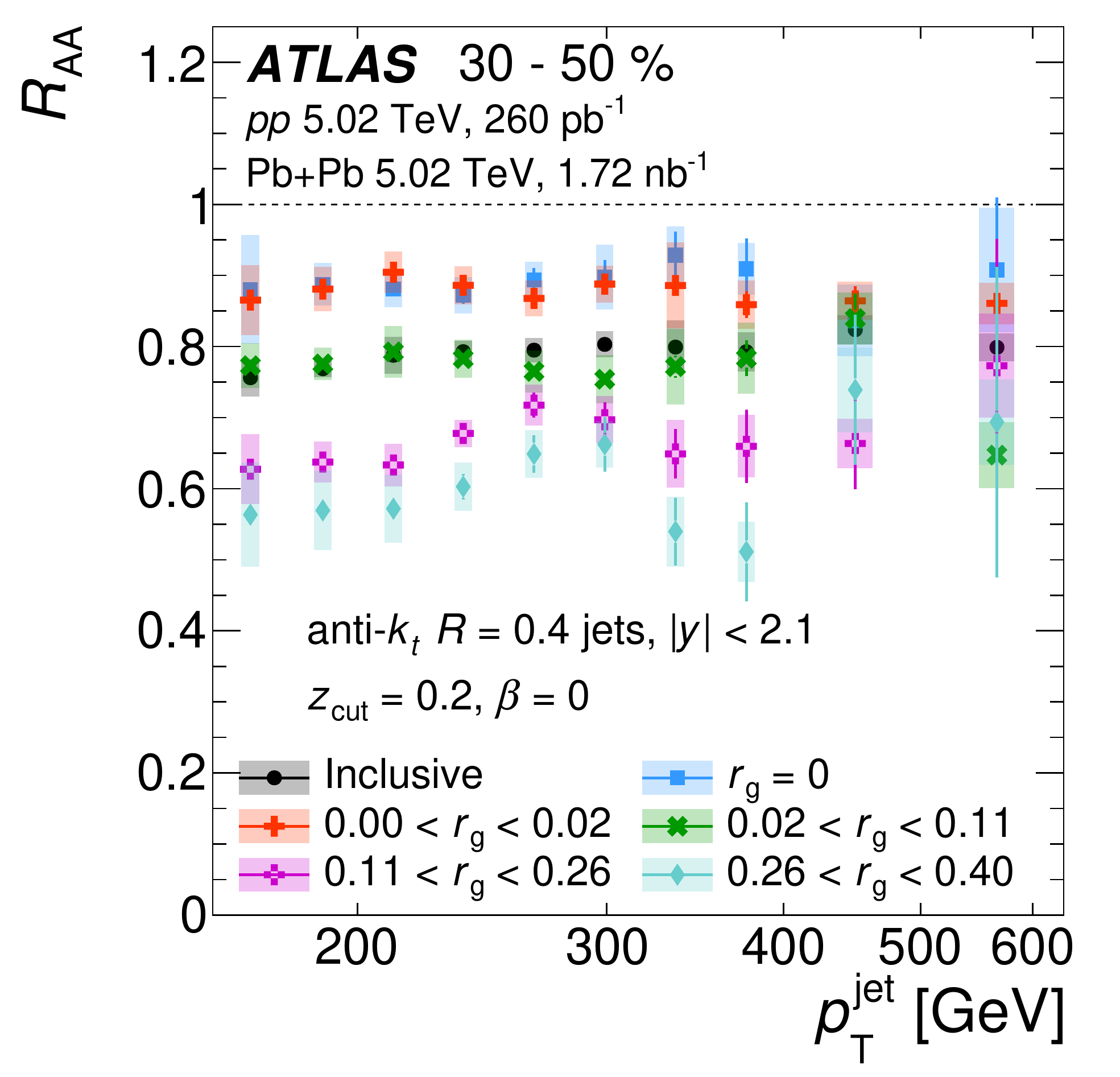}
\includegraphics[width=0.475\textwidth]{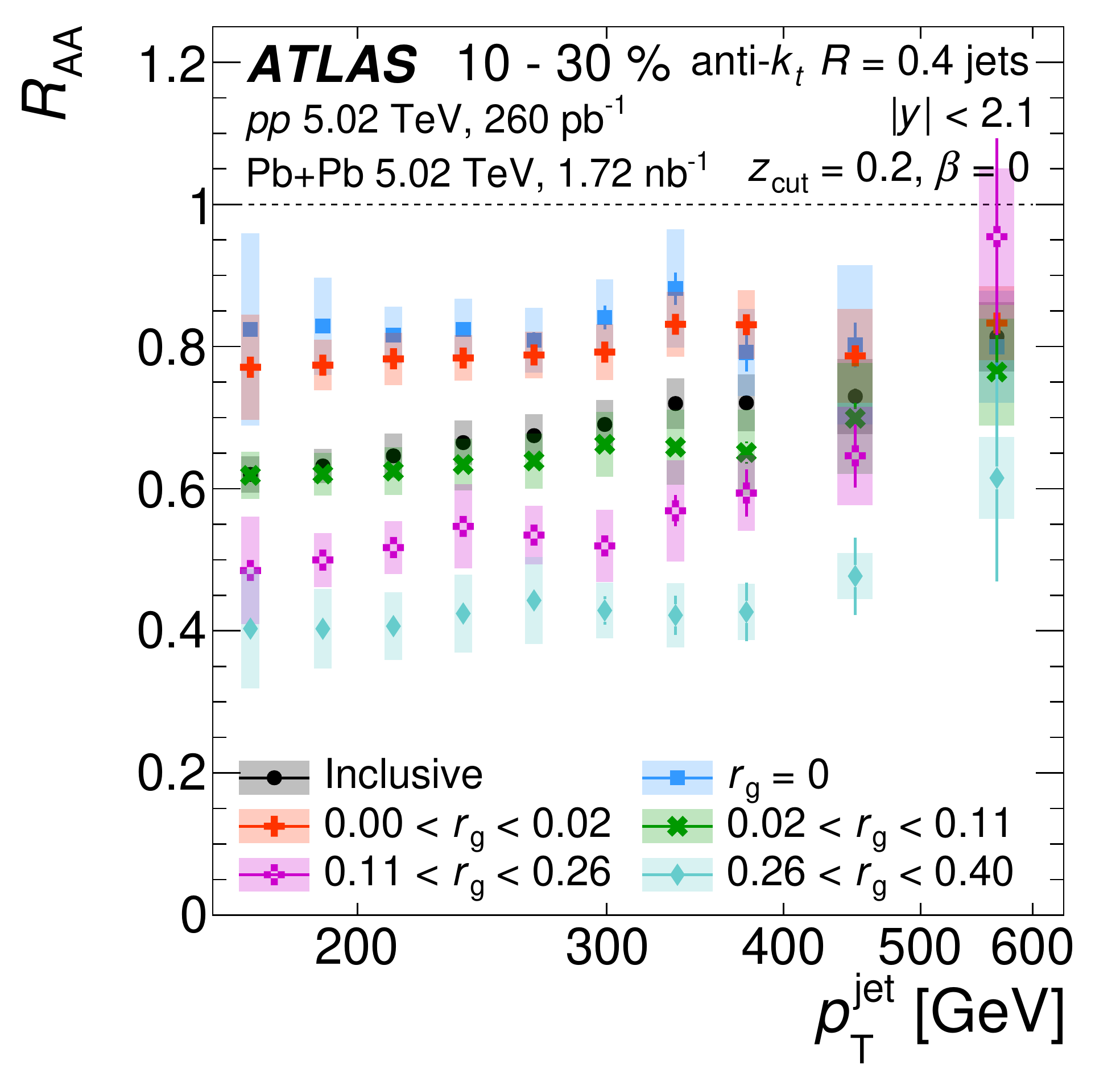}
\includegraphics[width=0.475\textwidth]{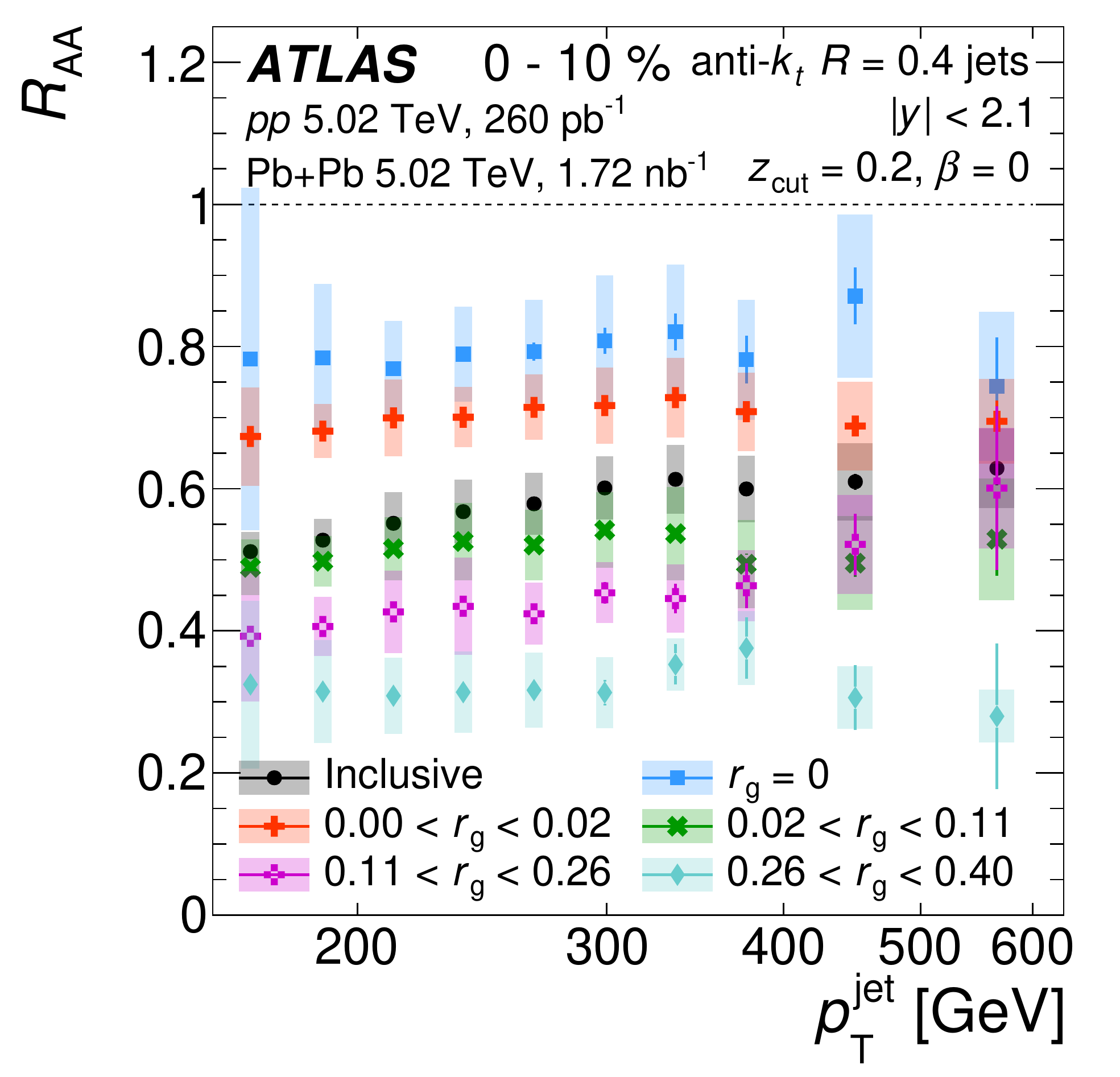}
\caption{Nuclear modification factor, \RAA,  as a function of \ptjet for \SoftDrop groomed jets with $|y| < 2.1$ in four centrality intervals and four intervals of \rg. Groomed jet \RAA values are compared with \RAA values of jets without significant splitting identified by the \SoftDrop procedure ($\rg = 0$) and jets without grooming (inclusive). The error bars represent statistical uncertainties while the shaded bars represent bin-wise correlated systematic uncertainties. The uncertainties in the \pp luminosity ($1.6\%$) and \avgTAA are not included, but are listed in Table~\ref{tab:lumi}.
}
\label{fig:jetraa_pt}
\end{figure}

\begin{figure}[ht]
\centering
\includegraphics[width=0.475\textwidth]{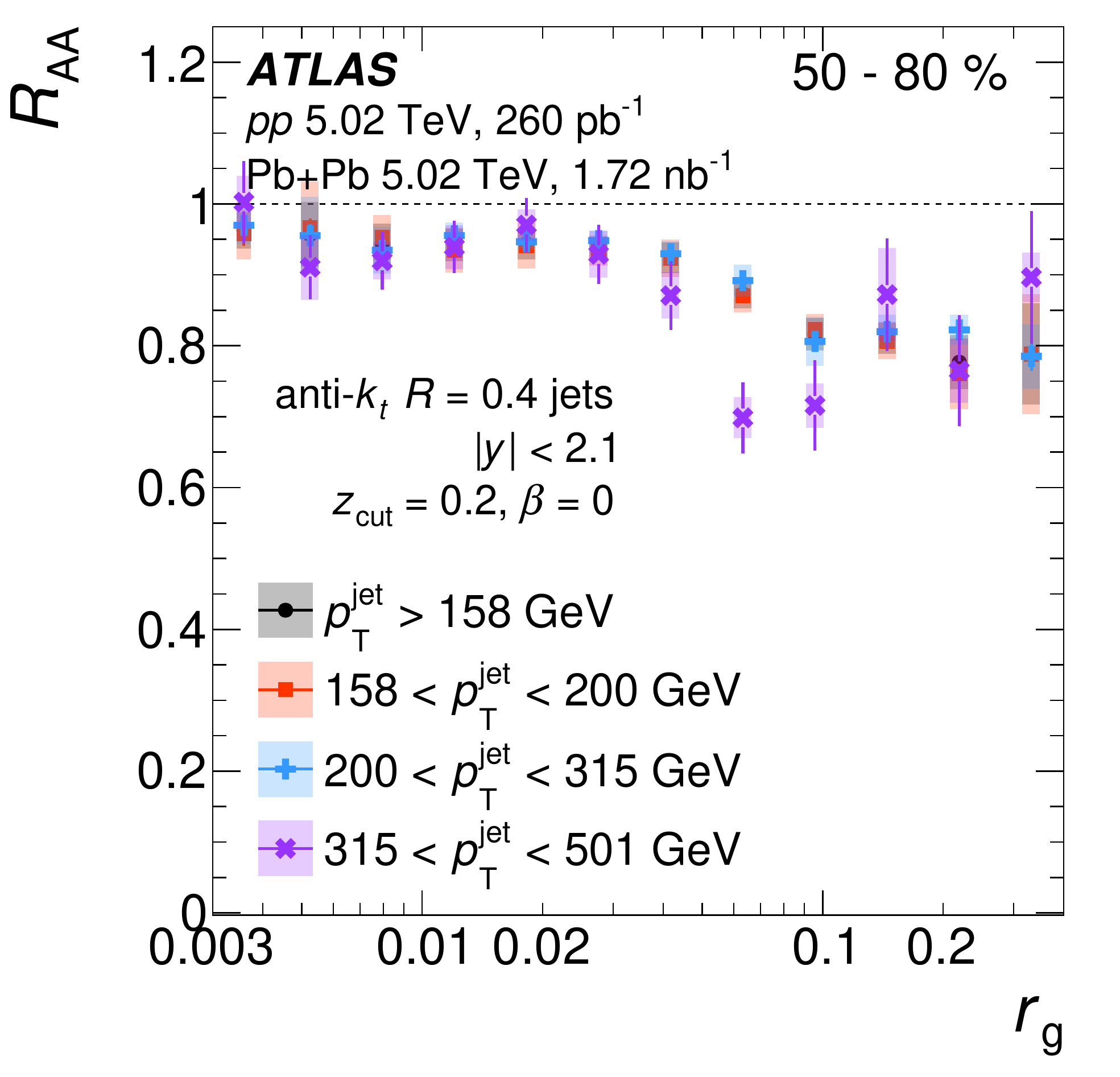}
\includegraphics[width=0.475\textwidth]{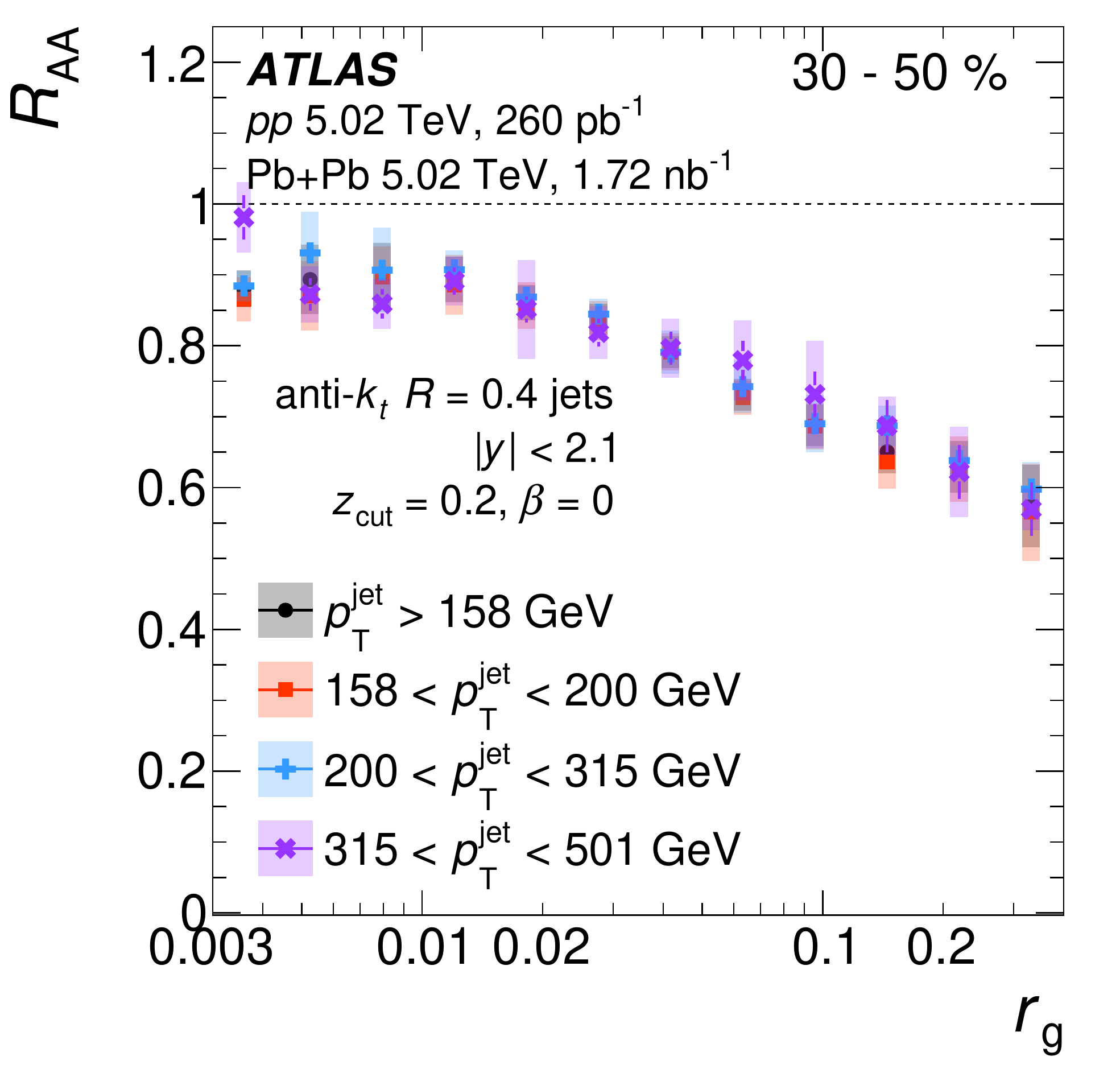}
\includegraphics[width=0.475\textwidth]{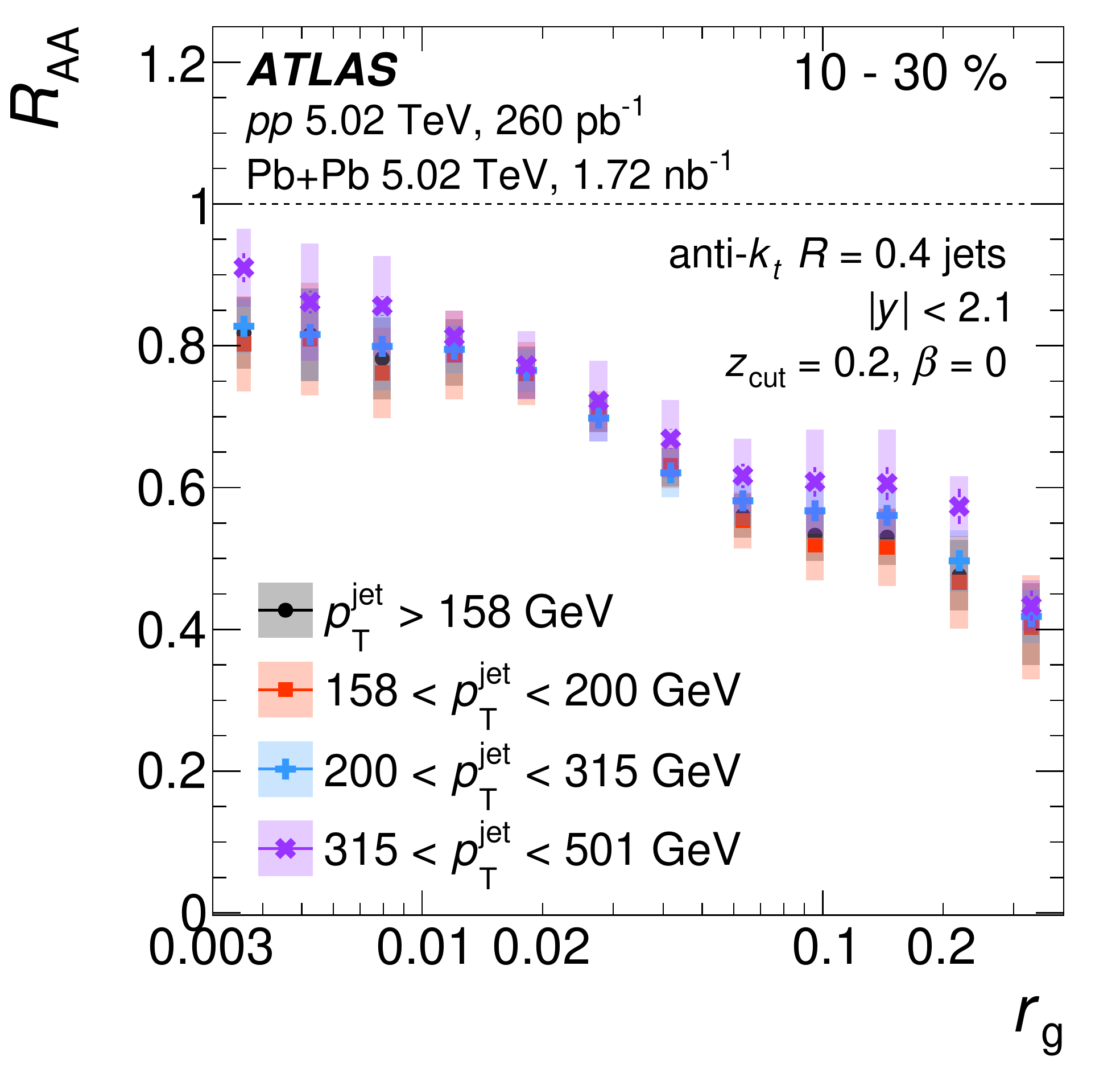}
\includegraphics[width=0.475\textwidth]{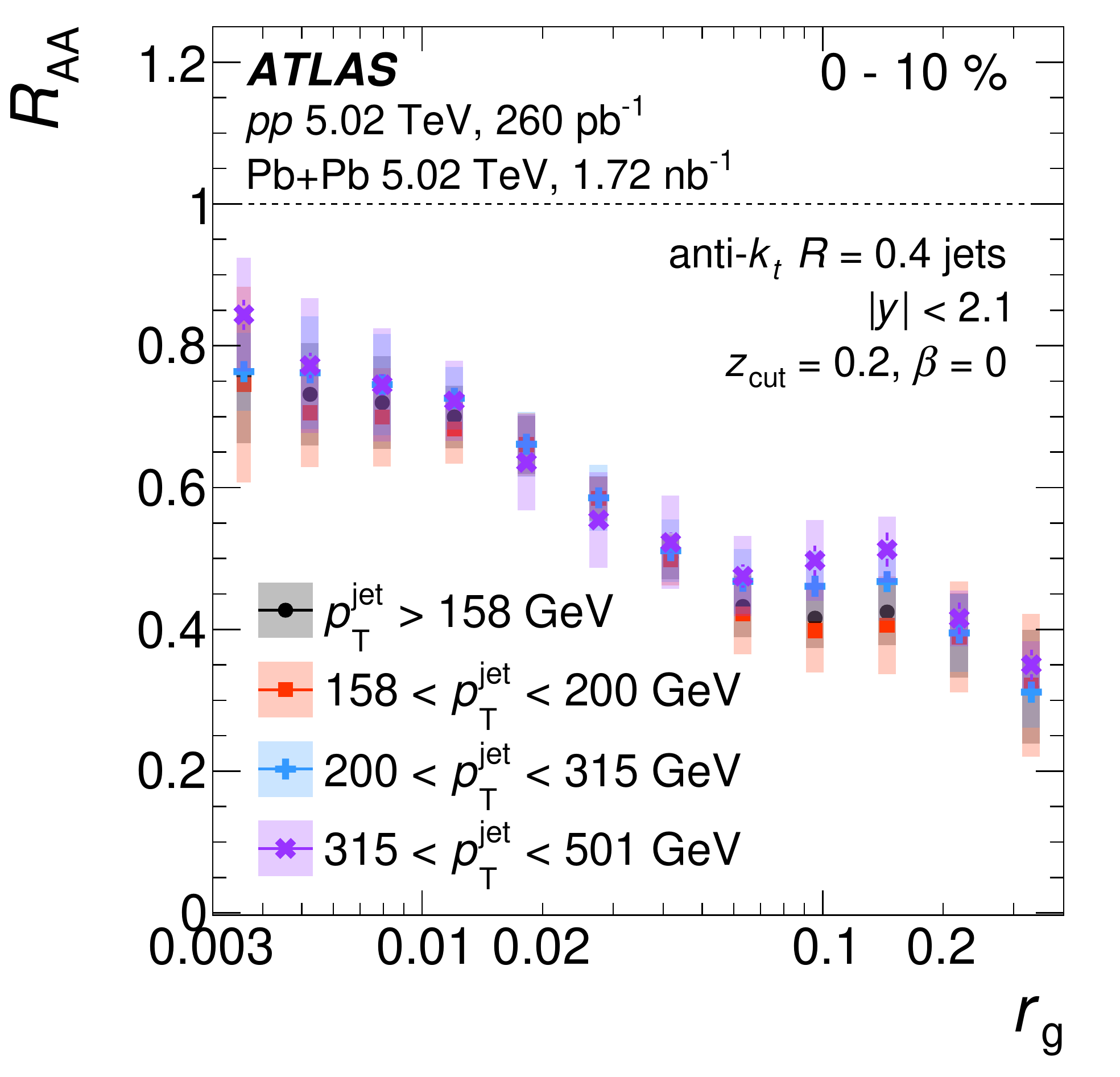}
\caption{Nuclear modification factor, \RAA,  as a function of \rg for \SoftDrop groomed jets with $|y| < 2.1$ in four centrality intervals and three intervals of \ptjet, in comparison with the \pt-inclusive results.  The error bars represent statistical uncertainties while the shaded bars represent bin-wise correlated systematic uncertainties. The uncertainties in the \pp luminosity ($1.6\%$) and \avgTAA are not included, but are listed in Table~\ref{tab:lumi}.
}
\label{fig:jetraa_thetag}
\end{figure}

The nuclear modification factor \RAA, which quantifies the jet suppression in \PbPb collisions relative to \pp collisions, is shown as a function of \ptjet in Figure~\ref{fig:jetraa_pt} and as a function of \rg in Figure~\ref{fig:jetraa_thetag} for four centrality intervals.  The \RAA value for inclusive jets is observed to vary from ${\approx}0.95$ for jets in the 50--80\% centrality interval to ${\approx}0.55$ for the most central collisions. The \RAA values for inclusive jets are also observed to have an upward slope with increasing \ptjet and are consistent with the results measured for jets using 2015 \PbPb data with a lower integrated luminosity of $0.49~\inb$~\cite{HION-2017-10}.
 
The values of \RAA are observed to depend significantly on \rg, with a clear ordering with respect to the splitting angle.
Jets that fail the \SoftDrop grooming procedure or have $\rg <0.02$ are observed to be the least suppressed with an \RAA value of ${\approx}0.75$ in central \PbPb collisions.
In contrast, jets with the widest splitting angle between their hardest subjets are observed to be suppressed with an \RAA value of ${\approx}0.3$. Notably, the difference in the \RAA values is largest between jets having $\rg$ values below or above $0.02$, supporting a coherence picture of jet quenching~\cite{Caucal:2019uvr, Caucal:2020uic}. The \RAA values for jets in different \rg regions are observed to have a weaker dependence on \ptjet in comparison with the \RAA behavior for the inclusive jets, especially in more central collisions.
The increase in \RAA for inclusive jets as a function of \ptjet may be explained as arising from tighter jet collimation with increasing \ptjet. Jets with lower \rg values in \PbPb collisions are significantly less suppressed than jets with large \rg values and they also constitute a larger fraction of the inclusive jets with increasing \ptjet, resulting in a rising trend for \RAA as a function of \ptjet.
As a function of \rg, \RAA decreases smoothly with increasing \rg (shown on logarithmic scale) in all centrality intervals and is again noted to not significantly depend on \ptjet in the measured kinematic range.
 
 
\FloatBarrier
 
The jet suppression measurements are compared with predictions from various theoretical models and frameworks including those from the JETSCAPE MC framework~\cite{Putschke:2019yrg, JETSCAPE:2022jer}, Caucal et al.~\cite{Caucal:2018dla, Caucal:2019uvr, Caucal:2020uic} and Ringer et al.~\cite{Ringer:2019rfk}. JETSCAPE provides a multistage model of jet evolution by combining several jet-quenching formalisms applied in its range of validity in shower energy and virtuality. The JETSCAPE predictions combine jet energy-loss calculations from the MATTER~\cite{Majumder:2013re} and LBT~\cite{He:2015pra, Cao:2016gvr} models at high and low virtualities of the parton shower in the QGP medium, respectively. The parameters used in generating the predictions, including the transport coefficient ($\hat{q}$) and the virtuality at which to switch parton energy loss from MATTER to LBT ($\mathrm{\textit{Q}}_\mathrm{sw}$), can be found in Ref.~\cite{JETSCAPE:2022jer}. JETSCAPE also includes a virtuality-dependent onset of coherence effects in order to model the inability of the medium to resolve narrow splittings of highly virtual partons at earlier timescales. The model from Caucal et al. presents a perturbative QCD (pQCD) picture of jet quenching dictated by the factorization of medium-induced parton branchings and vacuum-like emissions in a longitudinally expanding medium. An important aspect of this model is the onset of incoherent jet energy loss beyond a critical angle between the branches of the hardest jet splitting. It should be noted that the calculations from Caucal et al. are purely at the parton level and hadronization effects are not included. The analytic calculations for jet \RAA by Ringer et al. are based on medium-modified quark and gluon jet functions using a MC sampling approach. In conjunction with the quark and gluon differential jet energy loss, transverse momentum broadening is applied to all the subjets in this model, based on the transport coefficient parameter.
 
\begin{figure}[ht]
\centering
\includegraphics[width=0.42\textwidth]{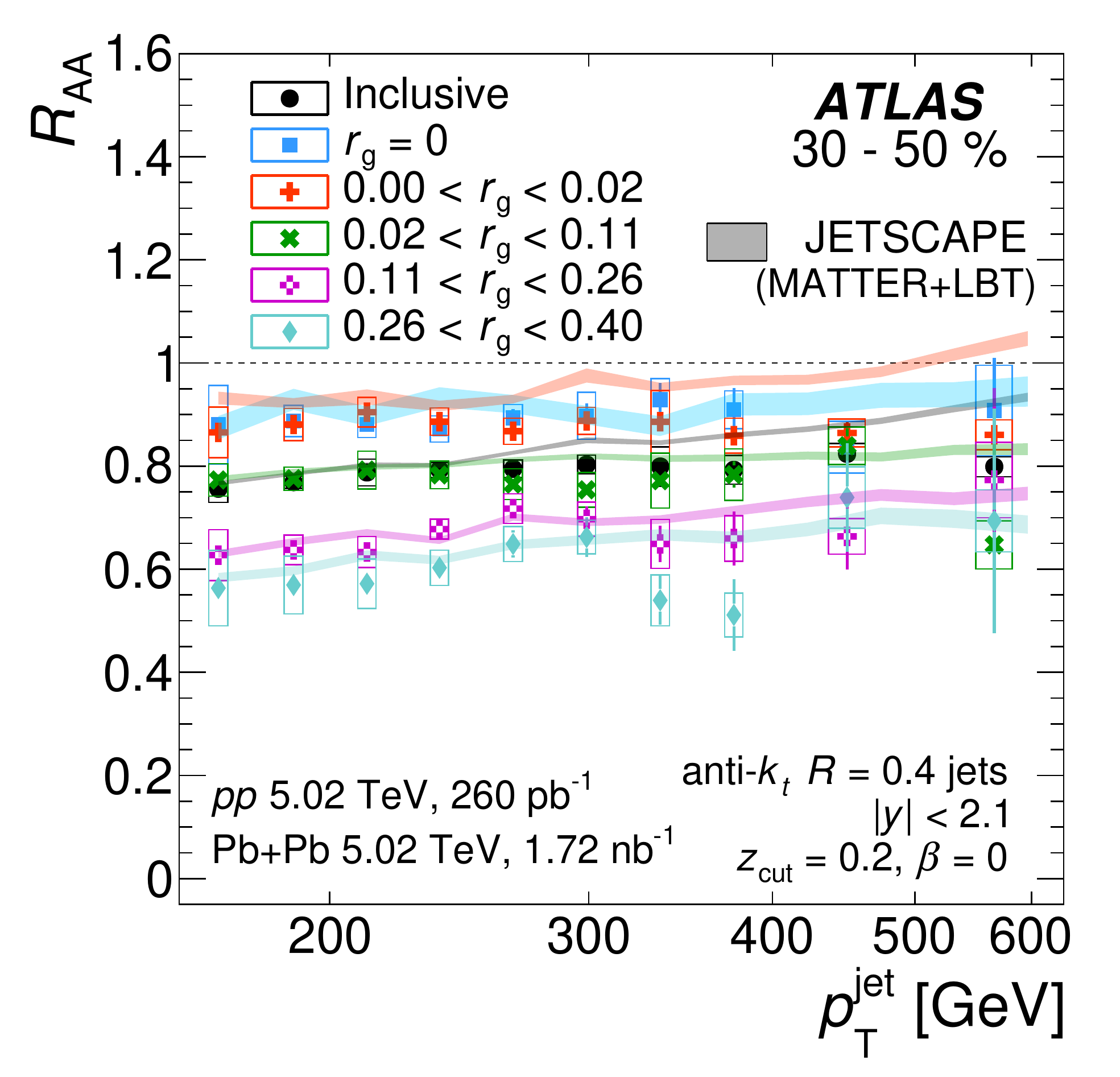}
\includegraphics[width=0.42\textwidth]{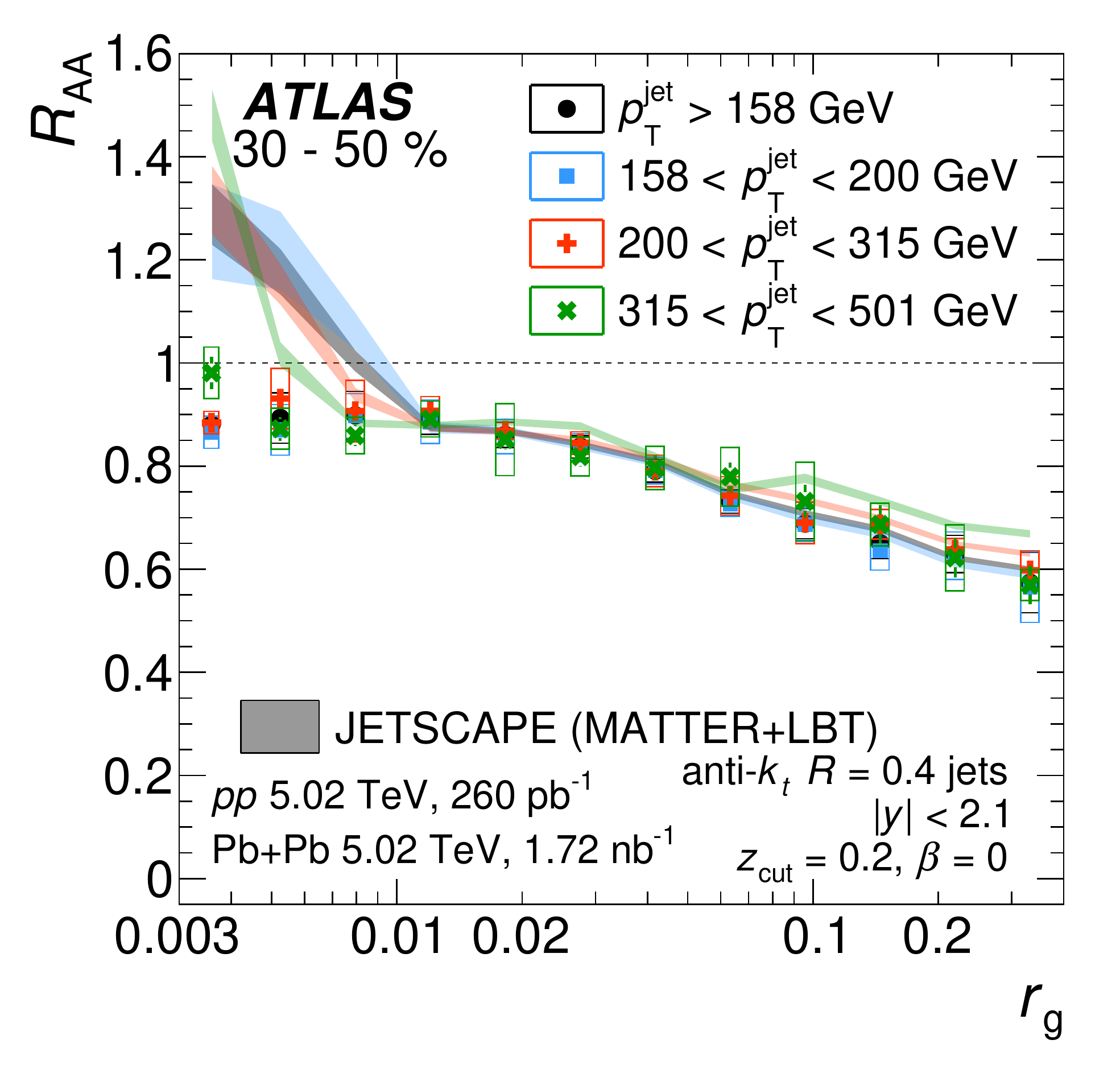}
\includegraphics[width=0.42\textwidth]{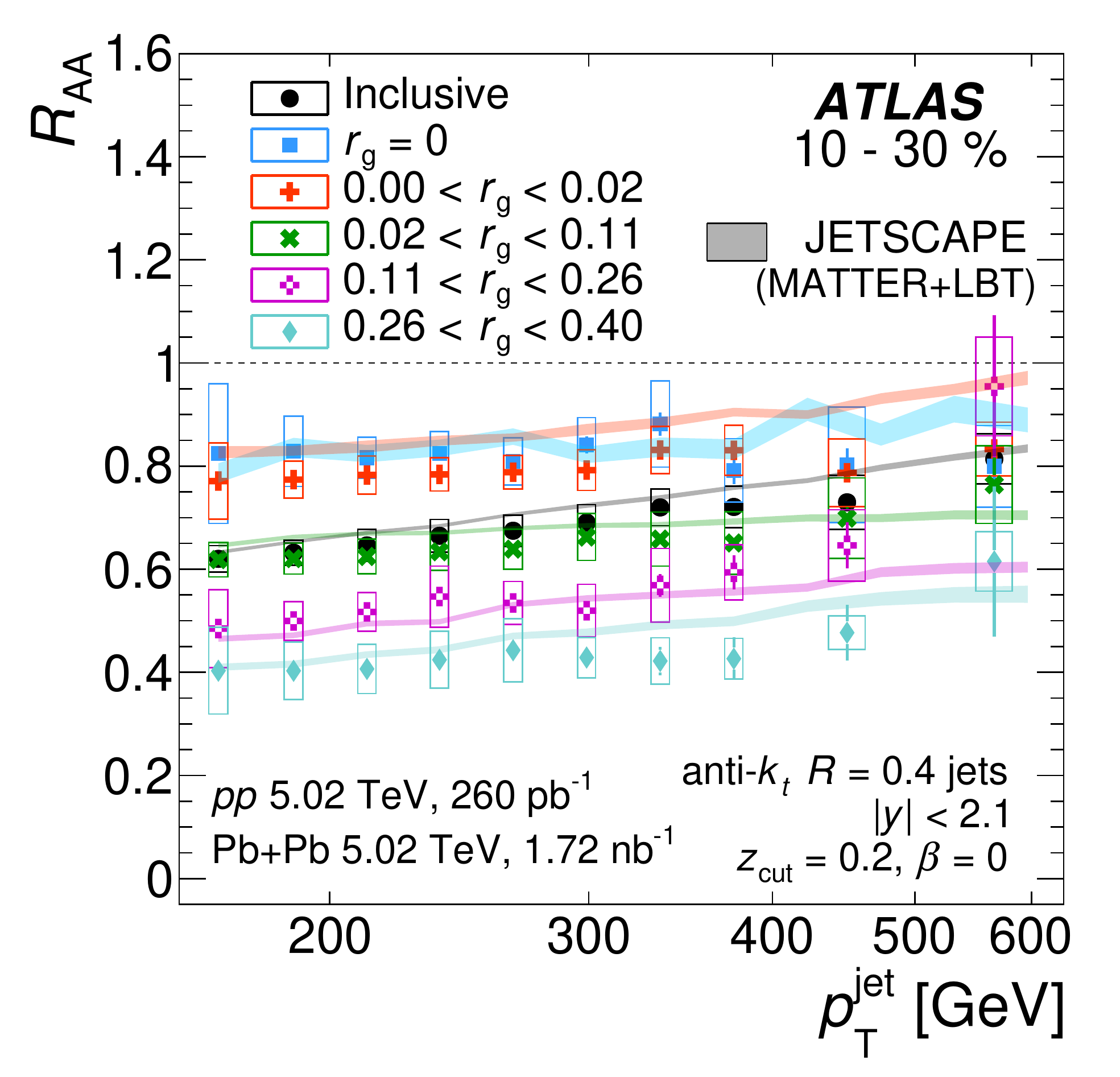}
\includegraphics[width=0.42\textwidth]{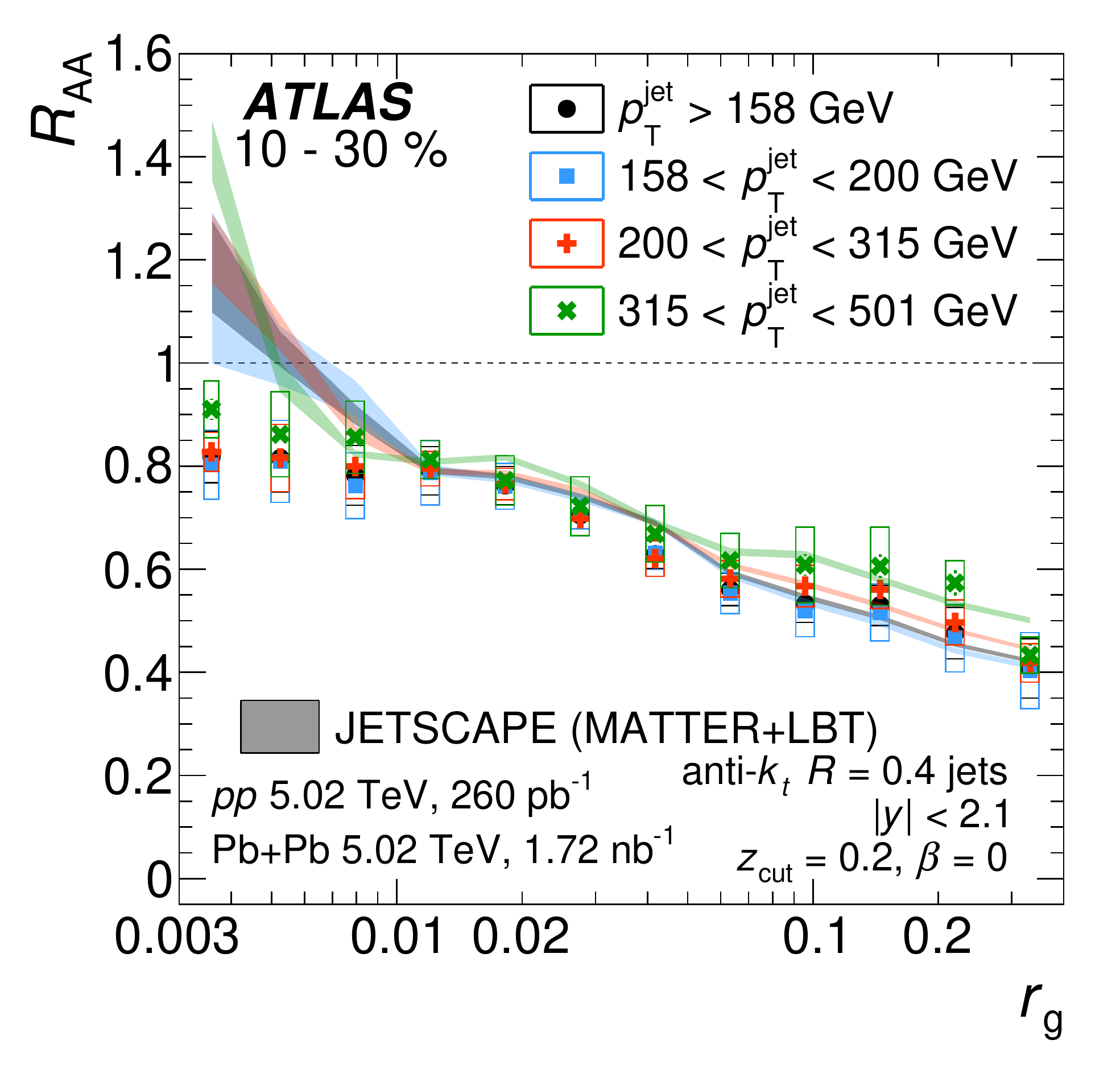}
\includegraphics[width=0.42\textwidth]{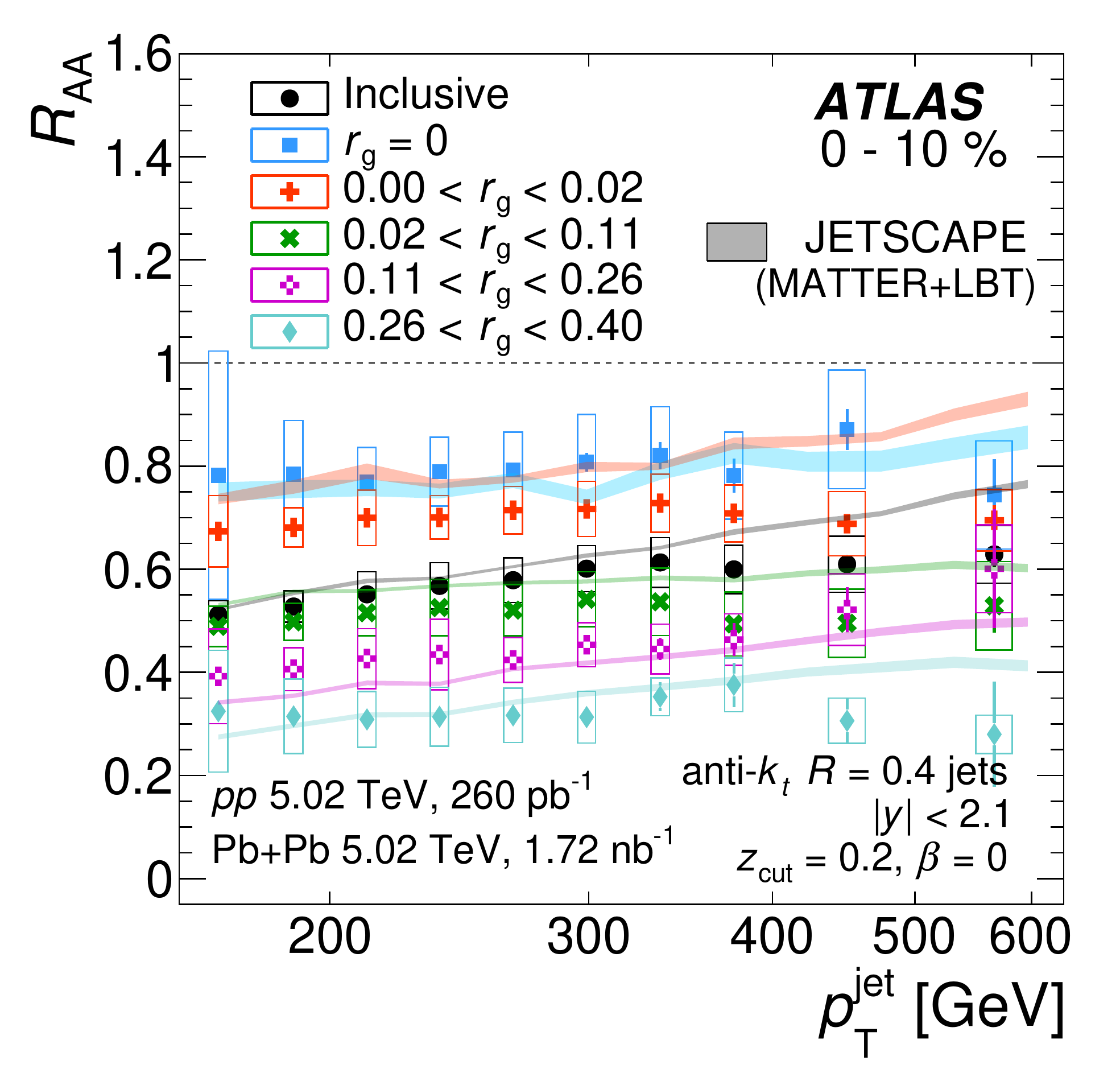}
\includegraphics[width=0.42\textwidth]{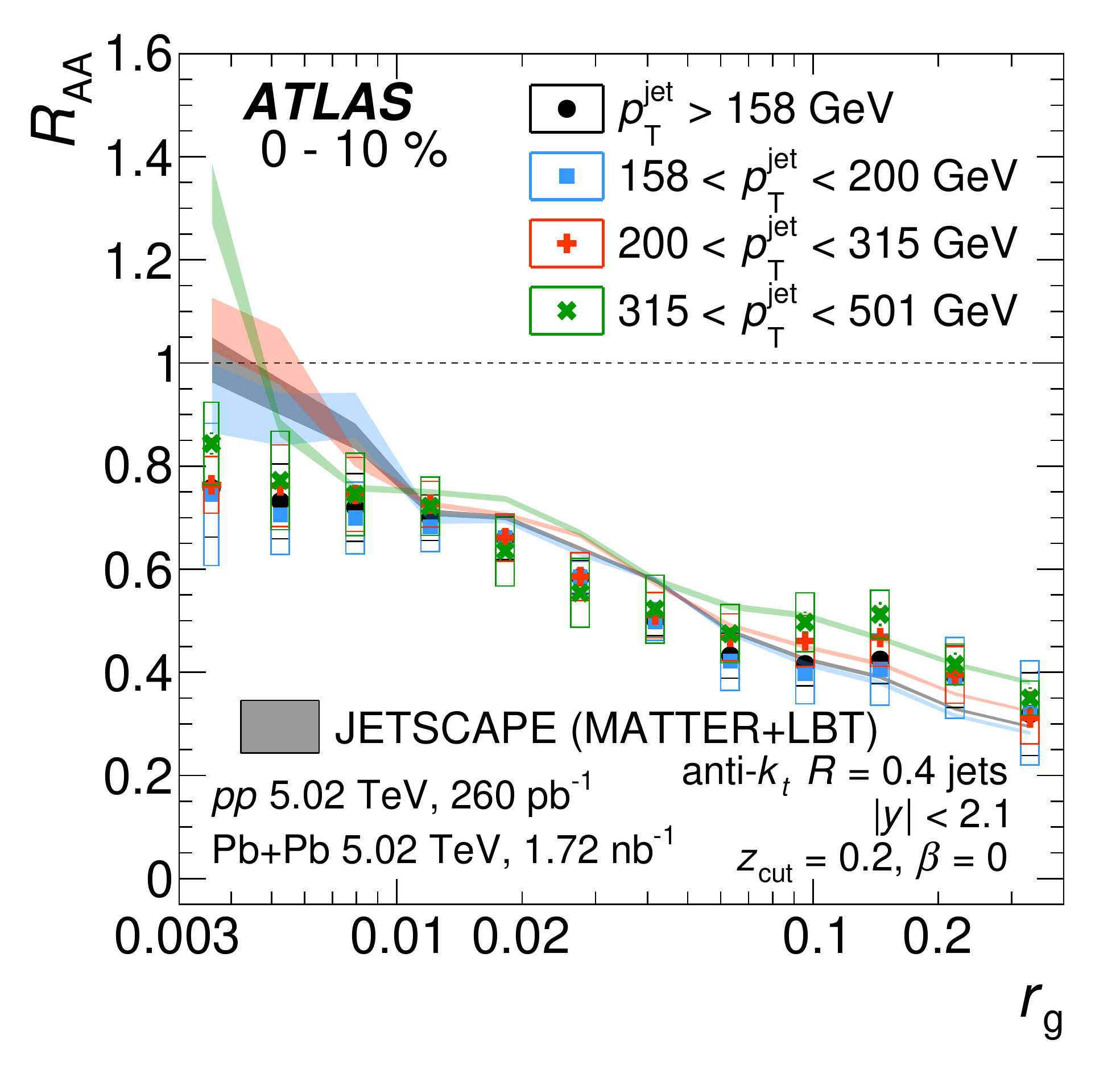}
\caption{Comparison of \RAA as a function of (left) \ptjet for inclusive jets and for four intervals of \rg, (right) \rg for inclusive jets and for three intervals of \ptjet, and in three centrality intervals of \PbPb events with predictions from the JETSCAPE framework~\cite{Putschke:2019yrg, JETSCAPE:2022jer}. The error bars and the open boxes around the data points represent the statistical and systematic uncertainties, respectively. The uncertainties in the \pp luminosity ($1.6\%$) and \avgTAA are not included, but are listed in Table~\ref{tab:lumi}. The widths of the bands representing the JETSCAPE predictions indicate their statistical uncertainties.}
\label{fig:jetscapetheorycomp_jetraa}
\end{figure}

\begin{figure}[ht]
\centering
\includegraphics[width=0.475\textwidth]{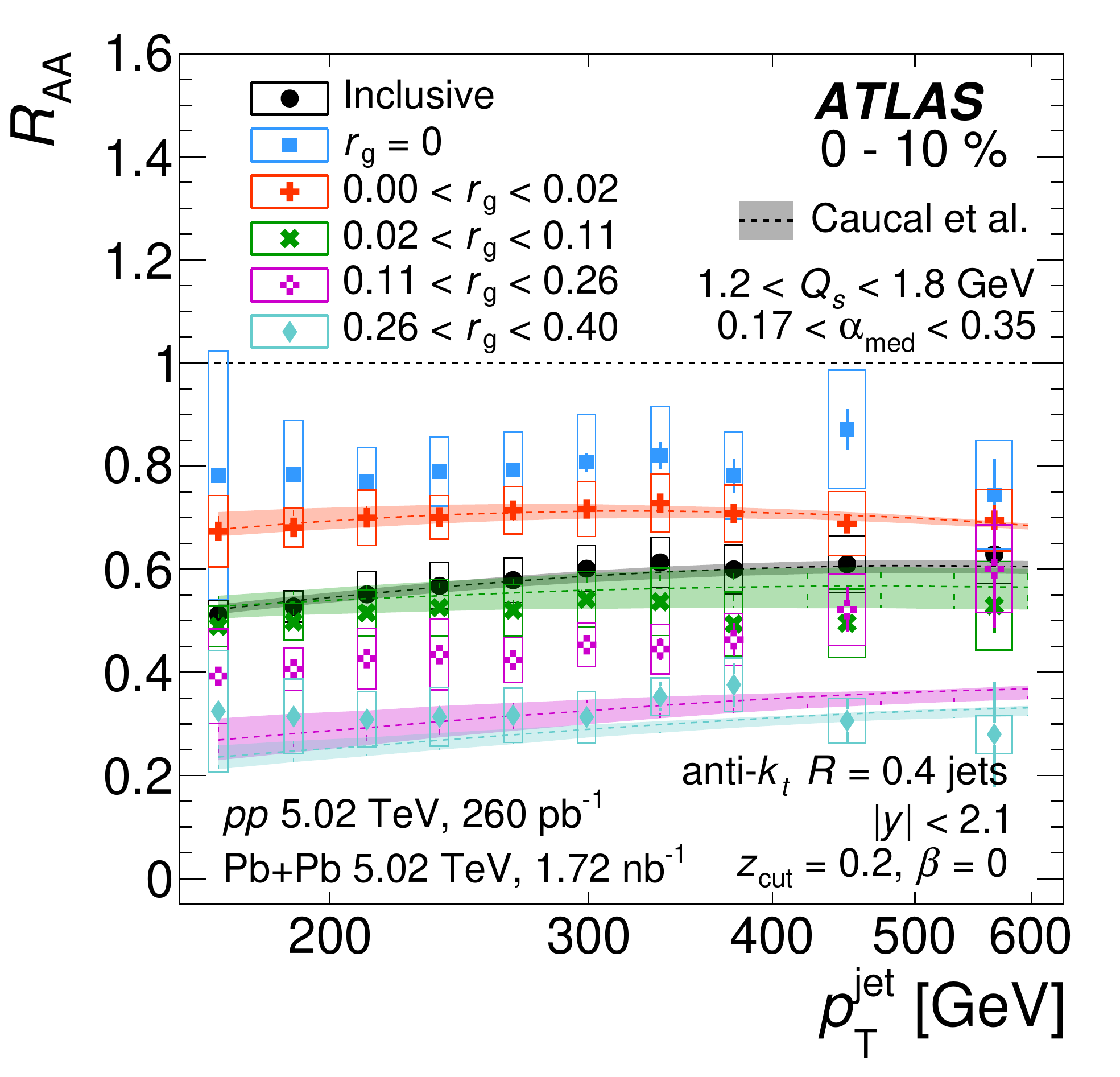}
\includegraphics[width=0.475\textwidth]{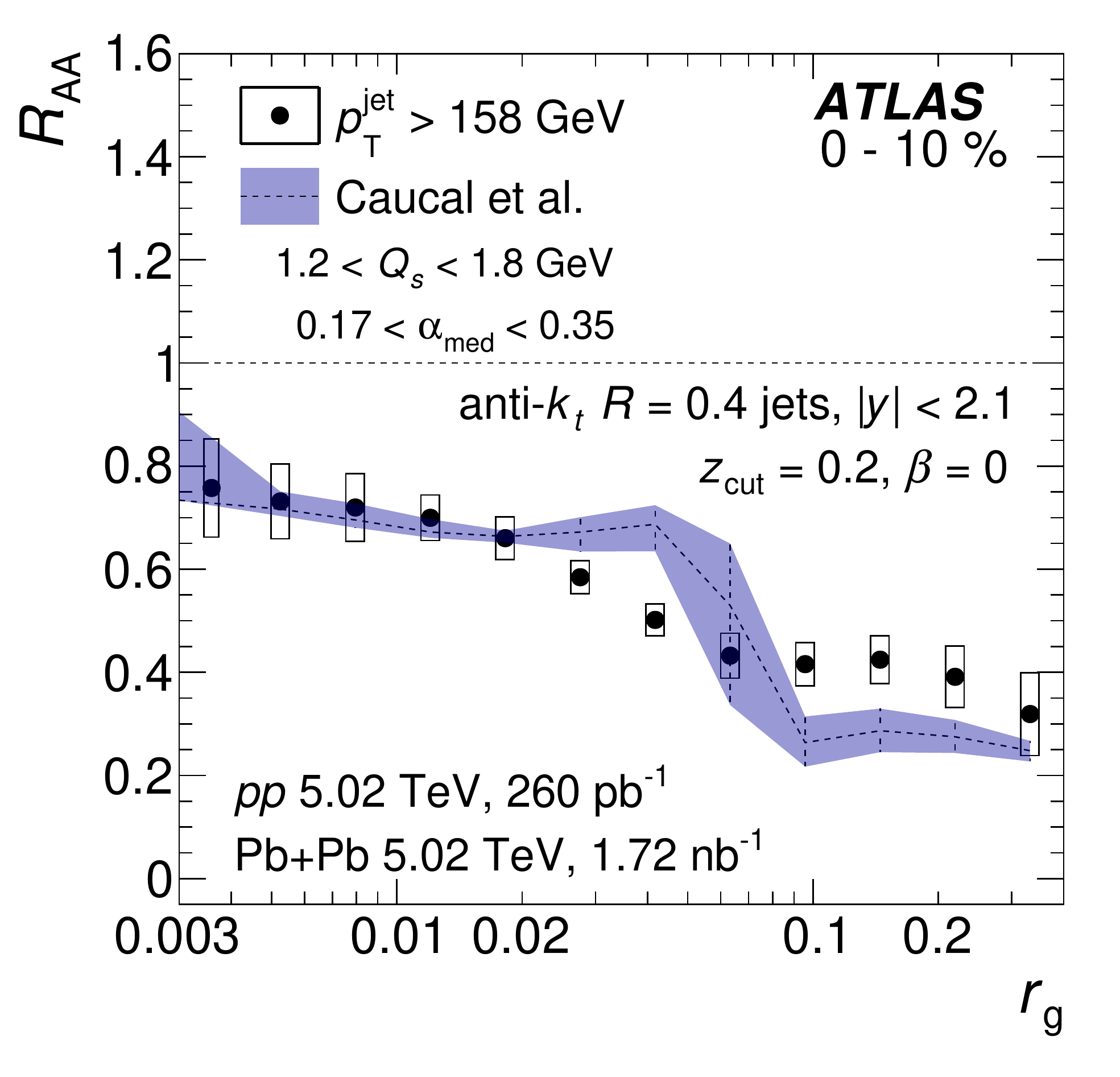}
\caption{Comparison of \RAA as a function of (left) \ptjet for inclusive jets and for four intervals of $r_\mathrm{g}$ and (right) as a function of \rg for inclusive jets in 0--10\% centrality \PbPb events with theoretical predictions from the pQCD framework of Caucal et al., described in Refs.~\cite{Caucal:2018dla, Caucal:2019uvr, Caucal:2020uic}. The error bars and the open boxes around the data points represent the statistical and systematic uncertainties, respectively. The uncertainties in the \pp luminosity ($1.6\%$) and \avgTAA are not included, but are listed in Table~\ref{tab:lumi}. The shaded areas around the theoretical predictions represent the uncertainties arising from the variation of the transport coefficient ($\hat{q}$) and shower cutoff parameters.}
\label{fig:pQCDtheorycomp_jetraa}
\end{figure}

\begin{figure}[ht]
\centering
\includegraphics[width=0.344\textwidth]{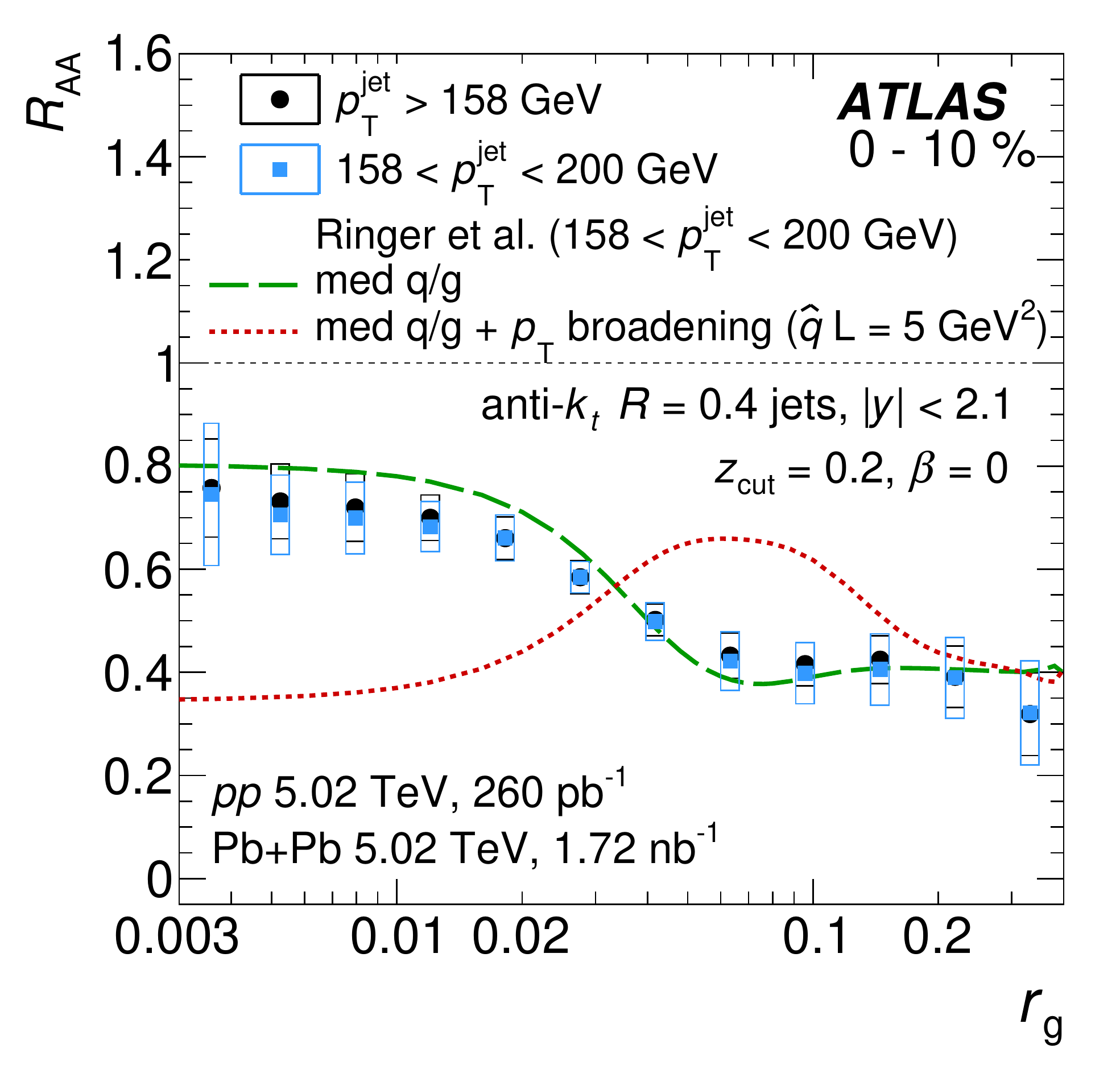}
\hspace{-0.44cm}
\includegraphics[width=0.344\textwidth]{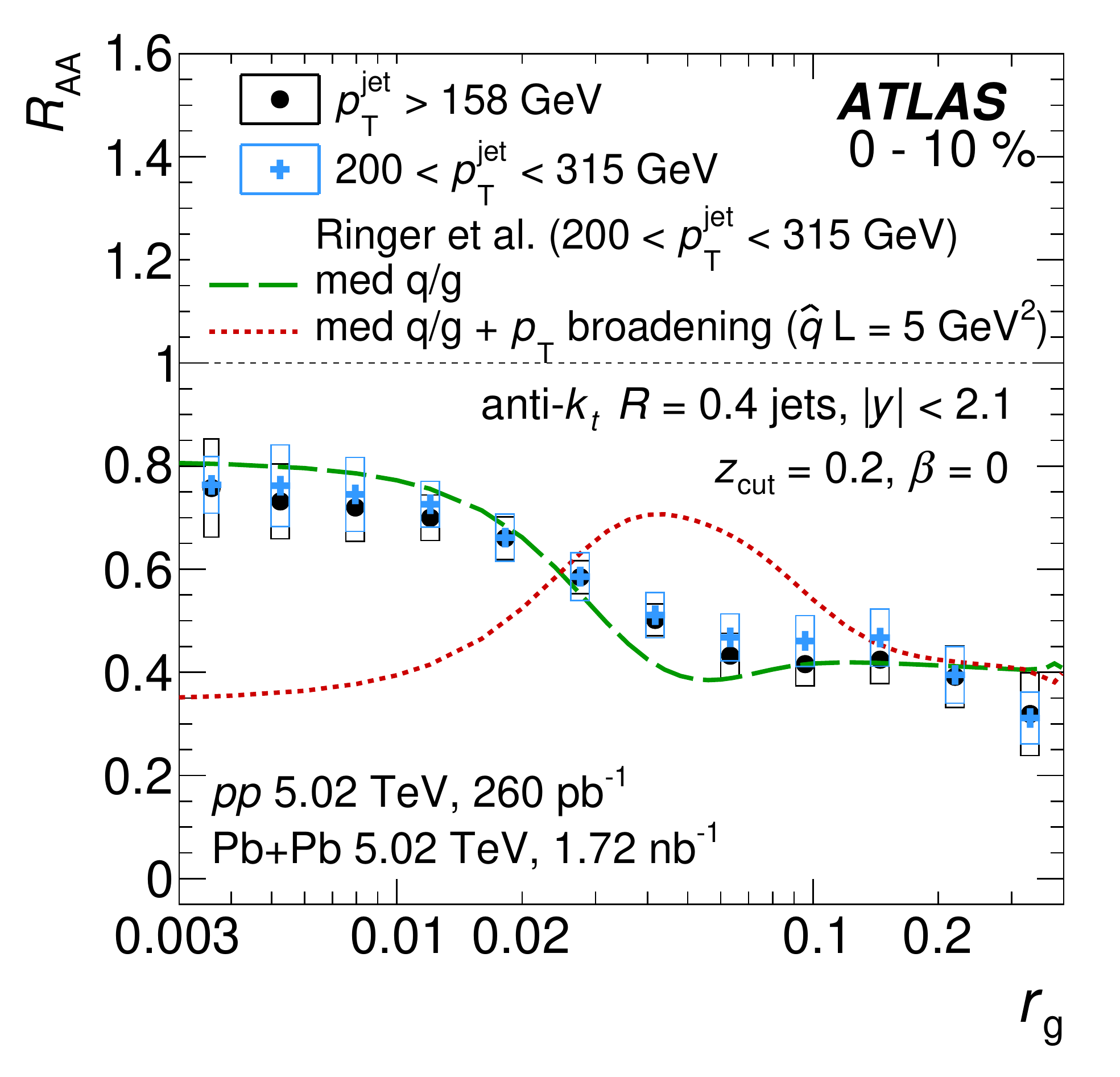}
\hspace{-0.44cm}
\includegraphics[width=0.344\textwidth]{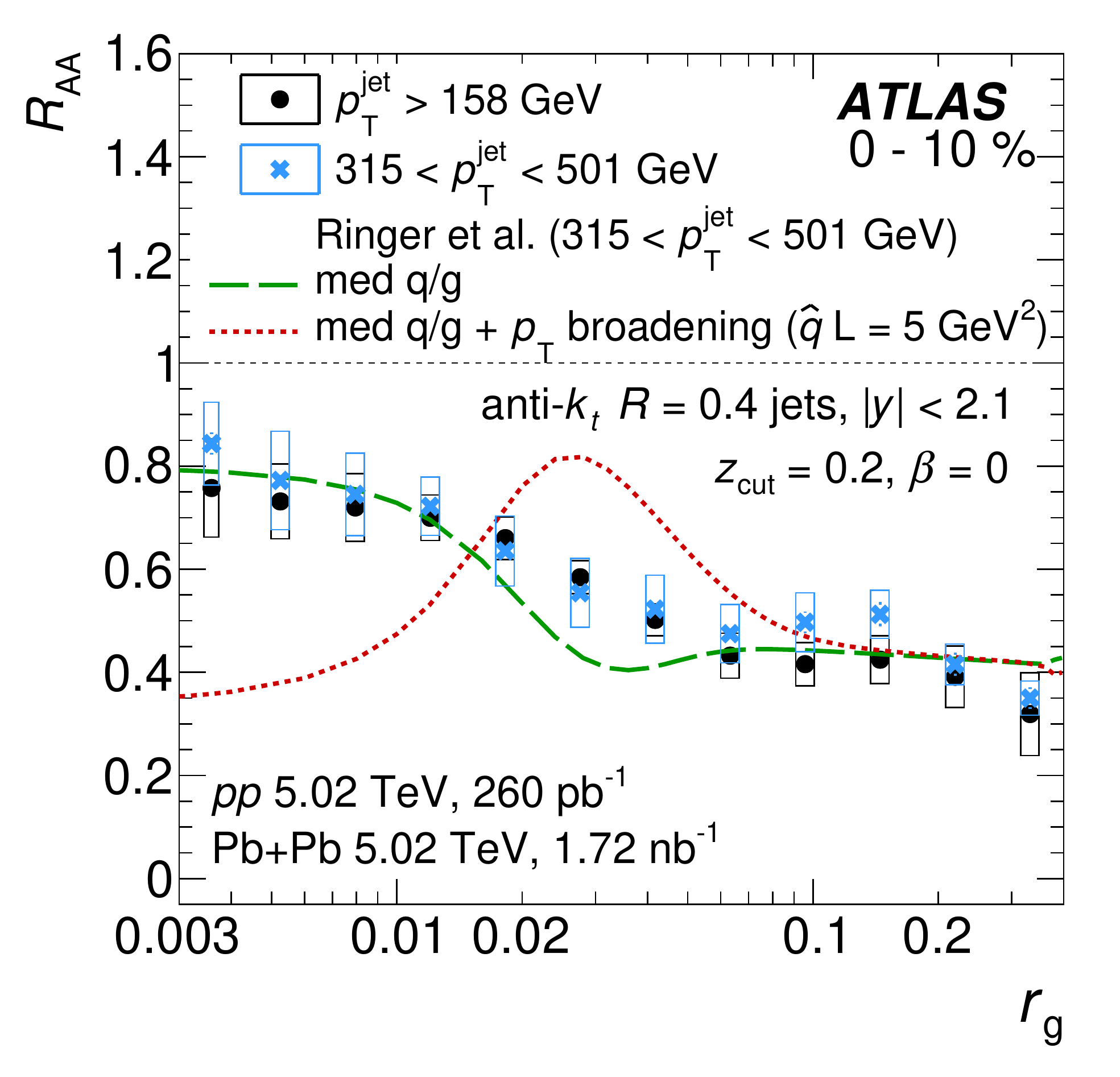}
\caption{Comparison of \RAA as a function of \rg for three intervals of $\ptjet$ in 0--10\% centrality \PbPb events with theoretical predictions from Ringer et al.~\cite{Ringer:2019rfk}. The error bars and the open boxes around the data points represent the statistical and systematic uncertainties, respectively. The uncertainties in the \pp luminosity ($1.6\%$) and \avgTAA are not included, but are listed in Table~\ref{tab:lumi}. The green curve (med q/g) represents the quark and gluon differential energy loss setting. The red curve (med q/g + \pt broadening) adds transverse momentum broadening effects for all subjets in the medium.}
\label{fig:qgtheorycomp_jetraa}
\end{figure}

The jet \RAA results are compared with predictions from the above frameworks and models as a function of \ptjet and \rg in Figures~\ref{fig:jetscapetheorycomp_jetraa}--\ref{fig:qgtheorycomp_jetraa}. Predictions from JETSCAPE (MATTER+LBT), shown in Figure~\ref{fig:jetscapetheorycomp_jetraa}, are able to capture the trend of \rg-dependent jet suppression in three centrality intervals of \PbPb collisions. However, JETSCAPE overestimates the \RAA values for jets in the low \rg (${<}0.01$) region, especially at higher \ptjet. The \RAA predictions from the pQCD picture modeled by Caucal et al., shown in Figure~\ref{fig:pQCDtheorycomp_jetraa}, qualitatively describe the \rg-dependent suppression trend observed in the most central collisions.
Compared to the data in Figure~\ref{fig:pQCDtheorycomp_jetraa}, the parton-level pQCD model calculations predict a sharper drop in \RAA around the critical angle beyond which incoherent jet energy loss sets in. The quark vs. gluon differential quenching effects modeled by Ringer et al., labeled `med q/g' in Figure~\ref{fig:qgtheorycomp_jetraa}, are able to describe the \rg-dependent jet suppression in three intervals of \ptjet. Transverse momentum broadening effects are included in the model calculations by adding \pt kicks perpendicular to the direction of each subjet by sampling from a simple Gaussian distribution, whose width is determined using the transport coefficient parameter~\cite{Ringer:2019rfk}.  The combined model calculations with quark and gluon differential jet energy loss and transverse momentum broadening effects are labeled as ‘med q/g + \pt broadening’ in Figure~\ref{fig:qgtheorycomp_jetraa}. Adding \pt-broadening effects to the model results in a significant suppression of jets at lower \rg values and a peak in \RAA at mid-\rg values, neither of which are observed in data.
 
\clearpage
\section{Conclusion}
Jet production in \PbPb and \pp collisions, both at $\sqn = 5.02~\TeV$, is measured with the ATLAS detector at the LHC, using integrated luminosities of $1.72~\inb$ and $260~\ipb$, respectively. The jets are reconstructed using the \antikt\ algorithm with $R=0.4$ and groomed using the \SoftDrop procedure with parameters $z_\mathrm{cut} = 0.2$ and $\beta=0$. The jet differential cross-sections and yields are presented as a function of \ptjet and \rg, the opening angle of the hardest splitting, and in intervals of collision centrality. Results unfolded to the particle level are presented for jets with $\ptjet > 158$~\GeV, $|y| < 2.1$, and $0 \leq \rg < 0.4$.
 
The \rg distributions obtained in \pp collisions are observed to peak at lower values of \rg with increasing \ptjet, indicating that higher-momentum jets are more collimated.
The differential cross-sections as a function of \ptjet and \rg are compared with results from MC event generators (\PYTHIA[8], \HERWIG, and \SHERPA), with the \PYTHIA[8] predictions best describing the shape of the \ptjet and \rg spectra seen in \pp collisions.
 
The jet energy loss and the resulting suppression in \PbPb collisions is quantified by the nuclear modification factor, \RAA. The \RAA values are observed to depend primarily on the jet \rg value and the event centrality. In the most central collisions, the \RAA value ranges between $\RAA\approx 0.75$ for the most collimated jets ($\rg<0.02$) and $\RAA\approx 0.3$ for the widest jets ($0.26< \rg < 0.4$). The \RAA values are observed to decrease monotonically with increasing \rg, and do not depend strongly on \ptjet for a selected \rg range. Although the energy loss is not measured directly and must be inferred from the \RAA values and slope of the differential cross-sections as a function of \ptjet, the \ptjet-dependence is observed to be similar for different \rg ranges. These features together with the dependence of \RAA on \rg indicate that jets with larger opening angles lose more energy. Predictions from theoretical models of jet quenching are compared with the \RAA measurements presented here and are observed to describe the \rg-dependent jet suppression features with varying accuracy. The results presented here are qualitatively consistent with a picture of jet quenching dictated by coherence and provide the most direct evidence supporting this approach so far.
 
\section*{Acknowledgments}

% The next lines are included from the .//acknowledgements/Acknowledgements.tex input file

We thank CERN for the very successful operation of the LHC, as well as the
support staff from our institutions without whom ATLAS could not be
operated efficiently.
 
We acknowledge the support of
ANPCyT, Argentina;
YerPhI, Armenia;
ARC, Australia;
BMWFW and FWF, Austria;
ANAS, Azerbaijan;
CNPq and FAPESP, Brazil;
NSERC, NRC and CFI, Canada;
CERN;
ANID, Chile;
CAS, MOST and NSFC, China;
Minciencias, Colombia;
MEYS CR, Czech Republic;
DNRF and DNSRC, Denmark;
IN2P3-CNRS and CEA-DRF/IRFU, France;
SRNSFG, Georgia;
BMBF, HGF and MPG, Germany;
GSRI, Greece;
RGC and Hong Kong SAR, China;
ISF and Benoziyo Center, Israel;
INFN, Italy;
MEXT and JSPS, Japan;
CNRST, Morocco;
NWO, Netherlands;
RCN, Norway;
MEiN, Poland;
FCT, Portugal;
MNE/IFA, Romania;
MESTD, Serbia;
MSSR, Slovakia;
ARRS and MIZ\v{S}, Slovenia;
DSI/NRF, South Africa;
MICINN, Spain;
SRC and Wallenberg Foundation, Sweden;
SERI, SNSF and Cantons of Bern and Geneva, Switzerland;
MOST, Taiwan;
TENMAK, T\"urkiye;
STFC, United Kingdom;
DOE and NSF, United States of America.
In addition, individual groups and members have received support from
BCKDF, CANARIE, Compute Canada and CRC, Canada;
PRIMUS 21/SCI/017 and UNCE SCI/013, Czech Republic;
COST, ERC, ERDF, Horizon 2020 and Marie Sk{\l}odowska-Curie Actions, European Union;
Investissements d'Avenir Labex, Investissements d'Avenir Idex and ANR, France;
DFG and AvH Foundation, Germany;
Herakleitos, Thales and Aristeia programmes co-financed by EU-ESF and the Greek NSRF, Greece;
BSF-NSF and MINERVA, Israel;
Norwegian Financial Mechanism 2014-2021, Norway;
NCN and NAWA, Poland;
La Caixa Banking Foundation, CERCA Programme Generalitat de Catalunya and PROMETEO and GenT Programmes Generalitat Valenciana, Spain;
G\"{o}ran Gustafssons Stiftelse, Sweden;
The Royal Society and Leverhulme Trust, United Kingdom.
 
The crucial computing support from all WLCG partners is acknowledged gratefully, in particular from CERN, the ATLAS Tier-1 facilities at TRIUMF (Canada), NDGF (Denmark, Norway, Sweden), CC-IN2P3 (France), KIT/GridKA (Germany), INFN-CNAF (Italy), NL-T1 (Netherlands), PIC (Spain), ASGC (Taiwan), RAL (UK) and BNL (USA), the Tier-2 facilities worldwide and large non-WLCG resource providers. Major contributors of computing resources are listed in Ref.~\cite{ATL-SOFT-PUB-2021-003}.

% End of text imported from the .//acknowledgements/Acknowledgements.tex input file

\printbibliography
 
\clearpage
\input{atlas_authlist}

\clearpage

\end{document}

%% file: atlas_authlist.tex
% ATLAS Collaboration author list
% Reference date of HION-2021-09 is 2022-03-16
% Author list last updated on date 01-DEC-23
% Data extracted on 01-Dec-2023 for paper reference HION-2021-09
% at 2:40pm
 
\begin{flushleft}
\hypersetup{urlcolor=black}
{\Large The ATLAS Collaboration}

\bigskip

\AtlasOrcid[0000-0002-6665-4934]{G.~Aad}$^\textrm{\scriptsize 101}$,
\AtlasOrcid[0000-0002-5888-2734]{B.~Abbott}$^\textrm{\scriptsize 119}$,
\AtlasOrcid[0000-0002-7248-3203]{D.C.~Abbott}$^\textrm{\scriptsize 102}$,
\AtlasOrcid[0000-0002-1002-1652]{K.~Abeling}$^\textrm{\scriptsize 55}$,
\AtlasOrcid[0000-0002-8496-9294]{S.H.~Abidi}$^\textrm{\scriptsize 29}$,
\AtlasOrcid[0000-0002-9987-2292]{A.~Aboulhorma}$^\textrm{\scriptsize 35e}$,
\AtlasOrcid[0000-0001-5329-6640]{H.~Abramowicz}$^\textrm{\scriptsize 150}$,
\AtlasOrcid[0000-0002-1599-2896]{H.~Abreu}$^\textrm{\scriptsize 149}$,
\AtlasOrcid[0000-0003-0403-3697]{Y.~Abulaiti}$^\textrm{\scriptsize 116}$,
\AtlasOrcid[0000-0003-0762-7204]{A.C.~Abusleme~Hoffman}$^\textrm{\scriptsize 136a}$
% ...
\bigskip
\\

% ...
$^{29}$Physics Department, Brookhaven National Laboratory, Upton NY; United States of America.\\
% ...
$^{35}$$^{(a)}$Facult\'e des Sciences Ain Chock, R\'eseau Universitaire de Physique des Hautes Energies - Universit\'e Hassan II, Casablanca;$^{(b)}$Facult\'{e} des Sciences, Universit\'{e} Ibn-Tofail, K\'{e}nitra;$^{(c)}$Facult\'e des Sciences Semlalia, Universit\'e Cadi Ayyad, LPHEA-Marrakech;$^{(d)}$LPMR, Facult\'e des Sciences, Universit\'e Mohamed Premier, Oujda;$^{(e)}$Facult\'e des sciences, Universit\'e Mohammed V, Rabat;$^{(f)}$Institute of Applied Physics, Mohammed VI Polytechnic University, Ben Guerir; Morocco.\\
% ...
$^{55}$II. Physikalisches Institut, Georg-August-Universit\"{a}t G\"ottingen, G\"ottingen; Germany.\\
% ...
$^{101}$CPPM, Aix-Marseille Universit\'e, CNRS/IN2P3, Marseille; France.\\
$^{102}$Department of Physics, University of Massachusetts, Amherst MA; United States of America.\\
% ...
$^{116}$Department of Physics, New York University, New York NY; United States of America.\\
% ...
$^{119}$Homer L. Dodge Department of Physics and Astronomy, University of Oklahoma, Norman OK; United States of America.\\
% ...
$^{136}$$^{(a)}$Departamento de F\'isica, Pontificia Universidad Cat\'olica de Chile, Santiago;$^{(b)}$Millennium Institute for Subatomic physics at high energy frontier (SAPHIR), Santiago;$^{(c)}$Instituto de Investigaci\'on Multidisciplinario en Ciencia y Tecnolog\'ia, y Departamento de F\'isica, Universidad de La Serena;$^{(d)}$Universidad Andres Bello, Department of Physics, Santiago;$^{(e)}$Instituto de Alta Investigaci\'on, Universidad de Tarapac\'a, Arica;$^{(f)}$Departamento de F\'isica, Universidad T\'ecnica Federico Santa Mar\'ia, Valpara\'iso; Chile.\\
% ...
$^{149}$Department of Physics, Technion, Israel Institute of Technology, Haifa; Israel.\\
$^{150}$Raymond and Beverly Sackler School of Physics and Astronomy, Tel Aviv University, Tel Aviv; Israel.\\
% ...
$^{a}$ Also Affiliated with an institute covered by a cooperation agreement with CERN.\\
$^{b}$ Also at An-Najah National University, Nablus; Palestine.\\
$^{c}$ Also at Borough of Manhattan Community College, City University of New York, New York NY; United States of America.\\
$^{d}$ Also at Bruno Kessler Foundation, Trento; Italy.\\
$^{e}$ Also at Center for High Energy Physics, Peking University; China.\\
$^{f}$ Also at Centro Studi e Ricerche Enrico Fermi; Italy.\\
% ...

\end{flushleft}

% Created with Glance <Atlas.Glance@cern.ch>